\immediate\write18{makeindex \jobname.nlo -s nomencl.ist -o \jobname.nls} 

\documentclass[a4paper,fleqn]{cas-sc}

\def\tsc#1{\csdef{#1}{\textsc{\lowercase{#1}}\xspace}}
\tsc{WGM}
\tsc{QE}
\tsc{EP}
\tsc{PMS}
\tsc{BEC}
\tsc{DE}

\usepackage{amssymb}
\usepackage{amsmath} 
\usepackage{amsfonts} 
\usepackage{mathrsfs}
\usepackage{graphicx}
\usepackage{xcolor}
\usepackage{pst-plot,pst-text,pst-grad,pst-node,pst-coil,pst-slpe}
\usepackage{pstricks,pst-tree,pst-plot,pstricks-add,pst-func}
\usepackage{threeparttable}
\usepackage{multirow}
\usepackage{subfigure}
\usepackage{epsfig}
\usepackage{epstopdf}
\usepackage{nomencl}
\usepackage{makeidx}
\usepackage{framed} 
\usepackage[displaymath,mathlines]{lineno}
\usepackage{caption} 
\usepackage{mathtools}
\usepackage{cuted}
\usepackage{lipsum}
\usepackage{mwe}
\usepackage{fp}
\usepackage{placeins}
\usepackage[square,comma,numbers,sort&compress]{natbib}
\usepackage[titletoc,toc,title]{appendix}
\usepackage{flushend}
\usepackage{multicol}
\usepackage{cuted}
\usepackage{enumerate}
\usepackage[inline,shortlabels]{enumitem}
\usepackage{geometry}
\usepackage{lscape}
\usepackage{csquotes}
\usepackage{longtable}
\usepackage{textcomp}
\usepackage{tikz}
\usetikzlibrary{arrows,positioning,shapes.geometric,fit,calc,decorations.pathmorphing,backgrounds,matrix}
\usepackage{pgfplots} 
\usepackage{grffile}
\usepackage{auto-pst-pdf} 

\newcommand{\abf}{\mbox{\bf a}}

\newcommand{\ebf}{\mbox{\bf e}}

\newcommand{\qbf}{\mbox{\bf q}}
\newcommand{\pbf}{\mbox{\bf p}}

\newcommand{\xbf}{\mbox{\bf x}}

\newcommand{\Abf}{\mbox{\bf A}}
\newcommand{\Cbf}{\mbox{\bf C}}

\newcommand{\Fbf}{\mbox{\bf F}}
\newcommand{\Hbf}{\mbox{\bf H}}

\newcommand{\Ibf}{\mbox{\bf I}}

\newcommand{\Sbf}{\mbox{\bf S}}
\newcommand{\Xbf}{\mbox{\bf X}}

\newcommand{\sigbf}{\mbox{\boldmath{$\sigma$}}}
\newcommand{\taubf}{\mbox{\boldmath{$\tau$}}}

\newgray{lightlightgray}{0.975}
\newgray{LightLightgray}{0.875}
\definecolor{darkpastelgreen}{rgb}{0.01, 0.75, 0.24}
\definecolor{darkgreen}{rgb}{0.0, 0.2, 0.13}
\definecolor{GreenLM}{HTML}{7DBD7D}
\definecolor{GreenMM}{HTML}{008000}
\definecolor{Azure}{HTML}{007FFF}
\usepackage{etoolbox}
\renewcommand\nomgroup[1]{%
  \item[\bfseries
  \ifstrequal{#1}{A}{Acronyoms}{}
  \ifstrequal{#1}{P}{Symbols}{}%
  \ifstrequal{#1}{S}{Super- and subscripts}{}%
  \ifstrequal{#1}{O}{Other Symbols}{}%
]}

\makenomenclature
\begin{document}

\let\WriteBookmarks\relax
\def\floatpagepagefraction{1}
\def\textpagefraction{.001}
\shorttitle{Model selection and sensitivity analysis in the biomechanics of soft tissues: a case study on the human knee meniscus}
\shortauthors{Elmukashfi et al}
%
%
\title[mode = title]{Model selection and sensitivity analysis in the biomechanics of soft tissues:~a case study on the human knee meniscus}  
       
\author[1]{Elsiddig Elmukashfi}


\author[2]{Gregorio Marchiori}
\author[3]{Matteo Berni}
\author[4]{Giorgio Cassiolas}
\author[4]{Nicola Francesco Lopomo}
\author [7] {Hussein Rappel}
\author [7,8]{Mark Girolami}
\author[1,5,6]{Olga Barrera}
\cormark[1]
\fnmark[1]
\ead{obarrera@brookes.ox.ac.uk, olga.barrera@ndorms.ox.ac.uk}

\address[1]{Department of Engineering Science, University of Oxford, Parks Road, Oxford OX1 3PJ, UK}
\address[2]{IRCCS Istituto Ortopedico Rizzoli, Surgical Sciences and Technologies, Via di Barbiano 1/10, 40136 Bologna, Italy}
\address[3]{IRCCS Istituto Ortopedico Rizzoli, Medical Technology Laboratory, Via di Barbiano 1/10, 40136 Bologna, Italy}
\address[4]{Dipartimento di Ingegneria dell'Informazione, Universit\'{a} degli Studi di Brescia, Italy}
\address[5]{Department of Medical Research, China Medical University Hospital, China Medical University, Taichung, Taiwan}
\address[6]{School of Engineering, Computing and Mathematics, Oxford Brookes University, UK}
\address[7]{The Alan Turing Institute, UK}
\address[8]{Department of Engineering, University of Cambridge, UK}
%
%
\begin{abstract}
Soft tissues - such as ligaments and tendons - primarily consist of solid (collagen, predominantly) and liquid phases.
Understanding the interaction between such components and how they change under physiological loading sets the basis for elucidating the essential link between their internal structure and mechanical behaviour.
In fact, the internal heterogeneous structure of this kind of tissues leads to a wide range of mechanical behaviours, which then determine their own function(s). Characterising these behaviours implies an important experimental effort in terms of tissue harvesting, samples preparation and implementation of testing protocols - which, often, are not standardised. These issues lead to several difficulties in both collecting and providing comparable and reliable information. In order to model the behaviours of heterogeneous tissues and identify material parameters, a large volume of reproducible experimental data is required; unfortunately, such an amount of information is often not available. In reality, most of the studies that are focused on the identification of material parameters, are largely based on small sets of experimental data, which present a large variability. Such a large variability opens on to uncertainties in the estimation of material parameters, as reported in the literature. Hence, the use of a rigorous probabilistic framework, that is able to address uncertainties due to paucity of data, is of paramount importance in the field of biomechanics; in this perspective, Bayesian inference represents a promising approach. 
This study was focused on the analysis of the knee meniscus as a paradigmatic example of human soft tissue. Indeed, the heterogeneous internal architecture of this structure is linked to functionally graded material properties, which enable this fibrocartilaginous tissue to perform a wide range of functions within the knee joint. More in detail, within this work we specifically addressed: (\emph{i}) the variability of parameters for the meniscal non-fibrous, fibrous solid phase, as well as for the liquid one, (\emph{ii}) the material models currently used to interpret experimental data, (\emph{iii}) a comparative finite element study on the knee joint in which the meniscus is modelled by using several material models, (\emph{iv}) an outlook on Bayesian inference for the identification of material parameters, and models selection and comparison.
Our findings suggest that an accurate descriptions of the time-independent, time-dependent and spatial variability of soft tissues, such as the human meniscus, are essential to correctly define and develop any modelling solution. This work is relevant to the description of the physiological biomechanics of human menisci, and paves the way to generalise this approach to different soft tissues. 

\end{abstract}
%
%
            
%
%
%
%
%
\maketitle
%
    %
    \nomenclature[A]{$\mathrm{ACL}$}{Anterior cruciate ligament}
    \nomenclature[A]{$\mathrm{ECM}$}{Extracellular Matrix}
    \nomenclature[A]{$\mathrm{LM}$}{Lateral meniscus}
    \nomenclature[A]{$\mathrm{MM}$}{Medial meniscus}
    \nomenclature[A]{$\mathrm{MRI}$}{Magnetic resonance imaging}
    \nomenclature[A]{$\mathrm{OA}$}{Osteoarthritis}
    \nomenclature[A]{$\mathrm{GAG}$}{Glycosaminoglycan}
    \nomenclature[A]{$\mathrm{PG}$}{proteoglycan}
    \nomenclature[A]{$\mathrm{CT}$}{Computer tomography}
    \nomenclature[A]{$\mathrm{FTIR}$}{Fourier Transform Infrared Spectroscopy}
    \nomenclature[A]{$\mathrm{DD}$}{Digital densitometry}
    \nomenclature[A]{$\mathrm{qPCR}$}{Quantitative polymerase chain reaction}
    \nomenclature[A]{$\mathrm{BI}$}{Bayesian inference}
    \nomenclature[A]{$\mathrm{MCMC}$}{Markov chain Monte Carlo}
    \nomenclature[A]{$\mathrm{MAP}$}{maximum a posteriori}
    \nomenclature[A]{$\mathrm{PPD}$}{posterior predictive distribution}
    \nomenclature[P]{$\Xbf, \, X_{i}$}{the Cartesian material coordinates}
    \nomenclature[P]{$\xbf, \, x_{i}$}{the Cartesian spatial coordinates}
    \nomenclature[P]{$\Fbf, \, F_{ij}$}{deformation gradient}
    \nomenclature[P]{$\Cbf, \, C_{ij}$}{right Cauchy-Green deformation tensor}
    \nomenclature[P]{$\sigbf, \, \sigma_{ij}$}{Cauchy stress tensor ($\mathrm{N}/\mathrm{m}^{2}$)}
    \nomenclature[P]{$\taubf, \, \tau_{ij}$}{Kirchhoff stress tensor ($\mathrm{N}/\mathrm{m}^{2}$)}
    \nomenclature[P]{$\Psi$}{strain energy density function}
    \nomenclature[P]{$U$}{volumetric part of strain energy density function}
    \nomenclature[P]{$J$}{Jacobian of the deformation gradient}
    \nomenclature[P]{$\lambda_{i}$}{principal stretches of the deformation gradient  ($i=1,2,3$)}
    \nomenclature[P]{$I_{i}$}{principal stretches of the deformation gradient  ($i=1,2,3, \cdots$)}
    \nomenclature[P]{$\mu$}{shear modulus ($\mathrm{N}/\mathrm{m}^{2}$)}
    \nomenclature[P]{$\nu$}{Poisson's ratio}
    \nomenclature[P]{$E$}{Young's modulus ($\mathrm{N}/\mathrm{m}^{2}$)}
    \nomenclature[P]{$k$}{Permeability ($\mathrm{m}^4/\mathrm{N} \, \mathrm{s}$)}
    \nomenclature[P]{$\bar{g}_{i}$}{shear modulus in viscoelastic law}
    \nomenclature[P]{$\bar{k}_{i}$}{bulk modulus in viscoelastic law}
    \nomenclature[P]{$\eta_{i}$}{characteristic retardation time constant in viscoelastic law ($\mathrm{s}$)}
    \nomenclature[P]{$\bar{g}_{i}$}{shear compliance in viscoelastic law}
    \nomenclature[P]{$\bar{k}_{i}$}{bulk compliance in viscoelastic law}
    \nomenclature[P]{$\textbf{z}$}{measurement set}
    \nomenclature[P]{$\textbf{p}$}{parameter set}
    \nomenclature[P]{$\pi(\cdot)$}{probability density function}
    \nomenclature[P]{$n_{m}$}{number of measurements}
    \nomenclature[P]{$n_{p}$}{number of parameters}
    \nomenclature[P]{$\mathcal{M}$}{model class}
    \nomenclature[S]{$\bar{\left(\bullet \right)}$}{isochoric property}
    \nomenclature[S]{${\mathrm{iso}}$}{isotropic property}
    \nomenclature[S]{${\mathrm{aniso}}$}{anisotropic property}
    \nomenclature[S]{$0$}{initial (instantenous) property}
    \nomenclature[S]{${\mathrm{eq}}$}{equilibrium property}
    \nomenclature[S]{${\mathrm{f}}$}{fibre property}
    %
%
%
\section{Introduction}

\begin{figure}[!ht]
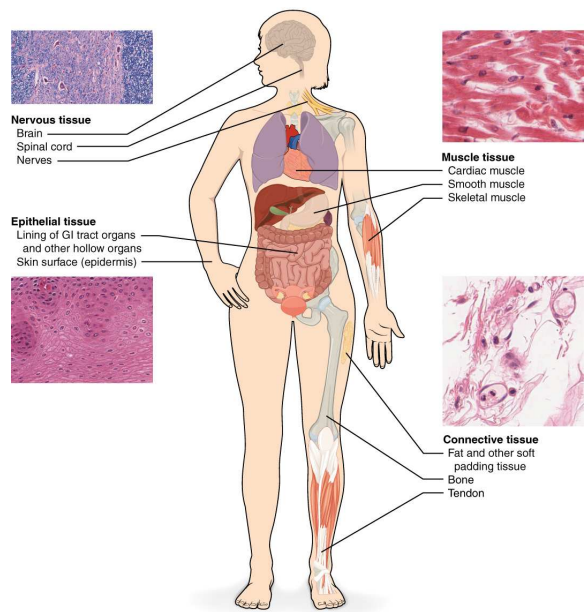

\begin{center}
\include{Types_of_Tissue}
\end{center}
\caption{Anatomy of the human body where the four primary soft tissue types are illustrated, i.e. nervous tissue, epithelial tissue, muscle tissue, and connective tissue (figure adapted from ~\cite{Biga2020}).}
\label{fig:Tissues}
\end{figure}

Within the human body, the role of soft tissues is to comprise, envelope, support and connect organs \cite{Biga2020} (see, for examples, Fig.~\ref{fig:Tissues}). From the anatomical point of view, soft tissues are in general classified into four groups based on the role they play in making the body works: \emph{connective tissue} - connects various parts of the body and provides support and protection to organs; \emph{muscle tissue} - allows the body to move; \emph{epithelial tissue} - forms the exterior surfaces of the body; \emph{nervous tissue} - propagates information. Due to these wide variety of roles, soft tissues often exhibit a non-unique structure allowing them to perform a range of different functions. The main constituents of these tissues comprise of a porous solid (predominantly composed of abductin, resilin, elastin and collagen) and a liquid phase. 

Our working hypothesis, which is largely supported by the constitutive modelling literature, is that the understanding of structure-function relationships of soft tissues is the key to make a breakthrough in both improving the knowledge of their inherent functional behaviour and in developing biological implants able to repair, maintain or improve tissue functions. Such a connection between the fine scales (i.e.~pores, holes, inclusions, channels, and other heterogeneities) can also be the basis to develop more informed phenomenological models at the tissue scale and, therefore, to further understand the functions of human tissues. 

Under this hypothesis, there is no ``one-size-fits-all'' approach able to model tissues at the organ scale. Therefore, the first question to ask concerns the choice of the ``quantity of interest'', i.e. the macroscopic observable quantity which is of use to the researcher or practitioner. For example, a neurosurgeon may be interested in the fracture resistance of certain blood vessels during an intervention, whilst an interventional radiologist may need to know how the position of the target within the liver is displacing due to the breathing motion of the patient. When performing laser ablation, the temperature field within the tissue is critical to decide upon the intensity of the laser and the duration of the pulse. Other treatments, such as the high intensity focused ultrasound, require the understanding of acoustic pressure waves within the organ. The simulation of medical imaging such as magnetic resonance may also necessitate the prediction of electromagnetic wave propagation within the body. An excellent review of these points is provided in Ref.~\cite{YohanPayanbook}.
Therefore, modelling tissues for medical applications requires taking into account a large number of physical phenomena and processes, including solid, fluid and poro-mechanics, heat transfer, acoustics and electromagnetics. Once the different types of ``physics'' have been identified, mathematical models have to be devised, so as to predict the behaviour of the tissue depending on boundary and initial conditions. It is worth noting that those conditions were reported, for selected cases in biomechanics, to be more impactful than the choice of the mathematical models themselves (see, for example, what is reported in Ref.~\cite{christianDuriezBC}). 

Selecting the most suitable model for a given quantity of interest demands understanding the physics of the problem as well as the clinical uses that are expected. For example, a linear elastic model might be sufficient as long as the deformation of the tissue remains within the small range. For larger deformations, phenomenological models such as those commonly used in hyperelasticity may be better suited. Phenomenological models take into account the physical phenomena present at the ``fine'' scale to describe the behaviour of the coarse scale in a mechanistic sense, e.g.~by adding required terms to a strain energy function. Focusing on biological tissues, literature reports the inclusion of fibre extension, bending, sliding within a hyperelastic model of arteries \cite{doi:10.1098/rsif.2005.0073} or of the skin \cite{Rolin2012}. Several microscopic phenomena can also be included in phenomenological models of the brain \cite{10.1371/journal.pone.0254512}.

Once the key mechanistic phenomena influencing the quantity of interest have been identified, the main difficulty associated with phenomenological models is the identification of the relevant parameters involved in expressing the constitutive model, such as - for example - Young's modulus and Poisson's ratio in elasticity, heat transfer coefficient in thermoelasticity, or permeability in poroelasticity.
Due to the large inter- and intra-subject variability in tissue behaviour, it is not feasible to use predefined sets of tissue parameters. Instead, the actual trend is to tackle this difficulty by using probabilistic approaches in the estimation of the distributions of each parameter, as we discuss in detail below. 

Alternatives to phenomenological models are multiscale models, which take into account the fine scale behaviour, and its evolution, explicitly \cite{GorielyKuhl,GORIELY201579}. The perceived advantage of such models is that complexity is built from the bottom up. Their underlying assumption is that material variability decreases as the scale decreases. Whilst this can indeed be true, geometrical complexity increases as the scale decreases, which makes simulations extremely demanding, as we can underline, for example, in the definition of finite element mesh to conform to discontinuities across material interfaces, voids, cracks, etc. Approaches based on advanced discretisation techniques are often used, among others, to simplify the treatment of geometrical complexities (see, for example, Refs.~\cite{https://doi.org/10.1002/nme.4823,BANSAL201945,Farina}).
What is more important, multiscale models may require the concurrent solution of a fine and coarse scale problem, which is indeed computationally demanding. The interested reader may refer to recent works which are focused on novel advances in multi-scale modelling Refs.~\cite{sonon2021advanced, aifantis2016internal, ostoja2016scaling}; the following review paper may also be of interest \cite{Talebi}. Further solutions, used to reduce computational expenses in those non-linear multi-scale models, include (Bayesian) model order reduction \cite{goury2016automatised,kerfriden2013partitioned} and machine learning approaches \cite{beex2011quasicontinuum,wilbrink2013discrete}.

Once a model has been selected, the most influential parameters must be identified. This step is usually done by implementing a methodological solution that is known as ``forward uncertainty quantification''. Following this method, each parameter is given a statistical distribution which has to be identified by using inverse uncertainty quantification, via either frequentist or Bayesian approach, as we discuss later in the manuscript. Forward Monte Carlo simulations are performed in order to elucidate the relative importance (and correlation) of the parameters. Scientific literature reports a large variety of methods used to accelerate such simulations, focusing on different applications, as - for example - to brain biomechanics, pathologies and models \cite{hauseux2017accelerating,hauseux2018quantifying,10.1371/journal.pone.0189994,10.1371/journal.pone.0254512}. The mathematical model is then discretised using numerical methods, and the discretisation error is measured in order to optimise the ``mesh'' for a given quantity of interest. The importance of error estimation in biomechanics was reported in several studies \cite{duprez2020quantifying,elouneg2021open}. In \citet{duprez2020quantifying}, error estimation techniques were put forward in order to optimise the (extended) finite element mesh in real time to minimise the error on quantities of interest relevant to needle insertion in the brain.

In this picture, meniscal tissue represents an example of soft tissue in which several physical phenomena take place, where geometrical and material complexity (i.e.~presence of channels, inclusions, pores, etc.) lead to functionally graded material properties. This type of tissue sits between tendons and cartilage and makes it an ideal test case to describe a modelling pipeline, we aim to reliably use in soft tissue biomechanics. Furthermore, we consider here the meniscal tissue in order to shed light on the process required to link internal architecture, mechanical behaviour and model selection of soft tissues. Their in-homogeneous internal structure, which varies spatially, and the observed hierarchical arrangement of the collagen material result in a mechanical behaviour that is complex to be mathematically modelled by using the ``appropriate'' set of parameters. Thus, the key is to relate the mechanical response to the microstructure and its evolution to set the basis for developing physically motivated constitutive models, including a suitable set of parameters. The identification of model parameters is then obtained from a set of experimental data by using several different methods. 
One of the most common approaches is the method of least squares \cite{A_Bjorck_1996}. Following this approach, the value of material parameters is obtained by minimising the sum of squared residuals, which represent the difference between experimental observations and model predictions. The unknown parameters are determined along with confidence intervals\footnote{An $n\%$ confidence interval is a range of values calculated from observed samples which is believed to contain the true parameter with $n\%$ probability. In other words, an $n\%$ confidence interval implies that if the estimation process is repeated again and again, then $n\%$ of the calculated intervals would contain the true parameter values. This probability level is a characteristic of the interval and \emph{not} the parameter itself which not considered to be a random variable \cite{B_Everitt_2010}.}, which are obtained assuming normally distributed data. The errors in the observations are implicitly symmetrically distributed around zero with constant variance. Hence, when the size of errors varies and depends on several unknown parameters, least squares methods are no longer appropriate \cite{Seber2003}.

An alternative approach is to use Bayesian inference. In a Bayesian framework, unknown parameters are treated as random variables with a joint probability distribution that provides the probabilities of the occurrence of various possible outcomes of an experiment \cite{J_Beck_2010}. Treating the unknowns as random variables allows the user to make probabilistic statements about the unknown parameters, based on the prior knowledge and the information contained in the measured data. For example, if the user's goal is to collect data from a mechanical experimental test and estimate Young's modulus $E$, the Bayesian framework allows him/her to make statements such as:~``There is a $n\%$ chance that the Young's modulus lies in the interval $[E_{1},E_{2}]$''. This interval is known as the credible interval \cite{W_Edwards_1963}. Making a similar statement in an identification framework based on least squares methods is not possible as the unknown parameters are treated as deterministic values and not random variables.
Additionally, Bayesian frameworks can assess how adequate different models are in relation to their compatibility with measurements. In other words, a Bayesian framework gives the possibility to address uncertainties due to insufficient information including identification of model parameters and model suitability. 

\begin{figure}[!ht]
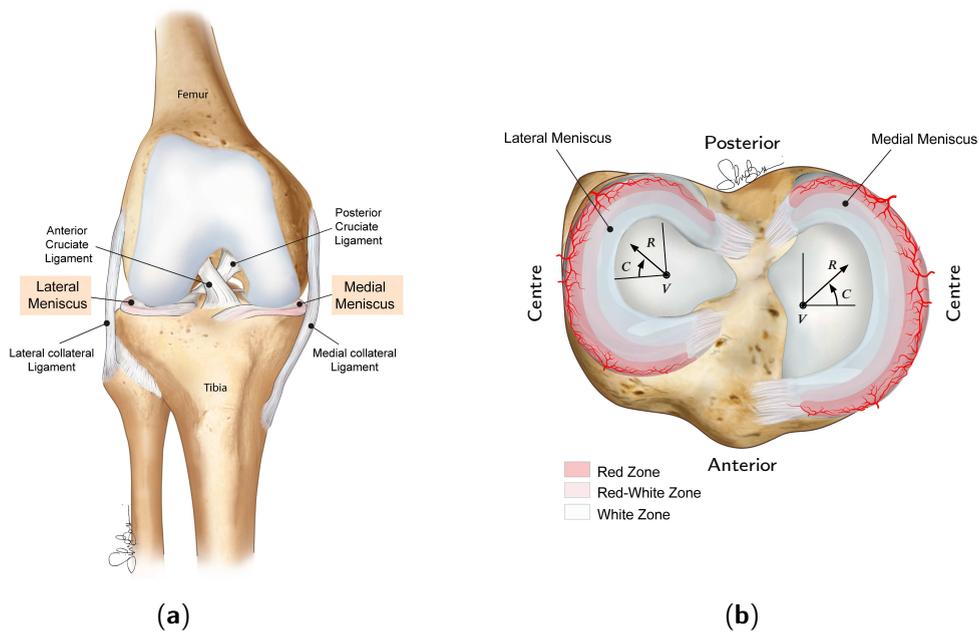

\begin{center}
\include{Knee_Joint_Anatomy}
\end{center}
\caption{Anatomy of the knee joint: \textbf{a} frontal view of the knee joint; and \textbf{b} top view of the menisci which can be divided into the anterior horn (Ant), central body (Cent) and posterior horn (Post) regions with their relative vertical (V), circumferential (C) and radial (R) directions. Along the radial direction: the outer zone: red zone (vascular), the intermediate region: red-white (semi vascular) and the internal region: white (avascular). }
\label{fig:KneeJoint}
\end{figure}

The meniscus can be anatomically identified as a tough crescent-shaped fibrocartilaginous soft tissue that conforms to the surfaces of the femur and the tibia within the synovial diarthrodial joint. In the human knee, there are two menisci, corresponding to the lateral and medial compartments of the tibiofemoral joint (see Fig.~\ref{fig:KneeJoint}). 
The meniscal architecture is functionally graded to accommodate loading and kinematic constraints which are not uniform throughout the tissue, also constituents exhibit a non-uniform arrangement along different directions leading to a macroscopic anisotropic behaviour. Therefore, a complete mechanical characterisation of these structures should investigate time, regional and directional dependent phenomena produced by the crosstalk of their phases. A comprehensive mechanical description of human menisci is still lacking, especially considering different tissue regions – {i.e.}~anterior horn, central body and posterior horn – and directions – {i.e.}~vertical, circumferential and radial - and vascularity- vascular (red), semi vascular (red-white) and avascular (white).
From the functional point of view, menisci have been widely reported to play a fundamental role in the knee biomechanics, since they are able to equalise the load distribution by increasing the contact area between the different articulating surfaces, and to contribute to the overall joint stability \citep{Fithian1990,Aagaard1999,Allen2000}. These functions are favoured by the inhomogeneous composition and structure of this tissue \citep{Danso2017}. 
%
%
From the composition perspective, each meniscus generally consists of solid and liquid phases. More in details, the liquid phase is mainly water ($60$-$70\%$), whereas the solid phase consists primarily of an extracellular matrix (ECM), which is principally composed of collagen type I ($15$-$25\%$), interposed with chondrocyte cells \citep{Fox2012}. Proteoglycans (PGs), glycoproteins and non-collagenous proteins account for the remaining dry weight. 
%
%
The crosstalk between these phases rules the time-dependent behaviour of the tissue \citep{Seitz2013}. For example, by characterising material parameters such as permeability and elastic modulus, the contributions of the fluid and solid components can be usually underlined \citep{Danso2015}. 
%
%
On the other hand, Finite Element Modelling (FEM) is an important tool for analysing knee biomechanics in a pre-clinical – not invasive – manner. There is a limited number of papers in the literature presenting finite element models of the knee joint which include the menisci \citep{Pena2005a,Naghibi2020}.  
%
%
In general, the specific behaviour of the meniscal tissue can be modelled as visco-elastic \citep{Chia2008} or explicitly as biphasic/poro-elastic \citep{Leroux2002}, when considering a biphasic composition ({i.e.}~a solid matrix interspersed by a liquid \citep{Leslie2000}). These characteristics are usually shared among cartilage and menisci \citep{Higginson1976,Moore2016,Taffetani2014}, though for the latter are less described in literature. However, menisci should require specific attention standing their peculiar fibro-cartilaginous structure \citep{Herwig1984}.
These are related to the fact that the material properties change spatially (from the posterior horn/central body/anterior horn) and along the three directions (vertical, radial and circumferential) shown in Fig.~\ref{fig:KneeJoint}\textbf{b}.
Therefore, a model of the meniscus should contain the following ingredients:
\renewcommand{\labelenumi}{\Roman{enumi}.}
\setlength\itemsep{0.5em}
\begin{enumerate}[leftmargin=*,align=left]
  \item The \emph{time-independent} nonlinear elastic stress-strain response.
  \item The \emph{time-dependent response} of the tissue and biphasic solid/fluid nature.
  \item Description of the structural organisation of the meniscus tissue in terms of the \emph{spatial and directional variability} of the material properties.
\end{enumerate}
It is indeed difficult to find comprehensive works in the literature which include the three points above. Some studies focused on tissue anisotropic behaviour but did not characterise the time-dependent phenomena \citep{Leslie2000,Chia2008}, others included the time-dependent behaviour only for the vertical direction \citep{Bursac2006,Moyer2013}. A number of findings is related to the medial meniscus \citep{Kessler2015,Quiroga2014}, others to the lateral one \citep{Yan2015}. Investigations of the behaviour of the tissue in the three main regions of the meniscus (central body, anterior and posterior horns) are reported separately in \citet{Sandmann2013} (central body) and in \citet{Leslie2000} and \citet{Fischenich2015} (anterior and posterior horns). Collecting and using data from different studies is not a straightforward process due to the variation in material parameters mainly due to the lack of standards for testing procedures and material models. 
%
\\
The goal of this paper is to investigate and discuss model selection and sensitivity analysis in biomechanics by using the meniscus as a case study through: 
\renewcommand{\labelenumi}{\Roman{enumi}.}
\setlength\itemsep{0.5em}
\begin{enumerate}[leftmargin=*,align=left]
  \item giving an up-to-date overview of the functions of the meniscus and its internal architecture;
  \item describing the current state of the art of the experimental characterisation of the meniscal tissue;
  \item investigating the impact of different material models for the menisci; and
  \item presenting an outlook on Bayesian inference for parameter identification and model comparison as an approach to address uncertainties (e.g.~uncertainties due to an insufficient volume of experimental data).
\end{enumerate}
Therefore, first, we highlight shortfalls and opportunities in obtaining a complete and coherent material mechanical description, especially focusing on time-dependency, inhomogeneity and anisotropy. Second, we examine the importance of the ingredients needed to model the meniscus mechanical behaviour. Hence, we consider the range of constitutive models that are commonly used in the literature and classify them into: (\emph{i}) linear vs nonlinear elastic; (\emph{ii}) isotropic vs anisotropic elastic; and (\emph{iii}) time-independent vs time-dependent models. The same set of experimental data from the literature \citep{Proctor1989} was adopted to estimate material parameters using a number of models. To compare these models, we analyse a subject specific knee under quasi-static and dynamic loading conditions using the finite element method. The knee model is evaluated in terms of the contact pressure and $1^{\mathrm{st}}$ principal stress of the tibial cartilage; the deformation of the menisci; and the normal strain to menisci tearing paths ({i.e.}~the factors affect knee osteoarthritis (OA)). 

The outline of the paper is as follows:~the first four sections provide a review of the meniscus functions, internal architecture and poro-mechanics experimental data. Sections~\ref{sec:PhyMod} and \ref{sec:ConstitDes} focus on the influence of the meniscus models. The computational model is described in Section~\ref{sec:NumMod} in which boundary conditions, interactions between the knee parts and their material models are discussed. The analyses of the different meniscus material models are presented and discussed in Section~\ref{sec:AnalyRes}. Our comments on the methods for determination of material models are given in Section~\ref{Conc}. Section~\ref{sec:ParamIden} presents a discussion on Bayesian inference as an approach based on probability logic for parameter identification and model comparison. Finally this contribution is closed with conclusions (Sec.~\ref{Discussion}).

%
%
\section{Background}
\label{sec:Bacgrnd}
%
%
\subsection{Meniscus functions}

The scientific literature reports that - due to its great resilience, endurance and wear resistance - the human meniscus evolved to fulfil specific biomechanical roles within the knee. Specifically: 
\renewcommand{\labelenumi}{\Roman{enumi}.}
\renewcommand{\labelenumii}{(\alph{enumii})}
\setlength\itemsep{0.5em}
\begin{enumerate}[leftmargin=*,align=left]
  \item \emph{Transmitting loads to the articular cartilage} - The mobile menisci transfers up to $90\%$ of the load over the knee joint, depending on the flexion angle, femoral translation and rotation with respect to tibia \citep{Aagaard1999}. This function has been hypothesised to be accomplished by the enlarged tibio-femoral contact area and decreased contact stress during load. This functional behaviour of the meniscus has been often qualitatively inferred by analysing the degenerative changes that follow its removal, leading to osteoarthritis. Specifically, \citet{Radin1984} demonstrated that the loads within the joint are well distributed only when the menisci are intact. Indeed, removing the medial meniscus leads up to $70\%$ reduction of the femoral condyle contact area and a $100\%$ increase in contact stress \citep{Fukubayashi1980}. Although these studies provide useful insight on the role of the meniscus in load transferring, they present several limitations since:
  \begin{enumerate}[leftmargin=*,align=left]
  \item they were often carried out only qualitatively;
  \item they were usually based on in-vitro analyses focused on measuring contact pressure by using pressure sensitive films, which presented several drawbacks including low precision;
  \item they lack information about measurements of the loads transmitted from the femoral condyles.
  \end{enumerate}
  Finite element modelling (FEM) represents indeed a good candidate to explore the extent of the load-bearing capacity that the meniscus provides. Unfortunately, modelling the mechanical behaviour of menisci is usually based on ``ideal'' hypotheses \citep{Andrews2017}, since models are developed by considering material properties that are experimentally estimated without considering their dependence on location, specimen size, and orientation. Moreover, material properties are highly scattered in the literature. Finally, these models are validated against imprecise measurements \citep{Andrews2017}. 
  \item \emph{Absorbing impacts} - The literature widely reports that the menisci are crucial in attenuating the shock waves generated by intermittent impact loading. Several works demonstrated that, in the absence of the menisci, the shock-absorbing capacity decreases by about $20\%$ \citep{Voloshin1983}. However, the shock absorbing function is still debated \citep{Bursac2009,Andrews2011,Andrews2017}. This argument is supported by experimental dynamic testing simulating physiological activities on human menisci, which reported a low dynamic phase angle at the frequencies associated with these specific activities, corresponding to a non-significant hysteretic behaviour of these structures ({i.e.}~energy absorbing). Furthermore, recent analyses of the literature highlighted that the menisci actually present little capacity in energy absorption \citep{Bursac2009,Andrews2011}. 
  \item \emph{Contributing to the overall joint stability} - The geometry of the meniscus plays an important role for maintaining joint congruity and stability. Furthermore, the literature reports an increase in both anterior-posterior and rotational knee laxity, when menisci are removed \citep{Allen2000}. In fact, menisci can support the constraining action performed by specific ligamentous structure, such as the anterior cruciate ligament (ACL); the joint stability is indeed extremely compromised in the absence of ACL \citep{Sandmann2013}. Scientific works demonstrated that, in the cases of ACL deficiency, the posterior horn of the medial meniscus can carry a force that is developed in the anterior direction of the tibia. Further studies showed that, in the cases of ACL-deficient knee, the resultant force in the medial meniscus increased by $52\%$ in full extension and by $197\%$ at $60^{\circ}$ of flexion \citep{Skaggs1994}. These studies highlighted that the large changes in kinematics is due to medial meniscectomy in ACL-deficient knees. Hence, while the synergy between menisci and ACL seems relevant to providing joint stability, it is not entirely clear what is the role of the meniscus in stabilising the knee joint.
  \item \emph{Nurturing and lubricating the synovial joint} - Human meniscus is believed to play an important role in nurturing and lubricating the knee joint. However, the inherent mechanics underlying these two functions remains unknown \citep{Fox2012}. The presence of synovial fluid may be important in reducing friction during loading. From the nurturing perspective, we can suppose that menisci contribute to compress synovial fluid into the articular cartilage. Thus, mechanical loading during physical activities might trigger nutrients and lubricant fluid transport in the meniscus. 
\end{enumerate}
%
%
%
\subsection{Meniscus internal architecture}

\begin{figure}[!ht]
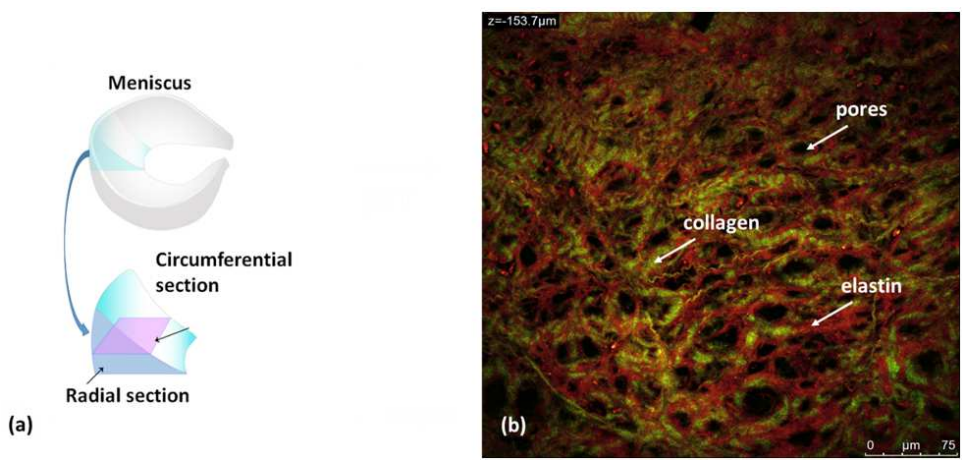

\begin{center}
\include{Meniscus_Structure}
\end{center}
\caption{The meniscus structure: (\textbf{a}) Radial and circumferential section of the meniscus (adapted from \citet{Vetri2019}), (\textbf{b}) multiphoton microscopy showing collagen (green signal), elastin (red signal) and pores (black) of dimensions $10$-$40~\mu\mathrm{m}$.}
\label{fig:MenSec}
\end{figure}

Proteoglycans (PGs) are hydrophilic molecules consisting of a large, highly negatively charged central protein and are attached to multiple glycosaminoglycan (GAG) chains. They attract and hold water, which is incompressible, and so are temporarily resistant to load before the water flows out the glycosaminoglycan chains. Therefore, their purpose is to keep the extracellular matrix (ECM) hydrated as well as to provide the load bearing capacity against compressive forces. Consequently, their concentration is the highest within both the ``radially inner'' and within the ``horns'' of the meniscus which are associated with the highest load \citep{Fox2012}. The purpose of the glycoproteins is to provide the adhesion between the different molecules within the extracellular matrix (ECM). Although there are multiple different types of molecules present, they are thought to each have different functions, which are presently still being discovered. Non-collagenous proteins within the dry phase include fibronectin and they help to facilitate repair of tissue, clotting of blood and adhesion of cells. Finally, the elastin within the dry phase is thought to provide the resilience of the tissue, which is crucial for the meniscus as it struggles to heal itself due to its minimal vascularisation \citep{Fox2012}.
\\
More than twenty-years ago Scanning Electron Microscopy (SEM) on human meniscus samples revealed, for the first time, the structure and orientation of the collagen fibers \citep{Petersen1998}. Recently, further studies based even on novel micro-imaging techniques ({e.g.}~light and electron microscopic and optical projection tomography) were conducted to elucidate the complex structure of the knee menisci \citep{Rattner2011,Andrews2013}. In particular, new insight on the arrangement of the circumferential and radial sections (see Fig.~\ref{fig:MenSec}) of fixed and dehydrated human meniscal samples has been highlighted \citep{Rattner2011}. Analysis of the radial section showed that collagen fibre bundles divide the body of the meniscus into a honeycomb network which is filled with collagen fibrils (whose diameter is about $5$~$\mu\mathrm{m}$), running in the circumferential direction (see Fig.~\ref{fig:MenSec}) \citep{Vetri2019}.
\\
Such preferential arrangement and orientation of collagen fibers was hypothesised to be the main feature ruling the load bearing capacity of the meniscus. Despite the significant and innovative findings reported in these papers, it is worth noting that the reported results are limited by the technique used for the analysis; more precisely, the sample preparation including fixation, mechanical and chemical peeling, dehydration and metallisation, could have altered the samples themselves, falsifying the obtained outcomes. These procedures have been indeed proven to change to a greater or lesser extent the tissue structure and unavoidably limit the quantitative aspect of the results \citep{Georgiadis2016}. Environmental Scanning Electron Microscopy (ESEM) and confocal Multi-Photon (MP) fluorescence/second harmonic generation microscopy have been recognised to be valuable methods to analyse the architecture of living tissues, without the need for tissue processing. MP microscopy enables, for example, an accurate microscale 3D reconstruction of the main structural proteins of living tissues, {i.e.}~collagen and elastin \citep{Zipfel2003,Brockbank2008,Schenke2008,Georgiadis2016,Vetri2019}. Despite the difficulty in distinguishing the collagen and elastin signals, ESEM technique has been demonstrated to be suitable as it allows 3D imaging of tissue structure down to the nanoscale \citep{Vetri2019}. Recently, a qualitative and quantitative experimental study, coupling detailed information from ESEM and MP microscopy on non-manipulated samples \citep{Vetri2019}, showed a novel meniscal micro-nano structure, which is essentially composed of a network of collagen channels through which the fluid flows when the knee is under physiological loading. The architecture has been also imaged on a similar scale by means of micro CT scans \citep{Agustoni2021}, revealing architectural features such as porosity, collagen channels diameters and tortuosity of these channels.
%
Summarising, the literature seems to suggest that:
\renewcommand{\labelenumi}{\Roman{enumi}.}
\setlength\itemsep{0.5em}
\begin{enumerate}[leftmargin=*,align=left]
  \item the meniscus is a highly porous "cushion" made of macro/micro/nano channels of collagen fascicles through which fluid flows \citep{Vetri2019, Agustoni2021};
  \item the tissue is functionally graded both in the three direction; graded mechanical parameters were linked with the grading architecture in the body region of the meniscus \citep{Bonomo2020,Maritz2020,Maritz2021}.
\end{enumerate}
These points are fundamental to understand the role of the pore pressure in carrying the load. 
%
%
%
\section{Review of experimental poro-mechanical data}
\label{sec:Expdata}
%
%
\subsection{Reviewing methods}

\begin{figure}[!ht]
\begin{center}
%
%
\begin{pspicture}[showgrid=false](0.0cm,0)(13.0cm,13.0cm)
%
%
\rput(6.0,0){
%
%
%
%
\rput(-2.250,11.50){\rnode{A}{\psframebox[fillstyle=none,fillcolor=orange!30,shadow=false,framesep=5pt,linewidth=0.25pt]{\parbox[c][1.75cm]{3.0cm}{\small \centering Records identified through database (PubM) searching ($n=44$)}}}}
\rput(2.25,11.50){\rnode{B}{\psframebox[fillstyle=none,fillcolor=orange!30,shadow=false,framesep=5pt,linewidth=0.25pt]{\parbox[c][1.75cm]{3.0cm}{\small \centering Records identified through database (WoS) searching ($n=56$)}}}}
\rput(0,9.00){\rnode{C}{\psframebox[fillstyle=none,fillcolor=orange!30,shadow=false,framesep=5pt,linewidth=0.25pt]{ \parbox[c][0.8cm]{5.50cm}{\small \centering Records after duplicates removed ($n=66$)}}}}
\rput(0,7.25){\rnode{D}{\psframebox[fillstyle=none,fillcolor=green!30,shadow=false,framesep=5pt,linewidth=0.25pt]{\parbox[c][0.8cm]{3.50cm}{\small \centering  Records screened ($n=66$)}}}} 
\rput(0,5.25){\rnode{E}{\psframebox[fillstyle=none,fillcolor=blue!50,shadow=false,framesep=5pt,linewidth=0.25pt]{\parbox[c][1.2cm]{3.50cm}{\small \centering  Full-text articles assessed for eligibility ($n=55$)}}}}
\rput(0,3.00){\rnode{F}{\psframebox[fillstyle=none,fillcolor=blue!50,shadow=false,framesep=5pt,linewidth=0.25pt]{\parbox[c][1.2cm]{3.50cm}{\small \centering  Studies included in mechanical synthesis ($n=24$)}}}}
\rput(0,0.75){\rnode{G}{\psframebox[fillstyle=none,fillcolor=blue!50,shadow=false,framesep=5pt,linewidth=0.25pt]{\parbox[c][1.2cm]{3.50cm}{\small \centering  Studies included in composition/sturcture synthesis ($n=15$)}}}}
\rput(5.0,7.25){\rnode{H}{\psframebox[fillstyle=none,fillcolor=blue!50,shadow=false,framesep=5pt,linewidth=0.25pt]{\parbox[c][0.80cm]{3.50cm}{\small \centering  Records excluded (Review) ($n=11$)}}}}
\rput(5.0,5.25){\rnode{I}{\psframebox[fillstyle=none,fillcolor=blue!50,shadow=false,framesep=5pt,linewidth=0.25pt]{\parbox[c][1.90cm]{3.50cm}{\small \centering Full-text articles excluded ($n=31$) because not testing mechanics of human meniscal tissue}}}}
\rput(0,-0.75){\rnode{EN}{}}
\ncangle[angleA=-90,angleB=90,loopsize=0.5,nodesepA=0,nodesepB=0,armB=0.50cm,linearc=.0]{->}{A}{C}
\ncangle[angleA=-90,angleB=90,loopsize=0.5,nodesepA=0,nodesepB=0,armB=0.50cm,linearc=.0]{->}{B}{C}
\ncline{->}{C}{D}
\ncline{->}{D}{E}
\ncline{->}{E}{F}
\ncline{->}{F}{G}
\ncline{->}{D}{H}
\ncline{->}{E}{I}
\rput{90.0}(-5.25,1.0){\psframebox[fillstyle=solid,fillcolor=cyan!50,shadow=false,framesep=5pt,linewidth=0.25pt,linearc=5.0pt,cornersize=absolute]{\parbox[c][0.5cm]{2.0cm}{\small \centering Included}}}
\rput{90.0}(-5.25,4.5){\psframebox[fillstyle=solid,fillcolor=cyan!50,shadow=false,framesep=5pt,linewidth=0.25pt,linearc=5.0pt,cornersize=absolute]{\parbox[c][0.5cm]{2.0cm}{\small \centering Eligibility}}}
\rput{90.0}(-5.25,8.0){\psframebox[fillstyle=solid,fillcolor=cyan!50,shadow=false,framesep=5pt,linewidth=0.25pt,linearc=5.0pt,cornersize=absolute]{\parbox[c][0.5cm]{2.0cm}{\small \centering Screening}}}
\rput{90.0}(-5.25,11.5){\psframebox[fillstyle=solid,fillcolor=cyan!50,shadow=false,framesep=5pt,linewidth=0.25pt,linearc=5.0pt,cornersize=absolute]{\parbox[c][0.5cm]{2.0cm}{\small \centering Identification}}}
}
%
\end{pspicture}
%
%
\end{center}
\caption{The flow chart of the systematic literature review. The chart shows the different steps of the reviewing process which includes identification, screening and checking eligibility of records in the databases and then reviewing the final collection of included records.}
\label{fig:FolChar}
\end{figure}
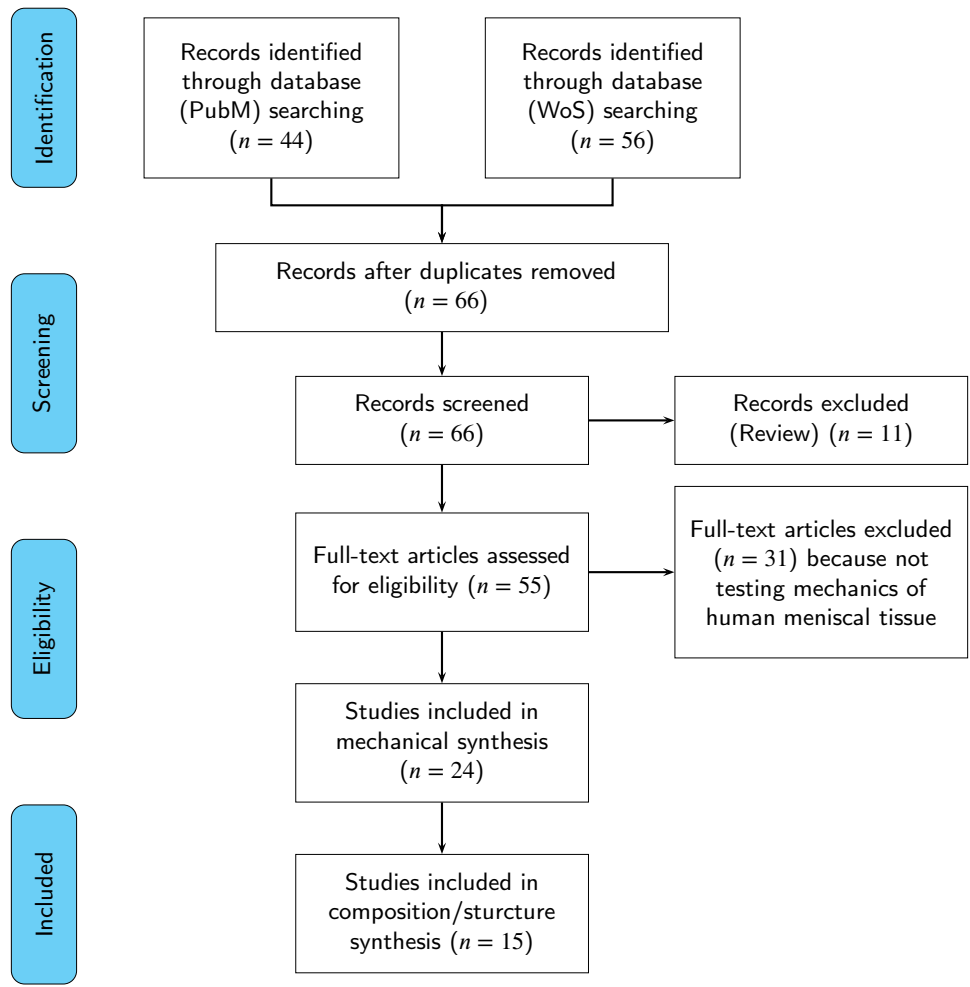

The search was performed through on-line databases, specifically, including Web of Science (WoS) and PubMed (PubM), and verified on SCOPUS. Search was conducted including the following search string: 
%
%
\begin{center}
\begin{minipage}[t]{0.9\textwidth}
\emph{[meniscus] AND [human] AND [viscous OR viscoelastic OR poro OR poroelastic OR biphasic]}.
\end{minipage}
\end{center}
%
%
\textcolor{black}{Authors performed the search}, independently, and a preliminary screening excluding Review articles was done. Then, articles were considered eligible if describing mechanical testing and explicitly considering human meniscus tissue because of the strong differences with respect to the animal meniscus \citep{Chia2008,Katsuragawa2010}.
\\
Mechanical synthesis of the studies included: (\emph{i}) Region of the meniscus that was mechanically tested: medial (Medial) or lateral (Lateral) meniscus; anterior horn (Ant), central body (Cent), posterior horn (Post); (\emph{ii}) Direction of the meniscus that was mechanically tested: circumferential (C), radial (R), vertical (V). 
\\
Material models used to describe the meniscus mechanical response, or implementable by the reported data, included: Strain investigation (Str), linear-elastic (LE), hyper-elastic (HE), visco-elastic (VE), linear-biphasic (LB), non-linear-biphasic (NLB) and fiber-reinforced-poro-elastic (FRPE). ``Elastic'' means considering reversible deformations, which are ``linear'' if considered ``small'' or ``hyper''/``non-linear'' if ``large''. Time-dependent phenomena are considered in ``visco'' models, and in ``biphasic''/``poro'' models that explicitly consider the contributions of solid and liquid phases. ``Fiber-reinforced'' is when the contribution of collagen fibers in the solid phase is explicitly considered. 
\\
When studies included analyses about composition and structure, methods and targets of analysis have been reviewed and related to mechanics. Fig.~\ref{fig:FolChar} shows the flowchart of the literature review. 
\\
%
%
\subsection{Review results}
A total of $24$ studies have been reviewed, analysed and summarised in Table~\ref{tab:Tab1}. 
%
%
Five studies were specific to the medial meniscus ($21\%$), one to the lateral meniscus ($4\%$), whereas eighteen studies investigated both lateral and medial menisci ($75\%$). The anterior region ({i.e.}~anterior horn) of the meniscus appeared in seventeen studies ($71\%$), the posterior ({i.e.}~posterior horn) in eighteen studies ($75\%$), the central ({i.e.}~body) in thirteen studies ($54\%$). Comparison studies between the three regions have been reported in twelve studies ($50\%$). 
\\
The vertical direction of the meniscus appeared in eighteen studies ($75\%$), the radial in three studies ($12\%$) and the circumferential in two studies ($8\%$). One study ($4\%$) compared radial and vertical directions and One study ($4\%$) compared all the tree directions. 
%
No study investigated specifically the circumferential direction of medial meniscus body; no study investigated the circumferential and radial directions of lateral meniscus body.
\\
The mechanical model of the meniscus was linear-elastic in six studies ($25\%$), hyper-elastic in two studies ($8\%$), visco-elastic in seven studies ($29\%$), linear-biphasic in three studies ($12\%$), non-linear biphasic in two studies ($8\%$) and fiber-reinforced-poro-elastic in three studies ($12\%$). 
\\
Tables~\ref{tab:Tab2} and~\ref{tab:Tab3} list the papers that present equilibrium modulus ``$E_{\mathrm{eq}}$'', permeability ``$k$'' and fibre modulus ``$E_{\mathrm{f}}$''. Material parameter ``$E_{\mathrm{eq}}$'' represents the tissue response to the load at the equilibrium, {i.e.}~when time-dependent phenomena are exhausted, and it is mainly referred to the non-fibrous solid phase contribution. ``$E_{\mathrm{f}}$'' instead refers to the fibrous contribution to the loading response. ``$k$'' is related to the flow of the fluid component. The fibre-reinforced-poro-elastic model is the most complex material model, but it is the only one able to describe all those contributions together. Tables~\ref{tab:Tab2} and~\ref{tab:Tab3} collect also the testing configurations used to elicit the tissue mechanical response - specifically ``Unconfined compression" (UC), ``Confined compression'' (CC) and ``Indentation'' (I) - in order to relate them with the investigated regions, directions and material parameters. These tests are typically applied to cartilaginous tissue \citep{Marchiori2019}. The main difference between indentation and compression lies in the fact that the first uses a tip to load the tissue, while the second uses a piston. It follows that the first can investigate a lower dimensional scale (nanoscale), analysing the tissue with a higher spatial resolution. 
$E_{\mathrm{eq}}$ results in the order of $10^{-2}$-$10^{-1}$~$\mathrm{MPa}$ for confined compression, $10^{-2}$-$10^{0}$~$\mathrm{MPa}$ for indentation, $10^{-2}$~$\mathrm{MPa}$ for unconfined compression. $k$ is in the order of $10^{-15}$-$10^{-14}$~$\mathrm{m}^{4}/\mathrm{N} \, \mathrm{sec}$ for confined compression, $10^{-17}$-$10^{-15}$~$\mathrm{m}^{4}/\mathrm{N} \, \mathrm{sec}$ for indentation. No study provided meniscus permeability by unconfined compression. $E_{\mathrm{f}}$ is in the order of $10^{-2}$-$10^{2}$~$\mathrm{MPa}$ for indentation. No study provided meniscus fibre modulus by compression. $E_{\mathrm{eq}}$ and $E_{\mathrm{f}}$ are in general higher for the medial meniscus and anterior horn; $k$ is in general higher in the lateral meniscus.
\\
Finally, fifteen studies coupled the mechanical analysis with a composition/structure investigation (Table~\ref{tab:Tab4}).
%
%
%
\begin{landscape}
\renewcommand{\arraystretch}{1.5}
\begin{center}
\begin{longtable}{p{3.5cm} | c  c  c  c  c  c  c  c  c | c  c  c  c  c  c  c  c c}
\caption{Summary of the reviewed studies. Anterior (Ant), central (Cent) and posterior (Post) regions of the meniscus; circular (C), radial (R) and vertical (V) directions; strain investigation (Str) and linear-elastic (LE), hyper-elastic (HE), visco-elastic (VE), linear-biphasic (LB), non-linear-biphasic (NLB) and fiber-reinforced-poro-elastic (FRPE) models.}
\\
\multirow{3}{*}{Ref.} & \multicolumn{9}{ c |}{Medial} & \multicolumn{9}{c}{Lateral}   \\
\cline{2-19}
 & \multicolumn{3}{c}{Ant} & \multicolumn{3}{c}{Cent} & \multicolumn{3}{c|}{Post} & \multicolumn{3}{c}{Ant} & \multicolumn{3}{c}{Cent} & \multicolumn{3}{c}{Post}   \\
\cline{2-19}
 & C & R & V & C & R & V & C & R & V & C & R & V & C & R & V & C & R & V   \\
\hline
\endfirsthead
\multirow{3}{*}{Ref.} & \multicolumn{9}{ c |}{Medial} & \multicolumn{9}{c}{Lateral}   \\
\cline{2-19}
 & \multicolumn{3}{c}{Ant} & \multicolumn{3}{c}{Cent} & \multicolumn{3}{c|}{Post} & \multicolumn{3}{c}{Ant} & \multicolumn{3}{c}{Cent} & \multicolumn{3}{c}{Post}   \\
\cline{2-19}
 & C & R & V & C & R & V & C & R & V & C & R & V & C & R & V & C & R & V   \\
\hline
\endhead
\hline
\endfoot
\citet{Kessler2015} &  &  &  &  &  &  &  &  & Str &  &  &  &  &  &  &  &  &  \\
\citet{Fischenich2015} &  &  & LE &  &  & LE &  &  & LE &  &  & LE &  &  & LE &  &  & LE  \\
\citet{Fischenich2015} and \citet{Choi2016} & \multicolumn{9}{c|}{LE (random region, V)} & \multicolumn{9}{c}{LE (random region, V)}  \\
\citet{Bursac2009} &  &  & LE &  &  & LE &  &  & LE &  &  & LE &  &  & LE &  &  & LE \\
\citet{Son2013} &  &  & LE &  &  & LE &  &  & LE &  &  & LE &  &  & LE &  &  & LE  \\
\citet{Fischenich2015} &  &  & LE &  &  &  &  &  & LE &  &  & LE &  &  &  &  &  & LE \\
\citet{Yan2015} &  &  &  &  &  &  &  &  &  & \multicolumn{9}{c}{LE ("Ring" samples from the inside to the outside)}   \\
\citet{Leslie2000} & HE & HE & HE &  &  &  & HE & HE & HE & HE & HE & HE &  &  &  & HE & HE & HE   \\
\citet{Chia2008} &  & HE & HE &  & HE & HE &  & HE & HE &  &  &  &  &  &  &  &  &    \\
\citet{Kobayashi2003}  & \multicolumn{9}{c|}{VE (mixed regions and directions)} & \multicolumn{9}{c}{VE (mixed regions and directions)}   \\
\citet{Quiroga2014} & VE &  &  &  &  &  & VE &  &  &  &  &  &  &  &  &  &  &    \\
\citet{Sandmann2013} &  &  &  &  & VE &  &  &  &  &  &  &  &  &  &  &  &  &    \\
\citet{Bursac2006} &  &  & VE &  &  & VE &  &  & VE &  &  & VE &  &  & VE &  &  & VE   \\
\citet{Moyer2013} &  &  & VE &  &  & VE &  &  & VE &  &  & VE &  &  & VE &  &  & VE   \\
\citet{Pereira2014} &  &  & VE &  &  & VE &  &  & VE &  &  & VE &  &  & VE &  &  & VE   \\
\citet{Moyer2012} &  &  & VE &  &  & VE &  &  & VE &  &  & VE &  &  & VE &  &  & VE   \\
\citet{Katsuragawa2010} &  &  & LB &  &  &  &  &  &  &  &  & LB &  &  &  &  &  &    \\
\citet{Joshi1995} &  &  &  &  &  &  &  &  & LB &  &  &  &  &  &  &  &  &    \\
\citet{Sweigart2004} &  &  & LB &  &  & LB &  &  & LB &  &  &  &  &  &  &  &  &    \\
\citet{Seitz2013} &  &  & NLB &  &  & NLB &  &  & NLB &  &  & NLB &  &  & NLB &  &  & NLB   \\
\citet{Joshi1995} & \multicolumn{9}{c|}{NLB (random region, V)} & \multicolumn{9}{c}{NLB (random region, V)} \\
\citet{Danso2015} &  &  & FRPE &  &  & FRPE &  &  & FRPE &  &  & FRPE &  &  & FRPE &  &  & FRPE   \\
\citet{Danso2017} &  &  & FRPE &  &  & FRPE &  &  & FRPE &  &  & FRPE &  &  & FRPE &  &  & FRPE   \\
\citet{Ala2017} &  &  & FRPE &  &  & FRPE &  &  & FRPE &  &  & FRPE &  &  & FRPE &  &  & FRPE   
\label{tab:Tab1}
\end{longtable}
\end{center}
%
%
%
\begin{center}
\begin{longtable}{p{2.5cm} | c | c  c  c  c  c  c  c  c  c }
\caption{The equilibrium modulus ($E_{\mathrm{eq}}$), permeability ($k$) and fibre modulus ($E_{\mathrm{f}}$) of the medial meniscus and in the anterior (Ant), central (Cent) and posterior (Post) regions of the meniscus; circular (C), radial (R) and vertical (V) directions; unconfined compression (UC), indentation (I) and confined compression (CC). $E^{*}_{\mathrm{eq}}$ is not properly the equilibrium modulus: it is compressive Young's modulus in \citet{Bursac2009}, calculated Young’s modulus in \citet{Leslie2000}, aggregate modulus in \citet{Katsuragawa2010}, \citet{Joshi1995}, \citet{Sweigart2004} and \citet{Seitz2013}, and non-fibrillar matrix modulus in \citet{Danso2015}. Values are indicated as Average(Standard Deviation), otherwise as range min-max. $E_{\mathrm{eq}}$ and $E_{\mathrm{f}}$ are expressed in $\mathrm{MPa}$, $k$ in $10^{-15}$~$\mathrm{m}^{4}/\mathrm{N} \, \mathrm{sec}$. In \citet{Bursac2009}, (+) states positive statistical difference respect to (-). In \citet{Danso2017}, values are not found and only statistical differences are indicated between anterior horn of the medial (MA) meniscus, middle site of medial (MM) meniscus, posterior horn of medial (MP) meniscus.}
\\
\multirow{2}{*}{Ref.}  & \multirow{2}{*}{Prop.} & \multicolumn{3}{c}{Ant} & \multicolumn{3}{c}{Cent} & \multicolumn{3}{c}{Post}   \\
\cline{3-11}
 &  &  C & R & V & C & R & V & C & R & V   \\
\hline
\endfirsthead
\multirow{2}{*}{Ref.} &  \multirow{2}{*}{Prop.} &  \multicolumn{3}{c}{Ant} & \multicolumn{3}{c}{Cent} & \multicolumn{3}{c}{Post}   \\
\cline{3-11}
 &  &  C & R & V & C & R & V & C & R & V   \\
\hline
\endhead
\hline
\endfoot
\citet{Bursac2009}      & $E_{\mathrm{eq}}^{*}$ ($\mathrm{UC}$) &  &  & 0.067(0.013) &  &  & 0.033(0.005) &  &  & 0.021(0.006)                                                         \\ 
\citet{Son2013}         & $E_{\mathrm{eq}}$ ($\mathrm{UC}$)     &  &  & 0.052(0.021) &  &  & 0.052(0.021) &  &  & 0.052(0.021)                                                         \\ 
\citet{Fischenich2015}  & $E_{\mathrm{eq}}$ ($\mathrm{I}$)      &  &  & 0.298(0.202) &  &  &  &  &  & 0.12(0.03)                                                                       \\ 
\citet{Leslie2000}      & $E_{\mathrm{eq}}^{*}$ ($\mathrm{UC}$) & 10 & 13 & 19 &  &  &  & 10 & 13 & 19                                                                                 \\ 
\citet{Chia2008}        & $E_{\mathrm{eq}}$ ($\mathrm{UC}$)     &  & 0.0418(0.0469) & 0.0373(0.034) &  & 0.0213(0.0092) & 0.0229(0.0152) &  & 0.0339(0.034) & 0.025(0.0446)            \\ 
\citet{Bursac2006}      & $E_{\mathrm{eq}}$ ($\mathrm{UC}$)     &  &  & 0.01-0.167 &  &  & 0.01-0.167 &  &  & 0.01-0.167                                                               \\ 
\citet{Moyer2013}       & $E_{\mathrm{eq}}$ ($\mathrm{I}$)      &  &  & 1.4(0.09) &  &  & 1.32(0.07) &  &  & 1.34(0.1)                                                                 \\ 
\citet{Moyer2012}       & $E_{\mathrm{eq}}$ ($\mathrm{I}$)      &  &  & 1.7(0.08) &  &  & 1.58(0.15) &  &  & 1.62(0.1)                                                                 \\ 
\citet{Katsuragawa2010} & $E_{\mathrm{eq}}^{*}$ ($\mathrm{CC}$) &  &  & 0.074(0.063) &  &  &  &  &  &                                                                                  \\ 
                        & $k$ ($\mathrm{CC}$)                   &  &  & 32(56) &  &  &  &  &  &                                                                                        \\ 
\citet{Joshi1995}       & $E_{\mathrm{eq}}^{*}$ ($\mathrm{CC}$) &  &  &  &  &  &  &  &  & 0.22(0.06)                                                                                   \\ 
                        & $k$ ($\mathrm{CC}$)                   &  &  &  &  &  &  &  &  & 2.1(0.7)                                                                                     \\ 
\citet{Sweigart2004}    & $E_{\mathrm{eq}}^{*}$ ($\mathrm{I}$)  &  &  & 0.15(0.03) &  &  & 0.1(0.03) &  &  & 0.11(0.02)                                                                \\ 
                        & $k$ ($\mathrm{I}$)                    &  &  & 1.84(0.64) &  &  & 1.54(0.71) &  &  & 2.74(2.49)                                                               \\ 
\citet{Seitz2013}       & $E_{\mathrm{eq}}^{*}$ ($\mathrm{CC}$) &  &  & 0.073(0.045) &  &  & 0.06(0.02) &  &  & 0.076(0.042)                                                           \\ 
                        & $k$ ($\mathrm{CC}$)                   &  &  & 3.31(2.09) &  &  & 3.78(2.25) &  &  & 5.37(4.58)                                                               \\ 
\citet{Joshi1995}       & $E_{\mathrm{eq}}^{*}$ ($\mathrm{CC}$) & \multicolumn{9}{c}{0.23(0.09)}                                                                                       \\ 
                        & $k$ ($\mathrm{CC}$)                   & \multicolumn{9}{c}{1.49(1.54)}                                                                                       \\ 
\citet{Danso2015}       & $E_{\mathrm{eq}}^{*}$ ($\mathrm{I}$)  &  &  & 0.076(0.049) &  &  & 0.079(0.037) &  &  & 0.08(0.039)                                                          \\ 
                        & $E_{\mathrm{f}}$ ($\mathrm{I}$)       &  &  & 0.1(0.17) &  &  & 0.09(0.087) &  &  & 0.06(0.07)                                                               \\ 
                        & $k$ ($\mathrm{I}$)                    &  &  & 0.06(0.06) &  &  & 0.05(0.05) &  &  & 0.11(0.09)                                                               \\ 
\citet{Danso2017}       & $E_{\mathrm{eq}}$ ($\mathrm{I}$)      &  &  & MA>MM &  &  & MM &  &  & MP>MM                                                                                    \\ 
                        & $E_{\mathrm{f}}$ ($\mathrm{I}$)       &  &  & MA>MM,LA,LM &  &  & MM &  &  & MP>MM                                                                                    \\ 
                        & $k$ ($\mathrm{I}$)                    &  &  & MA &  &  & MM &  &  & MP>MM                                                                                    \\ 
\citet{Ala2017}         & $E_{\mathrm{eq}}$ ($\mathrm{I}$)      &  &  & 0.055-0.6 &  &  & 0.056-0.29 &  &  & 0.027-0.32                                                                \\ 
                        & $k$ ($\mathrm{I}$)                    &  &  & 9-177 &  &  & 0.6-51 &  &  & 0.6-66                                                                            \\ 
\label{tab:Tab2}
\end{longtable}
\end{center}
%
%
%
%
%
\begin{center}
\begin{longtable}{p{3.5cm} | c | c  c  c  c  c  c  c  c  c }
\caption{The equilibrium modulus ($E_{\mathrm{eq}}$), permeability ($k$) and fibre modulus ($E_{\mathrm{f}}$) of the lateral meniscus in the anterior (Ant), central (Cent) and posterior (Post) regions of the meniscus; circular (C), radial (R) and vertical (V) directions; unconfined compression (UC), indentation (I) and confined compression (CC). $E^{*}_{\mathrm{eq}}$ is not properly the equilibrium modulus: it is compressive Young's modulus in \citet{Bursac2009}, calculated Young’s modulus in \citet{Leslie2000}, aggregate modulus in \citet{Katsuragawa2010}, \citet{Joshi1995}, \citet{Sweigart2004} and \citet{Seitz2013}, and non-fibrillar matrix modulus in \citet{Danso2015}. Values are indicated as Average(Standard Deviation), otherwise as range min-max: $E_{\mathrm{eq}}$ and $E_{\mathrm{f}}$ are expressed in $\mathrm{MPa}$, $k$ in $10^{-15}$~$\mathrm{m}^{4}/\mathrm{N} \, \mathrm{sec}$. In \citet{Bursac2009}, (+) states positive statistical difference respect to (-). In \citet{Danso2017}, values are not found and only statistical differences are indicated between anterior horn of the lateral (LA) meniscus, middle site of lateral (LM) meniscus, posterior horn of lateral (LP) meniscus.}
\\
\multirow{2}{*}{Ref.}  & \multirow{2}{*}{Prop.} & \multicolumn{3}{c}{Ant} & \multicolumn{3}{c}{Cent} & \multicolumn{3}{c}{Post}   \\
\cline{3-11}
 &  &  C & R & V & C & R & V & C & R & V   \\
\hline
\endfirsthead
\multirow{2}{*}{Ref.} &  \multirow{2}{*}{Prop.} &  \multicolumn{3}{c}{Ant} & \multicolumn{3}{c}{Cent} & \multicolumn{3}{c}{Post}   \\
\cline{3-11}
 &  &  C & R & V & C & R & V & C & R & V   \\
\hline
\endhead
\hline
\endfoot
\citet{Bursac2009}      & $E_{\mathrm{eq}}^{*}$ ($\mathrm{UC}$) &  &  & 0.037(0.005) &  &  & 0.049(0.008) &  &  & 0.06(0.004) \\
\citet{Son2013}         & $E_{\mathrm{eq}}$ ($\mathrm{UC}$)     &  &  & 0.052(0.021) &  &  & 0.052(0.021) &  &  & 0.052(0.021) \\
\citet{Fischenich2015}  & $E_{\mathrm{eq}}$ ($\mathrm{I}$)      &  &  & 0.27(0.13) &  &  &  &  &  & 0.28(0.37) \\
\citet{Leslie2000}      & $E_{\mathrm{eq}}^{*}$ ($\mathrm{UC}$) & 10 & 13 & 19 &  &  &  & 10 & 13 & 19 \\
\citet{Chia2008}        & $E_{\mathrm{eq}}$ ($\mathrm{UC}$)     &  &  &  &  &  &  &  &  &  \\
\citet{Bursac2006}      & $E_{\mathrm{eq}}$ ($\mathrm{UC}$)     &  &  & 0.01-0.167 &  &  & 0.01-0.167 &  &  & 0.01-0.167 \\
\citet{Moyer2013}       & $E_{\mathrm{eq}}$ ($\mathrm{I}$)      &  &  & 1.5(0.08) &  &  & 1.48(0.1) &  &  & 1.57(0.08) \\
\citet{Moyer2012}       & $E_{\mathrm{eq}}$ ($\mathrm{I}$)      &  &  & 1.39(0.11) &  &  & 1.61(0.12) &  &  & 1.54(0.07) \\
\citet{Katsuragawa2010} & $E_{\mathrm{eq}}^{*}$ ($\mathrm{CC}$) &  &  & 0.085(0.075) &  &  &  &  &  &  \\
                        & $k$ ($\mathrm{CC}$)                   &  &  & 60(77) &  &  &  &  &  &  \\
\citet{Joshi1995}       & $E_{\mathrm{eq}}^{*}$ ($\mathrm{CC}$) &  &  &  &  &  &  &  &  &  \\
                        & $k$ ($\mathrm{CC}$)                   &  &  &  &  &  &  &  &  &  \\
\citet{Sweigart2004}    & $E_{\mathrm{eq}}^{*}$ ($\mathrm{I}$)  &  &  &  &  &  &  &  &  &  \\
                        & $k$ ($\mathrm{I}$)                    &  &  &  &  &  &  &  &  &  \\
\citet{Seitz2013}       & $E_{\mathrm{eq}}^{*}$ ($\mathrm{CC}$) &  &  & 0.08(0.05) &  &  & 0.056(0.021) &  &  & 0.078(0.052) \\
                        & $k$ ($\mathrm{CC}$)                   &  &  & 3.86(2.32) &  &  & 4.6(3.63) &  &  & 3.37(1.95) \\
\citet{Joshi1995}       & $E_{\mathrm{eq}}^{*}$ ($\mathrm{CC}$) & \multicolumn{9}{c}{0.23(0.09)} \\
                        & $k$ ($\mathrm{CC}$)                   & \multicolumn{9}{c}{1.49(1.54)}  \\
\citet{Danso2015}       & $E_{\mathrm{eq}}^{*}$ ($\mathrm{I}$)  &  &  & 0.07(0.055) &  &  & 0.08(0.04) &  &  & 0.07(0.03) \\
                        & $E_{\mathrm{f}}$ ($\mathrm{I}$)       &  &  & 0.07(0.1) &  &  & 0.09(0.06) &  &  & 0.06(0.05) \\
                        & $k$ ($\mathrm{I}$)                    &  &  & 0.09(0.17) &  &  & 0.06(0.07) &  &  & 0.1(0.1) \\
\citet{Danso2017}       & $E_{\mathrm{eq}}$ ($\mathrm{I}$)      &  &  & LA &  &  & LM>MM &  &  & LP \\
                        & $E_{\mathrm{f}}$ ($\mathrm{I}$)       &  &  & LA &  &  & LM &  &  & LP \\
                        & $k$ ($\mathrm{I}$)                    &  &  & LA &  &  & LM &  &  & LP>MM \\
\citet{Ala2017}         & $E_{\mathrm{eq}}$ ($\mathrm{I}$)      &  &  & 0.055-0.6 &  &  & 0.056-0.29 &  &  & 0.027-0.32 \\
                        & $k$ ($\mathrm{I}$)                    &  &  & 9-177 &  &  & 0.6-51 &  &  & 0.6-66 
\label{tab:Tab3}
\end{longtable}
\end{center}
\renewcommand{\arraystretch}{1.0}
\end{landscape}
%
%
%
%
\renewcommand{\arraystretch}{1.5}
\begin{center}
\begin{longtable}{p{3.5cm} | p{5.50cm} | p{5.50cm} }
\caption{The human meniscus mechanical studies with coupled compositional/structural analyses. Abbreviations: MRI (Magnetic Resonance Imaging), GAG (Glycosaminoglycan), PG (proteoglycan), micro-CT (micro-computer tomography), qPCR (quantitative polymerase chain reaction), TEM (transmission electron microscopy), DD (digital densitometry) and FTIR (Fourier Transform Infrared Spectroscopy).}
\\
 Reference & Methods & Target   \\
\hline
\endfirsthead
 Reference & Methods & Target   \\
\hline
\endhead
\hline
\endfoot
\citet{Shriram2017}         & Histology	                                                          & Fibres organization                                                                           \\                      
\citet{Seitz2013}           & MRI	                                                              & Lamellar layer thickness                                                                      \\
\citet{Chia2008}            & Biochemical analysis	                                              & Water, GAG and collagen content                                                               \\
\citet{Leroux2002}          & MRI; biochemical analysis; histology	                              & T1${\rho}$, T2; Water, GAG and collagen content                                               \\
\citet{Danso2015}           & Biochemical analysis; histology	                                  & Water, GAG and collagen content; GAG and collagen distribution                                \\
\citet{Son2013}             & Histology		                                                      & GAG, PG content                                                                               \\
\citet{Katsuragawa2010}     & Biochemical analysis		                                          & GAG, collagen content                                                                         \\
\citet{Joshi1995}           & Histology		                                                      & GAG content and distribution                                                                  \\
\citet{Bursac2006}          & Micro-CT; flow cytometry; histology and histomorphometric analysis  & Porosity; cellular study and quantification                                                   \\
\citet{Moyer2012}           & Histology; qPCR; [3H]proline incorporation; TEM	                  & Sections, fibre bundles and cells; gene expression; collagen neosynthesis; ultrastructure     \\
\citet{Danso2017}           & Balance and lyophilization		                                  & Water content                                                                                 \\
\citet{Venalainen2014}      & Biochemical analysis		                                          & Collagen, PG content                                                                          \\
\citet{Meng2014}            & DD; microscope+FTIR; conventional light microscope		          & PG distribution; collagen content; collagen orientation                                       \\
\citet{Kessler2015}         & Optical Spectroscopy; histology		                              & Changes in structure and composition; PG content

\label{tab:Tab4}
\end{longtable}
\end{center}
\renewcommand{\arraystretch}{1.0}
%
%
%
\subsection{Critical view on the state of the art}
This review analysed the state of the art in mechanically tested human menisci, dealing with time-dependent response and searching for inhomogeneity and anisotropy at a millimetre-scale, which is the scale of interest at joint level ({e.g.}~knee). We specifically found that the vast majority of the investigations were related to inhomogeneity, whereas far fewer studies focused on anisotropy.
\\
In the identified studies, when performing a full regional investigation indentation prevailed as testing set-up, \citep{Joshi1995,Bursac2006,Katsuragawa2010,Moyer2012,Moyer2013,Seitz2013,Danso2015,Danso2017,Ala2017}. 
 A clear indication of inhomogeneity came from \citep{Bursac2006,Danso2017,Ala2017}. In particular, \citet{Ala2017} reported a statistically significant differences in the biomechanical parameters of samples obtained from the different anatomical locations of the menisci. Mechanical inhomogeneity can be related to structural and compositional aspects by micro-CT, microscopy, flow cytometry, histology and histomorphometric analysis \citep{Bursac2006,Vetri2019, Agustoni2021} and by measuring the water content \citep{Danso2017}. 
\\
Only two studies compared relationship between the sample's mechanical properties and different testing directions \citep{Leslie2000,Chia2008} using an unconfined compression set-up; however, the testing was not performed on the central body in Ref.~\citep{Leslie2000} and circumferential direction and lateral meniscus in Ref.~\citep{Chia2008}. A certain degree of anisotropy could be suggested - specifically related to the elastic modulus - whereas anisotropic permeability studies are still scarce. In particular, the authors underlined how overall mechanical characteristics are related to water, PGs and collagen content by biochemical analysis \citep{Chia2008}.
\\
This review thus underlines that a full regional-directional investigation is still lacking. Furthermore, we had more hints on how to perform a comprehensive assessment of the meniscus mechanical response. First, in order to limit the inter-sample variability, and thus the number of specimens, the same samples can be tested along the three directions. This requires specific sample geometry and testing set-up; the samples could be of cubic geometry and unconfined compression can be used in the directional studies \citep{Leslie2000,Chia2008}. However, looking at Tables~\ref{tab:Tab2} and~\ref{tab:Tab3}, it seems that unconfined compression cannot give information regarding fibre and liquid tissue components ({i.e.}~$E_{\mathrm{f}}$ and $k$). Actually, it is a matter of loading protocol and data analysis; with a specific procedure (i.e~Mach-1 Analysis - Extraction of Mechanical Parameters Following Unconfined Compression \citep{Mach,Mach1}), unconfined compression can be used to reveal solid (i.e.~both non-fibrous and fibrous) and liquid contributes under different levels of strain. Starting from these results, we could then model the meniscal tissue as a strain-dependent fiber-reinforced-poro-elastic material (i.e.~non-linear FRPE).
\\
In the context of modelling, finite element models of the knee were implemented highlighting the importance of the meniscus material model \citep{Mononen2013}. The study suggested that a fibre-reinforced-poro-visco-elastic (i.e.~FRPVE, an upgrade of FRPE) material model is required for meniscus as well as cartilage; however, they highlighted that realistic values of the FRPVE material parameters for meniscus has not been determined yet. However, in the most recent studies \citep{Venalainen2014,Meng2014}, the meniscus is modelled using FRPE model rather than FRPVE model. FRPVE model would probably represent the best description of the interplay between the various tissue components described above, in the most various loading scenario, acknowledging that not only the fluid, but also the matrix solid phase can undergo viscous, {i.e.}~time-dependent response. Nevertheless, upgrades in the material models, in terms of parameters, anisotropy and inhomogeneity, result in an increase in complexity and computational cost. Therefore, their real influence on the overall biomechanics on the joint should be further investigated. 
%
%
\section{Important issues to consider for meniscal tissue modelling}
In the light of our findings, in order to fully characterise the meniscal tissue and appropriately model it, the following steps are necessary:
\renewcommand{\labelenumi}{\Roman{enumi}.}
\setlength\itemsep{0.5em}
\begin{enumerate}[leftmargin=*,align=left]
  \item To test all meniscal regions (i.e.~anterior/posterior horns and central body) and directions (i.e.~radial, circumferential and vertical).
  \item To model the tissue as fibre-reinforced-poro-elastic (FRPE).
  \item To implement step~1 by different strain-levels such to obtain a non-linear FRPE.
  \item To implement step~2 by different strain-rates such to obtain a non-linear FRPVE.
\end{enumerate}
Steps~2,~3 and ~4 are essential to recover parameters for FEM simulations. However, it is to be determined yet what is the influence of (\emph{i}) linear versus non liner material, (\emph{ii}) time-dependent versus time-independent and (\emph{iii}) isotropic and anisotropic models. This will be presented in the subsequent sections.
%
%
\section{The patient-specific knee joint model}
\label{sec:PhyMod}

Implementation of a subject-specific anatomical model of the knee and calculation of the joint motion and loading from gait analysis have been described in \citet{Cassiolas2018}. Briefly, virtual models of proximal tibia and fibula, distal femur, menisci, femoral and tibial cartilages were segmented from MRI sagittal proton density sequences of a patient that had undergone ACL reconstruction. Bone models, anthropometric measures, ground reaction force and inferior limb motion during gait of the same patient served for an inverse dynamic analysis in order to obtain knee kinematics and loading. The knee comprises also other anatomical structures such as: (\emph{i}) patellar bone and cartilage, (\emph{ii}) ligaments that restrain the motion and (\emph{iii}) tendons that transmit the muscles forces to the joint, see Fig.~\ref{fig:KneeJoint}. In order to focus on tibio-femoral articulation and to save time during FEM simulations, the knee model was fixed in the position of maximum ground reaction force - only vertical movement was free - and loaded by its vertical displacement component. Therefore, ligaments, tendons and patella were excluded as in \citet{Mononen2015}. 
%
%
%
\section{Constitutive Descriptions}
\label{sec:ConstitDes}
In this section, we discuss the constitutive models that are mainly used to simulate the knee components as reported in Sec.~\ref{sec:PhyMod}. Further, we present the range of models that are commonly used to describe the meniscus tissue. We limited our considerations to the cases of linear elastic and hyperelastic models for the time-independent behaviour; and linear viscoelasticity for the time-dependent response. 
%
\subsection{Linear elastic model}
\label{subsec:LinElas}
The menisci and cartilages are often modelled using isotropic linear elastic material (e.g.~\citep{Trad2017}) assuming nearly incompressibility conditions. It follows that the Poisson's ratio is taken to be $\nu=0.49$ and Young's modulus, $E$, is obtained from fitting experimental data.
%
\subsection{Hyperelastic models}
\label{subsec:HyperElas}

Several models of strain energy functions have been introduced in literature. These models describe the elastic response of incompressible and compressible isotropic as well as anisotropic hyperelastic materials with a variety in efficiency. In this study, only strain energy models for incompressible isotropic and anisotropic hyperelastic materials will be considered.
\\
Consider a deformation that is defined by the deformation gradient $\Fbf$ and the associated Jacobian $J$ as
\begin{equation}
\Fbf\left(\Xbf,t \right) = \dfrac{\partial \xbf \left(\Xbf,t \right)}{\partial \Xbf}, \quad J\left(\Xbf,t \right) = \mathrm{det} \, \Fbf\left(\Xbf,t \right).
\label{eq:DefGI}
\end{equation}
The deformation gradient is decomposed into volume changing (dilatational) and volume preserving (distortional) parts according to $\Fbf = J^{\frac{1}{3}} \, \bar{\Fbf}$, where $J^{\frac{1}{3}} \, \Ibf$ is the dilatational part and $\bar{\Fbf}$ the distortional part with $ \mathrm{det} \, \bar{\Fbf}=1$. The deviatoric right Cauchy-Green deformation tensor is $\bar{\Cbf} = \bar{\Fbf}^{\mathrm{T}}\bar{\Fbf} = J^{-\frac{2}{3}} \Cbf$. 
\\
The initial free strain energy function, {i.e.}~the time-independent response, can be written in the decoupled form
\begin{equation}
\Psi_{0}\left( \Fbf \right) = U \left( J \right) + \bar{\Psi} \left( \bar{\Fbf} \right), 
\label{eq:Psi0I}
\end{equation}
where $U$ is a purely volumetric part and $\bar{\Psi}$ is a purely isochoric part. The time-dependent viscoelastic response is expressed by the total strain energy function
\begin{equation}
\Psi\left( \Fbf, t \right) = k_{\mathrm{R}}\left( t\right) \, U \left( J \right) + g_{\mathrm{R}}\left( t\right) \, \bar{\Psi} \left( \bar{\Fbf} \right), 
\label{eq:PsiI}
\end{equation}
where $k_{\mathrm{R}}$ and $g_{\mathrm{R}}$ are dimensionless stress relaxation functions for the volumetric and isochoric parts, respectively. 
It should be mentioned that the isochoric part of the free energy function can be written in terms of the principal stretches $\bar{\lambda}_{i}$, $i=1,2,3$, of $\bar{\Fbf}$ or the classical strain invariants, $\bar{I}_{i}$, $i=1,2,3$, of $\bar{\Cbf}$ (for further details see \citet{Holzapfel2000}). 
\\
For incompressible materials, the volumetric part of the strain energy function is then expressed in terms of a hydrostatic pressure, $p$, as
\begin{equation}
U \left( J \right) = p\left( J-1\right).
\label{eq:PsivolI}
\end{equation}
The initial stresses can then be obtained in terms of the second Piola-Kirchhoff stress tensor as
\begin{equation}
\Sbf_{0} = 2 \, \dfrac{\partial \Psi_{0}}{\partial \Cbf} = -p \, \Cbf^{-1} + 2 \, \dfrac{\partial \bar{\Psi}}{\partial \bar{\Cbf}}.
\label{eq:S0I}
\end{equation}
The initial Kirchhoff stress tensor is determined using a standard push-forward operation as $\taubf_{0} = \Fbf \, \Sbf_{0} \, \Fbf^{\mathrm{T}}$. The total Kirchhoff stress is written as the sum of initial and non-equilibrium stress parts. The volumetric and deviatoric parts of the non-equilibrium stress are obtained using the hereditary integral in the reference configuration. Thus, the volumetric and deviatoric parts of the Kirchhoff stress are respectively written as
\begin{align}
\taubf_{\mathrm{h}} \left(t \right) &= \taubf_{0}^{\mathrm{h}} + \int\limits_{0}^{t} \dot{k}_{\mathrm{R}} \, \taubf_{0}^{\mathrm{h}} \left( t - \tau \right) \, \mathrm{d}\tau,
\label{eq:TauhI} \\
\taubf_{\mathrm{d}}\left(t \right) &= \taubf_{0}^{\mathrm{d}} +\mathrm{dev} \left[ \int\limits_{0}^{t} \dot{g}_{\mathrm{R}} \, \bar{\Fbf}_{t}^{-1}\left( t - \tau \right) \, \taubf_{0}^{\mathrm{d}}\left( t - \tau \right) \, \bar{\Fbf}_{t}^{-\mathrm{T}}\left( t - \tau \right) \, \mathrm{d}\tau \right],
\label{eq:TaudevI}
\end{align}
where $\taubf_{\mathrm{h}} = \mathrm{tr}\left(\taubf \right)/3$ and $\taubf_{\mathrm{d}} = \mathrm{dev} \left(\taubf \right)$ are the volumetric and deviatoric parts, respectively, $\mathrm{dev} \left(\bullet \right) = \left(\bullet \right) - \mathrm{tr}\left(\bullet \right)/3$ and $\mathrm{tr}\left(\bullet \right)$ is the trace of $\left(\bullet \right)$.
It should be mentioned that the initial stress is in a configuration defined at time $t-\tau$ and the total stress is determined in the configuration at time $t$. Hence, the non-equilibrium stress part is determined by mapping the initial configuration using the \emph{relative deformation gradient}:
\begin{equation}
\Fbf_{t-\tau} \left( t \right) = \dfrac{\partial \xbf \left(t \right)}{\partial \xbf \left( t -\tau \right)}.
\label{eq:FrI}
\end{equation}
The Cauchy stress tensor is computed as $\sigbf = J^{-1} \, \taubf$.
%
%
\subsubsection{Isotropic hyperelastic models}
\label{subsubsec:IsoHyperElas}
Various models are devised to describe the isotropic deformation of rubber and rubber-like materials which have also successfully been extended to biological tissues. These model have been reviewed in \citep{Boyce2000,Vahapouglu2006}. Here, we limit our considerations to the most frequently used models in the literature. 
%
%
\begin{itemize}[label=I.,leftmargin=*,align=left]
  \item \emph{Neo-Hookean model} 
\end{itemize}
The neo-Hookean model is the simplest physically based isotropic model for rubber material. In this model, the rubber materials are modelled by a network of long flexible randomly distributed links of equal length using Gaussian chain elasticity. The strain energy function takes the form:
\begin{equation}
\bar{\Psi} \left( \bar{I}_{1}\right) = \frac{1}{2} \, \mu \, \left( \bar{I}_{1} - 3 \right),
\label{eq:NHI}
\end{equation}
where $\mu$ is the classical shear modulus.
%
%
\begin{itemize}[label=II.,leftmargin=*,align=left]
  \item \emph{Moony-Rivlin model} 
\end{itemize}
Mooney \citep{Mooney1940} used linear response observations of a simple shear experiment of rubber material and developed the basic formula of strain energy function using the mathematical basis considering the symmetry. The strain energy function is written in terms of the first and second invariants as: 
\begin{equation}
\bar{\Psi} \left( \bar{I}_{1}, \bar{I}_{2}\right) = c_{1} \, \left( \bar{I}_{1} - 3 \right) + c_{2} \, \left( \bar{I}_{2} - 3 \right),
\label{eq:MRI}
\end{equation}
where the constants $c_{1} = \mu_{1}/2$ and $c_{2} = \mu_{2}/2$ are material parameters that define the classical shear modulus as $\mu=\mu_{1}-\mu_{2}$. 
%
%
\begin{itemize}[label=III.,leftmargin=*,align=left]
  \item \emph{Yeoh model} 
\end{itemize}
Yeoh \citep{Yeoh1993} investigated a carbon-black filled vulcanised rubber and showed that the dependence of $\Psi_{\mathrm{iso}}$ on $\bar{I}_{1}$ is much larger compared with $ \bar{I}_{2}$. Therefore, he postulated a three terms reduced polynomial strain energy function independent of the second invariant. Yeoh model is written in the form:
\begin{equation}
\bar{\Psi} \left( \bar{I}_{1}\right) = c_{1} \, \left( \bar{I}_{1} - 3 \right) + c_{2} \, \left( \bar{I}_{1} - 3 \right)^{2}  + c_{3} \, \left( \bar{I}_{1} - 3 \right)^{3},
\label{eq:YHI}
\end{equation}
where the constants $c_{1}$, $c_{2}$ and $c_{3}$ are material properties. The classical shear modulus is defined as $\mu = 2 \, c_{1} + 4 \, c_{2} \left( \bar{I}_{1} - 3 \right)+ 6 \, c_{3} \left( \bar{I}_{1} - 3 \right)^{2}$ with $c_{1} > 0$, $c_{2} < 0$ and $c_{3} > 0$ to ensure convexity. 
%
%
\begin{itemize}[label=IV.,leftmargin=*,align=left]
  \item \emph{Ogden model} 
\end{itemize}
Ogden \citep{Ogden1972} postulated a sophisticated strain energy model for modelling incompressible rubber-like materials. This model is one of the most widely used models due to simplicity in expression as well as it can simulate a wide range of deformation. The strain energy function is written in terms of the principal stretches $\lambda_{a}$, $a=1,2,3$, and takes the following form,
\begin{equation}
\bar{\Psi} \left( \bar{\lambda}_{1}, \bar{\lambda}_{2}, \bar{\lambda}_{3}\right) = \sum\limits_{p = 1}^{N} \dfrac{\mu_{p}}{\alpha_{p}} \, \left( \bar{\lambda}_{1}^{\alpha_{p}} + \bar{\lambda}_{2}^{\alpha_{p}} + \bar{\lambda}_{3}^{\alpha_{p}} - 3 \right),
\label{eq:OgI}
\end{equation}
where the constants $\mu_{p}$ and $\alpha_{p}$ are material parameters and the consistency with the linear elasticity theory introduces the following stability condition $\mu = \sum\limits_{p = 1}^{N} \mu_{p} \alpha_{p}$ with $\mu_{p} \alpha_{p} > 0$ and $p = 1, \cdots, N$.

%
\subsubsection{Anisotropic hyperelastic models}
\label{subsubsec:AnisoHyperElas}
Different approaches have been used to describe the anisotropy of soft tissues, namely: (\emph{i}) strain-based; and (\emph{ii}) strain invariants-based \citep{Chagnon2015}. The strain-based approach uses the components of a deformation tensor ({e.g.}~the Green–Lagrange strain tensor) to define the contributions of different directions through a set of material parameters; in fact, the material parameters weight the contributions of the components, and therefore, the directions. The original model of this type was proposed by \citet{Tong1976} and later generalised by \citet{Humphrey1995}. The second approach adopts the strain invariants form of the strain energy function, which is additively split into isotropic and anisotropic contributions. A new set of invariants is then introduced to define the directional dependency and obtain the anisotropic part of the strain energy. The first development is attributed to \citet{Spencer1984}. Several models, based on both physical and phenomenological approaches, have been introduced in the literature and comprehensively reviewed by \citet{Chagnon2015}. In this study, we consider the second approach, i.e.~the phenomenological approach, and limit our selection to the Holzapfel-Gasser-Ogden (HGO) model \citep{Holzapfel2000b,Gasser2006}.
%
%
\begin{itemize}[label=I.,leftmargin=*,align=left]
  \item \emph{Holzapfel-Gasser-Ogden (HGO) model} 
\end{itemize}
HGO model is a physically-based model that determines the passive mechanical response of arterial tissue \citep{Holzapfel2000b,Gasser2006}. An arterial layer, based on the histological structure, is comprised of an embedded collagen fibres in an isotropic matrix. They postulated a hyperelastic free energy function that is motivated by the anisotropic structure of arterial tissue using the second approach described above. Therefore, the isochoric strain energy function in Eq.~(\ref{eq:Psi0I}) is split into a part associated with isotropic deformations, $\bar{\Psi}_{\mathrm{iso}}$, and a part associated with anisotropic deformations, $\bar{\Psi}_{\mathrm{aniso}}$, as,
\begin{equation}
\bar{\Psi} = \bar{\Psi}_{\mathrm{iso}} \left( \bar{I}_{1},\bar{I}_{2}\right)+\bar{\Psi}_{\mathrm{aniso}} \left( \bar{I}_{1},\bar{I}_{2},\bar{I}_{4},\bar{I}_{5}, \cdots \right),
\label{eq:PsibarI}
\end{equation}
where $\bar{I}_{4},\bar{I}_{5}, \cdots,$ are pseudo-invariants that are defines the anisotropy of the tissue. 
\\
Neo-Hookean model in Eq.~(\ref{eq:NHI}) is assumed to determine the isotropic matrix response in Eq.~(\ref{eq:PsibarI}). The collagen fibre families are characterised by their unit direction vectors in the reference configuration, {i.e.}~$\abf_{0i}$ with $\left|\abf_{0i}\right|=1$ where $i$ denotes the fibre family. A symmetric second order structural tensor $\Abf_{i}= \abf_{0i}\otimes\abf_{0i}$, is introduced such that the pseudo-invariants are:
\begin{equation}
\begin{aligned}
\bar{I}_{4} \left( \bar{\Cbf},\abf_{01}\right)  &= \bar{\Cbf}:\Abf_{1}, \quad \bar{I}_{5} \left( \bar{\Cbf},\abf_{01}\right)  = \bar{\Cbf}^{2}:\Abf_{1}, \quad \bar{I}_{6} \left( \bar{\Cbf},\abf_{02}\right)  = \bar{\Cbf}:\Abf_{2}, \\ \bar{I}_{7} \left( \bar{\Cbf},\abf_{02}\right)  &= \bar{\Cbf}^{2}:\Abf_{2}, \quad \cdots.
\label{eq:PsdoInvI}
\end{aligned}
\end{equation}
The transversely anisotropic free-energy function for the $i~\mathrm{th}$ family of collagen fibres is assumed to take the form:
\begin{equation}
\bar{\Psi}_{\mathrm{aniso}} = \dfrac{\kappa_{1}}{2 \, \kappa_{2}} \, \Big[ \exp \left(\kappa_{2} \, \bar{E}_{i}^{2} \right) -1 \Big],
\label{eq:PsianisoI}
\end{equation}
where 
\begin{equation}
\bar{E}_{i} = \Hbf_{i}:\bar{\Cbf}-1, \quad \Hbf_{i} = \kappa \, \Ibf + \left(1-3 \, \kappa \right) \, \Abf_{i},
\label{eq:PsianisoII}
\end{equation}
$\kappa_{1}>0$ is a stress-like material parameter, $\kappa_{2}>0$ is a dimensionless parameter and $\kappa$ is a single dispersion parameter. The multiplication of the generalised structure tensor, $\Hbf_{i}$, in Eq.~(\ref{eq:PsianisoII}) results in $\bar{I}_{1} = \Ibf:\bar{\Cbf}$ and the pseudo-invariants $\Abf_{i}:\bar{\Cbf}$ as per Eq.~(\ref{eq:PsdoInvI}). It should be noted that $\kappa$ is between $0$ for perfectly aligned fibres and $1/3$ for the randomly oriented fibres. In this study, we consider one family of fibres, and assume perfect alignment along the circumferential direction, {i.e.} $\kappa = 0$ and $\Hbf_{i} = \Abf_{i}$.
%
%
\subsubsection{Viscoelastic model}
\label{subsubsec:ViscoElas}
A generalized Maxwell model (Prony series) is assumed to describe the full viscoelastic response. The model includes Maxwell elements, such that each element consists of a combination of spring and dashpot connected in series. Hence, the relaxation functions in Eq.~(\ref{eq:PsiI}) become
\begin{align}
k_{\mathrm{R}} \left( t \right) &= 1 - \sum\limits_{i = 1}^{M} \bar{k}_{i} \, \left( 1 - e^{-\frac{t}{\tau_{i}}} \right), \label{eq:kRI} \\
g_{\mathrm{R}} \left( t \right) &= 1 - \sum\limits_{i = 1}^{M} \bar{g}_{i} \, \left( 1 - e^{-\frac{t}{\tau_{i}}} \right), \label{eq:gRI}
\end{align}
where $\bar{k}_{i}$ and $\bar{g}_{i}$ are the fractions of the time dependent bulk and shear moduli, respectively, $\tau_{i}$ are the characteristic relaxation times and $M$ is the number of Maxwell elements ($i=1,\cdots, M$). The creep compliance can be characterised more easily using the generalised Voigt (Kelvin) model, which consists of $Q$ Voigt elements ({i.e.} a combination of spring and dashpot connected in parallel) that are connected in series. The compliance functions are given by:
\begin{align}
j_{\mathrm{C}} \left( t \right) &= 1 + \sum\limits_{i = 1}^{Q} \bar{j}_{i} \, \left( 1 - e^{-\frac{t}{\eta_{i}}} \right), \label{eq:DRI} \\
d_{\mathrm{C}} \left( t \right) &= 1 + \sum\limits_{i = 1}^{Q} \bar{d}_{i} \, \left( 1 - e^{-\frac{t}{\eta_{i}}} \right), \label{eq:JRI}
\end{align}
where $j_{\mathrm{C}}$ and $d_{\mathrm{C}}$ are the dimensionless creep compliance functions for the volumetric and isochoric parts, respectively, $\bar{j}_{i}$ and $\bar{d}_{i}$ are the fractions of the time dependent bulk and shear compliance, respectively, $\eta_{i}$ are the characteristic retardation times for the creep compliance functions. It should be noted that the parameters in the generalised Maxwell and Voigt models can be chosen such that the models are mathematically equivalent \citep{Park1999,Schapery1999}. An interconversion method is provided in Appendix~\ref{sec:EstVisco}.
%
%
\section{Finite Element model}
\label{sec:NumMod}

The details of the finite element model are given in this section. In particular, the FE mesh, applied loading and boundary conditions, interactions between knee parts, material models adopted for each part of the knee and experimental data are presented. 
%
%
\subsection{Finite element mesh}
\label{subsec:FEMesh}

\begin{figure}[!ht]
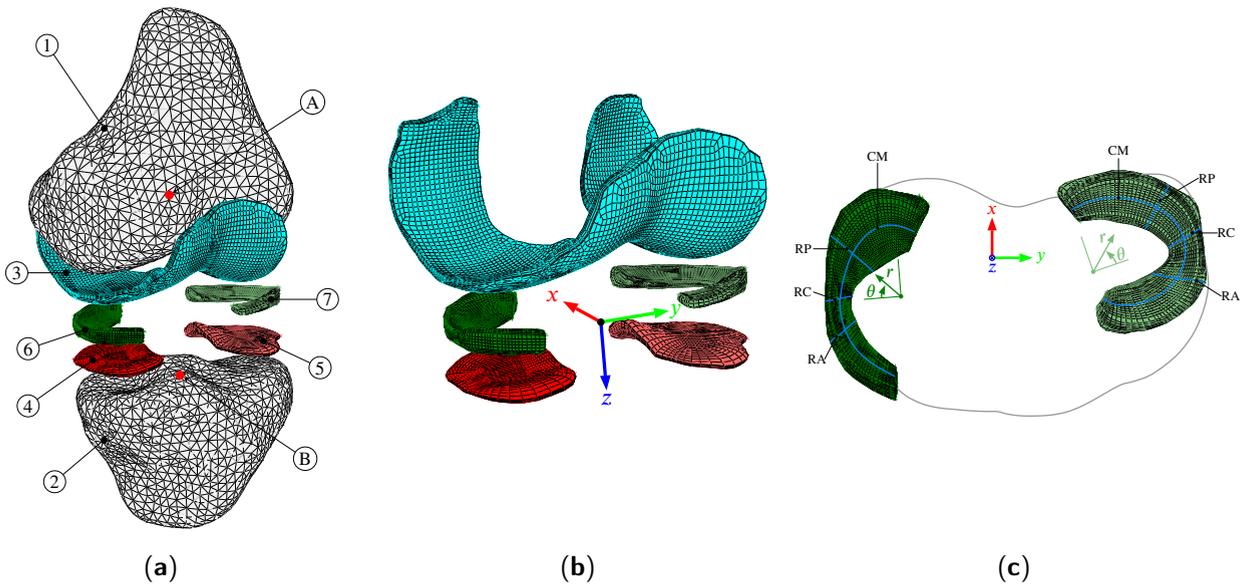

\begin{center}
\include{FE_Model}
\end{center}
\caption{The Finite Element model of the knee joint: \textbf{a} the details of the mesh of the different parts, where $1$ and $2$ are the femoral and tibial bones, $3$ is the femoral cartilage, $4$ and $5$ are the medial and lateral tibial cartilages, and $6$ and $7$ are the medial and lateral menisci, respectively; \textbf{b} the details of the parts of the knee joint and the Cartesian coordinates system $\left(x,y,z\right)$, and \textbf{c} the medial and lateral menisci (left and right, respectively) , the local polar coordinate systems $\left(r,\theta,z\right)$ and the radial anterior, central, posterior and circumferential tearing paths ({i.e.} $\mathrm{RA}$, $\mathrm{RC}$, $\mathrm{RP}$ and $\mathrm{CM}$, respectively). The reference points $\mathrm{A}$ and $\mathrm{B}$ in \textbf{a} are used to apply the boundary conditions and loading. }
\label{fig:FEModel}
\end{figure}

The finite element model is generated from the triangular surfaces of the segmented structures that are obtained in Sec.~\ref{sec:PhyMod}, by using a meshing algorithm provided by \citet{Rodriguez2017}. The algorithm generates an initial hexahedral low-resolution mesh using a sweeping algorithm and then expands the mesh to fit the original triangular surfaces of the structures. In the final step, the mesh is refined and optimised to achieve higher quality. Further, an ad hoc section was added to the Matlab code \citep{Matlab2020} in order to make the generated mesh compatible with the Finite Element software Abaqus \citep{ABAQUS2016}. 
\\
Figures~\ref{fig:FEModel}(\textbf{a})-(\textbf{c}) illustrate the FE model which is comprised of $5$ parts: (\emph{i}) the femoral bone, (\emph{ii}) tibial bone, (\emph{iii}) femoral cartilage, (\emph{vi}) medial/lateral tibial cartilages and (\emph{v}) medial/lateral menisci. Common Cartesian and cylindrical coordinate systems for the reference and deformed configurations $\left(x,y,z\right)$ and $\left(r,\theta,z\right)$, respectively, are assumed. The origin of the Cartesian coordinate system is located at the centre point of the knee joint, whereas the origins of the local cylindrical coordinate systems are located at the centre of each meniscus. It should be noted that the medial/lateral menisci lay on $x \, y$- and $r \, \theta$-planes. The cylindrical coordinates can be used to describe the radial and circumferential directions of the menisci, whereas $z$-direction corresponds to the vertical axis ({i.e.} axial direction) of the tibia. The femoral and tibial bones are considered rigid bodies; therefore, their triangular surface meshes are used to generate the rigid surfaces by using the $3$-node linear rigid elements (R3D3). The other parts, {i.e.} parts (\emph{iii})-(\emph{v}), are discretised using hexahedral meshes, {i.e.} $8$-node linear (C3D8H) hexahedral elements. It is worth mentioning that hybrid finite element formulation is used to ensure the incompressibility conditions (prevent volume strain locking). The mesh has $55846$ elements in total, of which $49042$ linear hexahedral elements of type C3D8H and $6804$ linear triangular elements of type R3D3. The femoral and tibial bones have $3592$ and $3212$ elements, respectively. The linear hexahedral elements are divided into $22320$ for femoral cartilage, $3906$ for lateral tibial cartilage, $3816$ for medial tibial cartilage, $7600$ for lateral meniscus and $11400$ for medial meniscus. A convergence study on the mesh refinement of the menisci was carried out for Ogden model of the circumferential middle specimen (CM) material parameters and walking loading case, see Secs.~\ref{subsec:LoadBCs} and~\ref{subsec:MenExpda}. We found that the mesh size is sufficient to obtain converged solutions.
\\
Clinically, the meniscal tears are classified into - based on the location and growth pattern and direction - radial, oblique, horizontal and longitudinal tears \cite{Greis2002,Lee2006}. \citet{Metcalf1996} reported that $81\%$ of tears were oblique or vertical longitudinal. The oblique tear often takes place in the junction of the anterior and/or posterior to middle third of the meniscus, whereas the vertical longitudinal tear occurs approximately in the mid section parallel to the circumference of the meniscus. Hence, we focus on studying the stain components perpendicular to the tears' paths. We choose four paths, namely, the radial anterior, central and posterior; and the circumferential paths ({i.e.} $\mathrm{RA}$, $\mathrm{RC}$, $\mathrm{RP}$ and $\mathrm{CM}$, respectively), as illustrated in Fig.~\ref{fig:FEModel}(\textbf{c}). The tear paths are chosen to match the reported data in the literature. 
%
%
\subsection{Loading and boundary conditions}
\label{subsec:LoadBCs}

The kinematics and external forces on the knee joint are usually defined in terms of translational and rotational degrees of freedom and/or normal force and moments with respect to the femur and tibia. Thus, assuming the femur and tibia are both rigid, any arbitrary reference points in the femoral and tibial bones can be used to define the loads and boundary conditions. Hence, to accurately model the knee motion and deformation of the moving parts within the joint, we need to apply the right combinations of the forces and moments, and define the correct kinematics. In other words, the combinations of the loads and boundary conditions should produce a physiologically accurate movement. Therefore, such loads and boundary conditions may be determined using computational methods ({e.g.}~inverse dynamics) or direct measurements in vivo ({e.g.}~using implants and telemetry systems). In this study, we consider a simple case of a one-dimensional movement in $z$-direction, see Sec.~\ref{sec:NumMod} and Fig.~\ref{fig:FEModel}. This choice replicates a constraint movement and ignores the contributions of the muscular forces. Femur and bone interface of the femoral cartilage are rigidly bounded to the reference point $\mathrm{A}$ of Fig.~\ref{fig:FEModel}. Tibia, bone interface of the tibial cartilages and meniscal horns are rigidly bounded to the reference point $\mathrm{B}$ of Fig.~\ref{fig:FEModel}. Femur is positioned with respect to the tibia as in the moment of maximum axial load during stance. Tibia is fixed, while the femur is free to move vertically along the tibial axis. \emph{First step} (static) is displacement controlled in order to bring the soft tissues in contact wherein a displacement of $u_{z}=-0.8~\mathrm{mm}$ is applied to the tibia at point $\mathrm{B}$ vertically upward. \emph{Second step} (static or dynamic) is load controlled in order to apply the axial force to the femur at point $\mathrm{A}$ vertically downward.
\\
Several studies have attempted to estimate the knee forces ({i.e.}~intersegmental reaction and muscle forces) during various activities, {e.g.}~walking, jogging, cycling, etc. They have used different approaches to relate knee kinematics and external forces to internal joint contact forces. Some of the commonly used approaches are inverse dynamics, forward dynamics, and static body analyses (see Ref.~\citep{DLima2012}). Additionally, the knee forces have also been measured in vivo after knee arthroplasty, which serves as valuable validation of computational predictions (see Ref.~\citep{Komistek2005}). These studies have shown that knee joint contact forces are related to the activity and can be related to the body weight. 
\\
Using the definition of Fourier series, we postulate that a force component transmitted across the knee joint for a periodic physical activity can by expressed in the following general form 
\renewcommand{\arraystretch}{1.5}
\begin{equation}
F\left(t \right) =  \left\{\begin{array}{l c l}
F_{0} + \sum\limits_{n=1}^{\infty} \Big[ \bar{F}_{n} \, \cos \left( \omega_{n} \, t \right) + \hat{F}_{n} \, \sin \left( \omega_{n}  \, t \right) \Big] & \quad \quad & \text{If} \ t \in \left[ 0 - T/2 \right], \\
F_{0} & \quad \quad & \text{Otherwise},
\end{array} \right.
\label{eq:FI}
\end{equation}
\renewcommand{\arraystretch}{1.0}
where $F_{0}$ is the force in the knee at rest, obtained by ensuring that the knee is in full contact, $\omega_{n} = 2 \, \pi \, n / T$ are the frequencies, $T$ is the periodic time of the activity, and $\bar{F}_{n}$ and $\hat{F}_{n}$ are force parameters that depend on the physical activity. It should be noted that this postulate stems from the fact that a cycle of an arbitrary periodic function can be approximated by Fourier series. Now, we consider the following simple form of the force in Eq.~(\ref{eq:FI}) for the interval $t \in \left[ 0 - T/2 \right]$ as:
\renewcommand{\arraystretch}{1.50}
\begin{equation}
F\left(t \right) = F_{0} + \hat{F} \, \sin \left( \omega  \, t \right),
\label{eq:FII}
\end{equation}
\renewcommand{\arraystretch}{1.0}
where $\hat{F}$ is the average peak force that is exerted by the body weight and muscle contraction during the activity and $\omega$ is the average frequency of the activity. The average peak force depends on the physical activity and takes the form $\hat{F} = \alpha \, W$, where $\alpha$ is a load factor that depends on the physical activity and $W$ is the body weight. Table~\ref{tab:LoadPara} summarises the values of the parameters $\alpha$, $\omega$ and $v$ for some physical activities, as reported in the literature \citep{Komistek2005,DLima2008,Stetter2019}.
\renewcommand{\arraystretch}{1.5}
\begin{table}[!ht]
\caption{The load factor, speed and frequency for different physical activities.}
\label{tab:LoadPara}
\begin{center}
\begin{threeparttable}
\begin{tabular}{c c c c}
\hline
Physical activity & Load factor, $\alpha$~[$-$] & Speed , $v$~[$\mathrm{mile}/\mathrm{hr}$] & Frequency$^{\dagger}$, $\omega$~[$\mathrm{Hz}$]  \\
\hline
Walking        &  $1.8-2.5$   & $1.0-3.0$   & $0.3-0.9$     \\
Power walking  &  $2.4-3.2$   & $3.0-4.0$   & $0.9-2.5$      \\
Running        &  $3.0-4.5$   & $4.0-5.0$   & $1.3-3.3$      \\
\hline
\end{tabular}
     \begin{tablenotes}
     \item[$\dagger$] The average frequency is obtained from human locomotion at different speeds data \citep{Nilsson1987}, {i.e.} $\omega=1/T_{\mathrm{s}}$ where $T_{\mathrm{s}}$ is duration of the support phase in which a full contact with the ground takes place.
     \end{tablenotes}
\end{threeparttable}
\end{center}
\end{table}
\renewcommand{\arraystretch}{1.0}
%
%
\subsection{Interactions}
\label{ssubsec:KneeIntr}

The knee model in Fig.~\ref{fig:FEModel} includes different interactions between parts' interfaces: (\emph{i}) the cartilage-bone, (\emph{ii}) the cartilage-cartilage, and (\emph{iii}) the cartilage-meniscus interfaces. The cartilage-bone interface is assumed to be coherent, as reported in \citet{Venalainen2016a}, and therefore, multi-point constraining (MPC) is applied between the contact surfaces. The cartilage-cartilage and cartilage-meniscus interfaces are modelled with hard pressure-overclosure relationship using surface to surface discretisation. The lubrication of the joint provides nearly frictionless contact between articulating surfaces (\citet{Mow1992}), therefore, frictionless formulation for tangential contact behaviour is adopted. Meniscal horns are fixed to the tibia in the central point (\citet{Mootanah2014}). Moreover, the peripheral rim of the medial meniscus is attached to the tibial plateau to simulate attachment to the deep medial collateral ligament and joint capsule.
%
%
\subsection{Material models}
\label{subsec:ConstitDes}

In this section, we present the material models, introduced in Sec.~\ref{sec:ConstitDes}, used to simulate the knee parts. The knee parts are categorised into three groups: (\emph{i}) the femoral and tibial bones, (\emph{ii}) medial/lateral menisci, and (\emph{iii}) the femoral and medial/lateral tibial cartilages. The models that are adopted for the different groups are discussed in the subsequent sections.  
%
%
\subsubsection{The femoral and tibial bones}
\label{subsubsec:Bones}

The femoral and tibial bones, in the studies that concern the deformation of cartilages and menisci, are usually assumed to be rigid due to the huge difference in their elastic moduli, {i.e.} typical Young's modulus of human bones is $~15$-$20$~$\mathrm{GPa}$ \cite{Reilly1974,Ashman1988} and for cartilages and menisci is $~0.1$-$10$~$\mathrm{MPa}$ \cite{Proctor1989,Kazemi2014}. However, using deformable models might influence the contact and deformation of the cartilages and menisci as reported by Ven{\"{a}}l{\"{a}}inen and co-workers \cite{Venalainen2016}. They have shown that the elastic modulus and bone density have significant effects on the contact and the cartilages deformation. Nevertheless, for higher values of Young's modulus ({i.e.}~$>5$~$\mathrm{GPa}$), the rigid and deformable assumptions agree. Hence, in this investigation, we assume that the bones are rigid, and we focus on studying the deformation of the cartilages and menisci. 
%
%
\subsubsection{The cartilage}
\label{subsubsec:Carti}

The cartilaginous tissue is described through an isotropic hyperelastic neo-Hookean model, assuming nearly incompressibility conditions \citep{Robinson2016}. The material parameter in Eq.~(\ref{eq:NHI}), i.e.~the shear modulus, is taken to be $\mu=1.67~\mathrm{MPa}$.
%
%
\subsubsection{The menisci}
\label{subsubsec:Menis}

The purpose of this study is to investigate the material modelling choice for the menisci. Therefore, we consider the models that are commonly used in the literature, see Sec.~\ref{sec:Expdata}. Accordingly, we limit our considerations to the isotropic linear elastic and isotropic and anisotropic hyperelastic models. We further explore the time-dependent behaviour by considering a finite strain viscoelasticity model. It is worth mentioning that the spatial dependency of the material properties is limited to the anisotropy case. The critical evaluation of the models is presented in the Sec.~\ref{sec:AnalyRes}. 

%
\subsection{Experimental data}
\label{subsec:MenExpda}
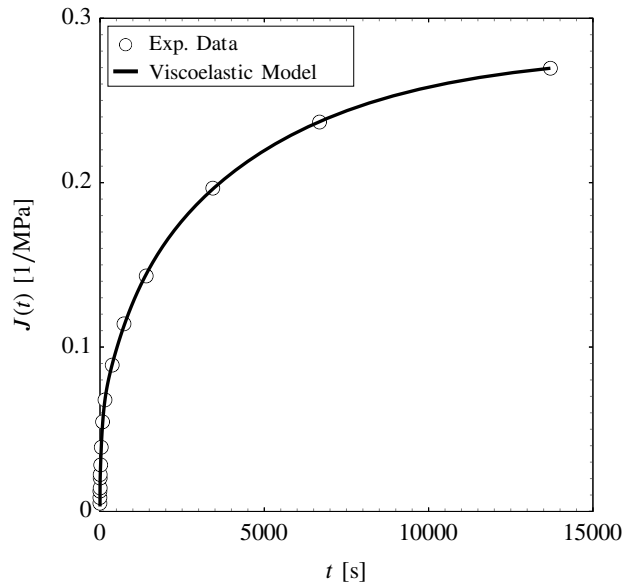
\begin{figure} 
\begin{center}
\readdata{\DataI}{Confined_Compression_Creep_Compliance_Experimental_Data.txt}
\readdata{\DataII}{Confined_Compression_Creep_Compliance_Fitting_Results.txt}
\begin{pspicture}[showgrid=false](0,0)(9.0,8.0)
%
%
\rput{0.0}(-1.0mm,4.0mm){
\rput{0.0}(15mm,8.5mm){
%
\psset{xunit=21.66668mm, yunit=21.66668mm}
{\small
\psaxes[axesstyle=frame,linewidth=0.0pt,tickstyle=inner,ticksize=0 1.5pt,subticksize=0.75,xsubticks=10,ysubticks=10,xlabelsep=-3pt,ylabelsep=-3pt,
        xLabels={$$0$$,$$5000$$,$$10000$$,$$15000$$},yLabels={$$0$$,$$0.1$$,$$0.2$$,$$0.3$$,}](3.0,3.0)
}
\psset{xunit=6.5mm, yunit=6.5mm}
\psaxes[linewidth=0.4pt,axesstyle=frame,tickstyle=inner,ticksize=0 1.5pt,subticksize=0.75,xsubticks=0,ysubticks=0,
        Dx=5000,dx=3.3333,Ox=0.0,Dy=0.10,dy=3.3333,Oy=0.0,labels=none]{->}(0.0,0.0)(10.0,10.0)
%
\pstScalePoints(0.0006666667,33.3333333){}{}
%
\rput{0.0}(0.0,-8.0mm){
\rput{0.0}(5.0,0.0){\small $t$~$[\mathrm{s}]$}
}
\rput{0.0}(-10.0mm,0.0){
\rput{90.0}(0.0,5.0){\small $J(t)$~$[1/\mathrm{MPa}]$}
}
%
%
\listplot[linecolor=black,fillcolor=white,plotstyle=dots,dotstyle=o,dotsize=5.5pt,linewidth=2.0pt]{\DataI}
\listplot[plotNoX=1,plotNo=1,plotNoMax=1,plotstyle=line,linecolor=black!100,linewidth=1.25pt,linestyle=solid,xStart=0,xEnd=200000]{\DataII}
%
%
\psset{xunit=10mm, yunit=10mm}
\rput[tl]{0.0}(0.10cm,5.60cm){
\psframe[linewidth=0.25pt,fillstyle=solid,fillcolor=white](0.0,0.0)(3.25,0.80) 
\psline[linewidth=1.5pt,linestyle=solid,linecolor=black!100]{-}(0.075,0.20)(0.40,0.20)
\rput[l](0.55,0.20){\footnotesize $\mathrm{Viscoelastic} \ \mathrm{Model}$}
\psdots[linecolor=black,fillcolor=white,dotstyle=o,dotscale=1.35](0.2375,0.55)
\rput[l](0.55,0.55){\footnotesize $\mathrm{Exp.} \ \mathrm{Data}$}
}
}
}
%
\end{pspicture}
%
%
\end{center}
\caption{The comparison between the confined creep test data and the viscoelastic model. The Kelvin viscoelastic model provides a good description of the material time-dependent behaviour.}
\label{fig:FitResIII}
\end{figure}

In this study, we adopt experimental data provided by \citet{Proctor1989}. They have investigated $14$ bovine menisci that were obtained from skeletal mature animals ($\sim 2$ years old) within $24$~$\mathrm{h}$ of slaughter. Firstly, a series of uniaxial experiments were conducted to determine the stress-strain behaviour and its variations with respect to the structural organisation of the tissue. The menisci were initially divided into two segments by separating the anterior and posterior regions, {i.e.} central-anterior and central-posterior regions, then sliced parallel to and from the femoral surface of the meniscus, and finally cut along the circumferential or the radial direction. We designate a specimen by its location ($\mathrm{A}\equiv$~anterior and $\mathrm{P}\equiv$~posterior), depth ($\mathrm{S}\equiv$~surface, $\mathrm{M}\equiv$~middle and $\mathrm{D}\equiv$~deep) and orientation ($\mathrm{C}\equiv$~circumferential and $\mathrm{R}\equiv$~radial directions). The different stress-strain data are illustrated in Figs.~\ref{fig:FitResI} and~\ref{fig:FitResII}. Secondly, the time-dependent response of the biphasic tissue was measured by a confined compression creep test. The specimens were of a cylindrical disc shape and harvested from four locations and two depths, {i.e.}~they were taken from posterior, central-posterior, central-anterior, and anterior regions and then sliced parallel to the femoral surface producing two specimens. The confined creep strain-time data are illustrated in Fig.~\ref{fig:FitResIII}.
\\
The experimental data provide a comprehensive data set that can be used to characterise the time-dependent and independent behaviour, anisotropy and spatial dependency of meniscus. Moreover, the number of specimens provides statistically significant data. 
%
%
\section{Numerical analyses and results}
\label{sec:AnalyRes}

The purpose of this study is to investigate the impact of the choice of a material model for the menisci on the knee modelling. We focus on studying factors that promote knee osteoarthritis (OA), which are widely believed to be the contact stresses in the cartilages. Further, the tearing of menisci is of interest because of their contribution to the acceleration of OA \citep{Englund2007}. Hence, we focus on studying the strain components perpendicular to the tears' paths discussed in Sec~\ref{sec:NumMod}. Therefore, the knee model is evaluated in terms of: (\emph{i}) the contact pressure and $1^{\mathrm{st}}$ principal stress of the tibial cartilage; (\emph{ii}) the deformation of the menisci; and (\emph{iii}) the perpendicular strain components on menisci tearing paths. In the subsequent sections, we start by exploring the isotropic linear elastic vs isotropic hyperelastic models. We then investigate the spatial dependency by comparing the isotropic vs anisotropic hyperelasticity. In the last part, we study the knee response for a range of physical activities to explore the time-dependent viscolelastic behaviour. 
%
%
\subsection{Linear vs nonlinear elasticity}
\label{subsec:LinVsNon}

\begin{figure} 
\begin{center}
%
\readdata{\DataI}{Stress_Stretch_Experimental_Data_Circumferential_Slice5.txt}
\readdata{\DataII}{Stress_Stretch_Fitting_Results_Circumferential_Slice5_Isotropic_Model.txt}
%
%
\begin{pspicture}[showgrid=false](0,0)(9.0,8.0)
%
\rput{0.0}(-1.0mm,4.0mm){
\rput{0.0}(15mm,8.5mm){
%
\psset{xunit=16.25mm, yunit=16.25mm}
{\small
\psaxes[axesstyle=frame,linewidth=0.0pt,tickstyle=inner,ticksize=0 1.5pt,subticksize=0.75,xsubticks=10,ysubticks=10,xlabelsep=-3pt,ylabelsep=-3pt,        xLabels={$$1.0$$,$$1.04$$,$$1.08$$,$$1.12$$,$$1.16$$},yLabels={$$0$$,$$10$$,$$20$$,$$30$$,$$40$$,}](4.0,4.0)
}
\psset{xunit=6.5mm, yunit=6.5mm}
\psaxes[linewidth=0.4pt,axesstyle=frame,tickstyle=inner,ticksize=0 1.5pt,subticksize=0.75,xsubticks=0,ysubticks=0,        Dx=0.04,dx=2.5,Ox=0.0,Dy=10.0,dy=2.5,Oy=0.0,labels=none]{->}(0.0,0.0)(10.0,10.0)
%
\pstScalePoints(62.5,0.25){1 sub}{}
%
\rput{0.0}(0.0,-8.0mm){
\rput{0.0}(5,0.0){\small $\lambda_{\theta}$~$[\mathrm{-}]$}
}
\rput{0.0}(-9.0mm,0.0){
\rput{90.0}(0.0,5.0){\small $\sigma_{\theta}$~$[\mathrm{MPa}]$}
}
%
\listplot[linecolor=black,fillcolor=white,plotstyle=dots,dotstyle=o,dotsize=5.5pt,linewidth=2.0pt]{\DataI}
\listplot[plotNoX=1,plotNo=1,plotNoMax=5,plotstyle=line,linecolor=black!80,linewidth=1.25pt,linestyle=solid,yStart=0,yEnd=40]{\DataII}
\listplot[plotNoX=1,plotNo=4,plotNoMax=5,plotstyle=line,linecolor=red!80,linewidth=1.25pt,linestyle=dashed,dash=0.5pt 0.75pt 2.0pt 0.75pt,yStart=0,yEnd=40]{\DataII}
\listplot[plotNoX=1,plotNo=5,plotNoMax=5,plotstyle=line,linecolor=blue!80,linewidth=1.0pt,linestyle=dashed,dash=2.0pt 1.0pt,yStart=0,yEnd=40]{\DataII}
%
%
\psset{xunit=10mm, yunit=10mm}
\rput[tl]{0.0}(0.10cm,4.90cm){
\psframe[linewidth=0.25pt,fillstyle=solid,fillcolor=white](0.0,0.0)(2.65,1.50)
\psline[linewidth=1.5pt,linestyle=dashed,dash=2.0pt 1.0pt,linecolor=blue!80]{-}(0.075,0.20)(0.40,0.20)
\rput[l](0.55,0.20){\footnotesize $\mathrm{Ogden} \ \mathrm{Model}$}
\psline[linewidth=1.5pt,linestyle=dashed,dash=0.5pt 0.75pt 2.0pt 0.75pt,linecolor=red!80]{-}(0.075,0.55)(0.40,0.55)
\rput[l](0.55,0.55){\footnotesize $\mathrm{Yeoh} \ \mathrm{Model}$}
\psline[linewidth=1.5pt,linestyle=solid,linecolor=black!100]{-}(0.075,0.90)(0.40,0.90)
\rput[l](0.55,0.90){\footnotesize $\mathrm{Elastic} \ \mathrm{Model}^{\dagger}$}
\psdots[linecolor=black,fillcolor=white,dotstyle=o,dotscale=1.35](0.2375,1.25)
\rput[l](0.55,1.25){\footnotesize $\mathrm{Exp.} \ \mathrm{Data}$}
}
}
}
%
\end{pspicture}
%
\end{center}
\caption{Comparison between the circumferential uniaxial true stress-stretch data of CM specimen and the different material models. $\dagger$ The elastic model, neo-Hookean and Mooney-Rivlin yield the same fitting curve. The fitting results imply that Ogden model with two parameters is the best fit.}
\label{fig:FitResI}
\end{figure}
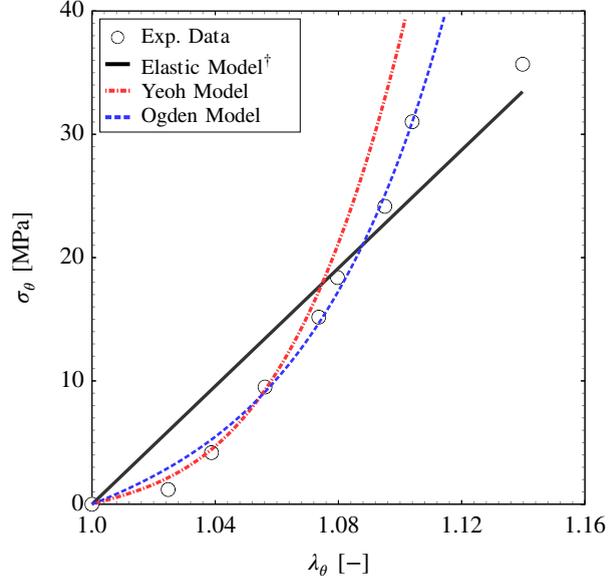

\begin{figure}
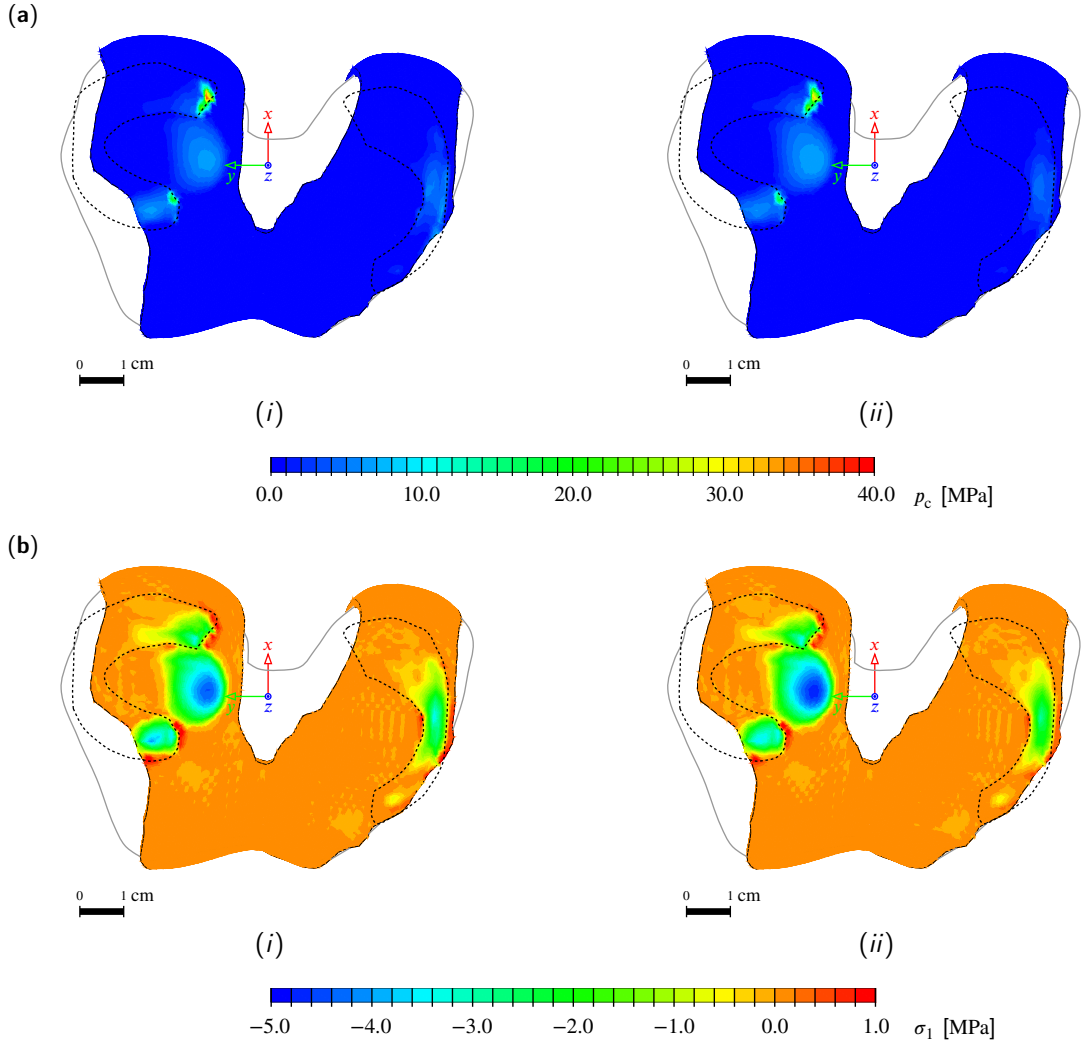
 
\begin{center}
\include{Tibial_Results_I}
\end{center}
\caption{Comparison between the linear elastic and hyperelastic Ogden models in (\emph{i}) and (\emph{ii}), respectively. The different distributions are: \textbf{a} the contact pressure $p_{\mathrm{c}}$; and \textbf{b} the $1^{\mathrm{st}}$ principal stress $\sigma_{1}$. The dashed lines indicate the current locations of the lateral and medial menisci ({i.e.}~left and right, respectively). The results imply that the femoral cartilage is in contact with lateral meniscus in its anterior and posterior horns, in the middle of medial meniscus and with the tibial cartilage in the vicinity of the lateral meniscus. The distributions are similar for both models, with a higher value of the maximum contact pressure in the case of linear elastic model. The distribution of the $1^{\mathrm{st}}$ principal stress complies with the contact pressure such that it is compressive and takes its minimum value within the contact area and its maximum value occurs in the region between contact areas of the femoral cartilage and lateral meniscus' anterior horn.}
\label{fig:CPTibI}
\end{figure}

\begin{figure*}
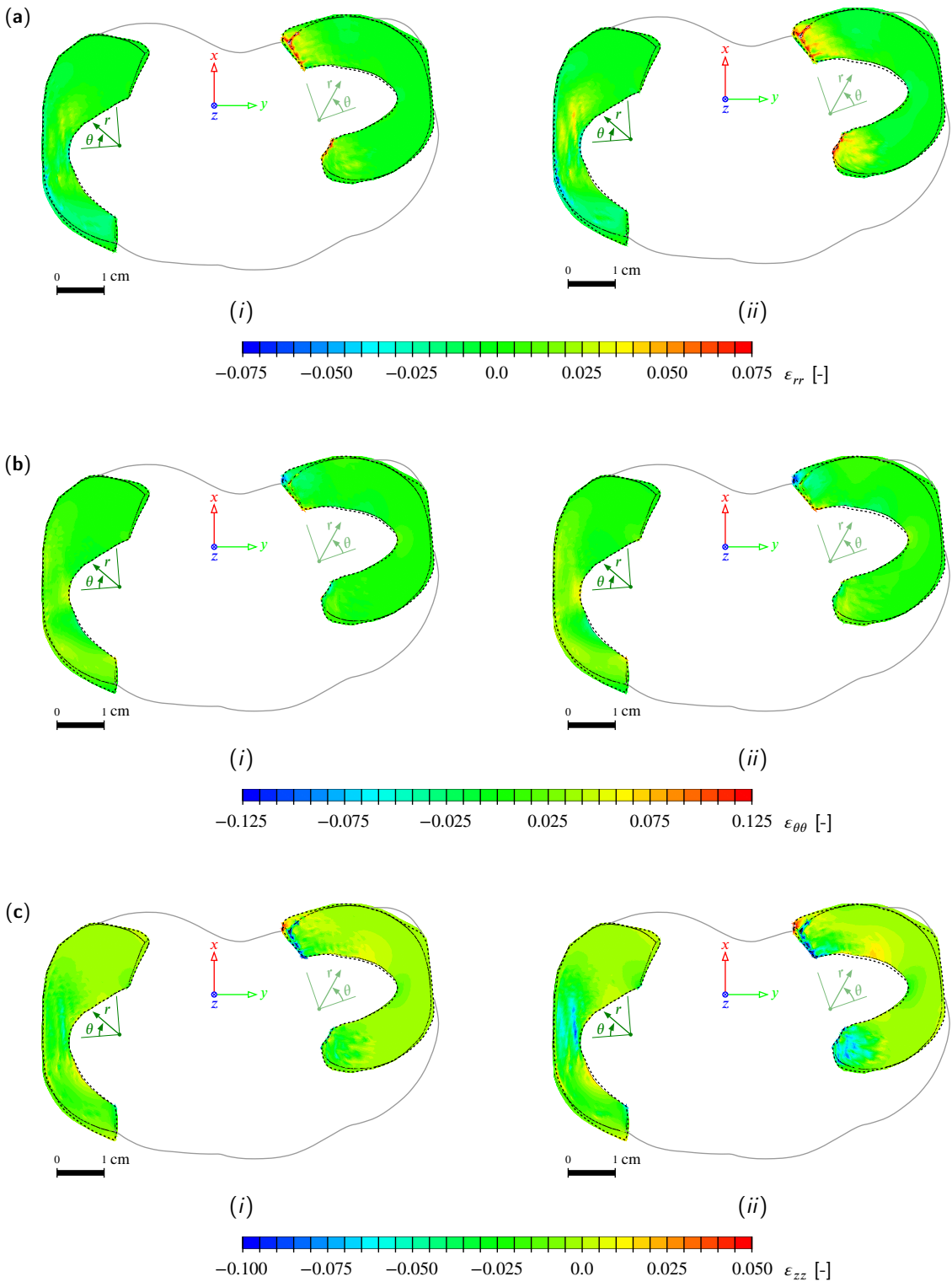
 
\begin{center}
\include{Menisci_Results_I}
\end{center}
\caption{Comparison between the linear elastic and hyperelastic Ogden models in (\emph{i}) and (\emph{ii}), respectively. The different distributions are the strain in: \textbf{a} $r$-direction; \textbf{b} $\theta$-direction and \textbf{c} $z$-direction. The dashed lines indicate the reference configuration of the medial and lateral menisci ({i.e.}~left and right, respectively). The results imply that the deformation is greater in the lateral meniscus (right) than the medial meniscus (left) with a larger values in the case of Ogden model. The comparison between reference and current configurations suggests that both menisci move radially and tangentially with a larger movement in the case of Ogden model. The meniscus movement is localised due the constraint by its attachment to the tibial plateau.}
\label{fig:MenResI}
\end{figure*}

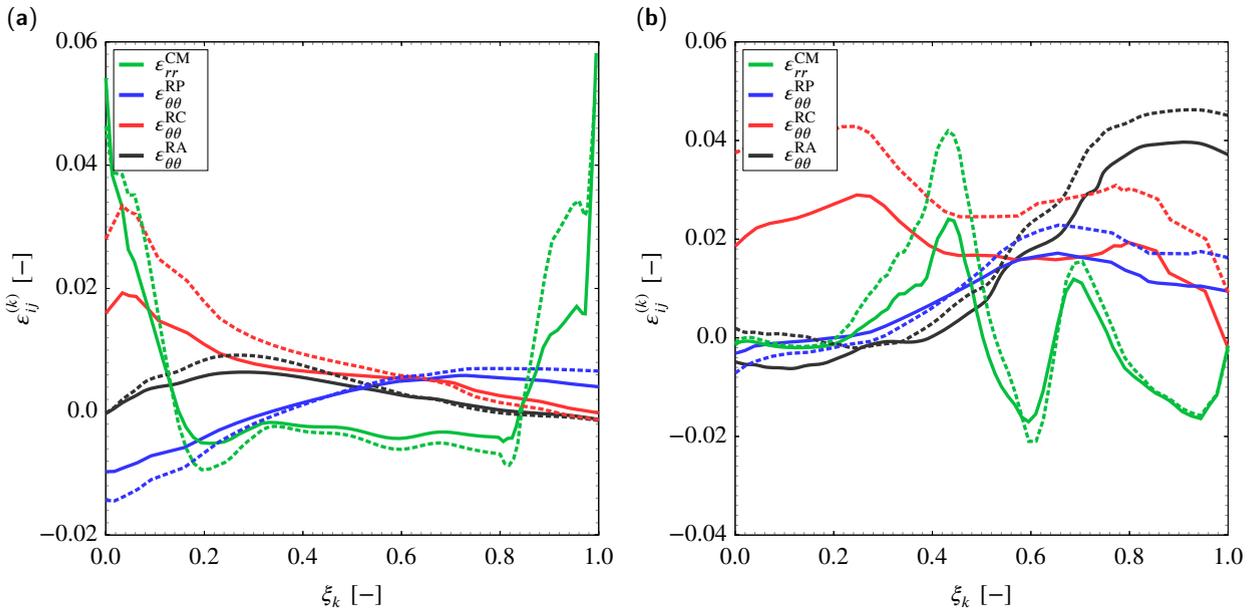
\begin{figure} 
\begin{center}
%
\readdata{\DataIa}{HK_Circum_Anterior_LM_Results.txt}
\readdata{\DataIb}{HK_Circum_Central_LM_Results.txt}
\readdata{\DataIc}{HK_Circum_Posterior_LM_Results.txt}
\readdata{\DataId}{HK_Circum_Circumferential_LM_Results.txt}
\readdata{\DataIe}{HK_Circum_Anterior_MM_Results.txt}
\readdata{\DataIf}{HK_Circum_Central_MM_Results.txt}
\readdata{\DataIg}{HK_Circum_Posterior_MM_Results.txt}
\readdata{\DataIh}{HK_Circum_Circumferential_MM_Results.txt}

\readdata{\DataIIa}{Ogden_Circum_Anterior_LM_Results.txt}
\readdata{\DataIIb}{Ogden_Circum_Central_LM_Results.txt}
\readdata{\DataIIc}{Ogden_Circum_Posterior_LM_Results.txt}
\readdata{\DataIId}{Ogden_Circum_Circumferential_LM_Results.txt}
\readdata{\DataIIe}{Ogden_Circum_Anterior_MM_Results.txt}
\readdata{\DataIIf}{Ogden_Circum_Central_MM_Results.txt}
\readdata{\DataIIg}{Ogden_Circum_Posterior_MM_Results.txt}
\readdata{\DataIIh}{Ogden_Circum_Circumferential_MM_Results.txt}
%
\begin{pspicture}[showgrid=false](0,0)(16.0,8.0)
%
\rput{0.0}(-3.0mm,-3.0mm){
\rput{0.0}(-1.0mm,4.0mm){
\rput{0.0}(15mm,8.5mm){
%
\psset{xunit=13.0mm, yunit=16.25mm}
{\small
\psaxes[axesstyle=frame,linewidth=0.0pt,tickstyle=inner,ticksize=0 1.5pt,subticksize=0.75,xsubticks=10,ysubticks=10,xlabelsep=-3pt,ylabelsep=-3pt,
        xLabels={$$0.0$$,$$0.2$$,$$0.4$$,$$0.6$$,$$0.8$$,$$1.0$$},yLabels={$$-0.02$$,$$0.0$$,$$0.02$$,$$0.04$$,$$0.06$$,$$0.08$$,$$1.00$$,$$1.20$$,}](5.0,4.0)
}
\psset{xunit=6.5mm, yunit=6.5mm}
\psaxes[linewidth=0.4pt,axesstyle=frame,tickstyle=inner,ticksize=0 1.5pt,subticksize=0.75,xsubticks=0,ysubticks=0,
        Dx=0.2,dx=2.0,Ox=0.0,Dy=0.02,dy=2.5,Oy=0.0,labels=none]{->}(0.0,0.0)(10.0,10.0)
%
\pstScalePoints(10.0,125.0){}{-0.02 sub}
%
\rput{0.0}(0.0,-8.0mm){
\rput{0.0}(5,0.0){\small $\xi_{k}$~$[\mathrm{-}]$}
}
\rput{0.0}(-11.0mm,0.0){
\rput{90.0}(0.0,5.0){\small $\varepsilon_{ij}^{(k)}$~$[\mathrm{-}]$}
}
\listplot[plotNoX=2,plotNo=2,plotNoMax=2,plotstyle=curve,linecolor=black!80,linewidth=1.25pt,linestyle=solid,xStart=0.0,xEnd=1.0,yStart=-0.04,yEnd=40]{\DataIa}
\listplot[plotNoX=2,plotNo=2,plotNoMax=2,plotstyle=line,linecolor=red!80,linewidth=1.25pt,linestyle=solid,xStart=0.0,xEnd=1.0,yStart=-0.04,yEnd=40]{\DataIb}
\listplot[plotNoX=2,plotNo=2,plotNoMax=2,plotstyle=line,linecolor=blue!80,linewidth=1.25pt,linestyle=solid,xStart=0.0,xEnd=1.0,yStart=-0.04,yEnd=40]{\DataIc}
\listplot[plotNoX=2,plotNo=2,plotNoMax=2,plotstyle=line,linecolor=darkpastelgreen,linewidth=1.25pt,linestyle=solid,xStart=0.0,xEnd=1.0,yStart=-0.04,yEnd=40]{\DataId}
\listplot[plotNoX=2,plotNo=2,plotNoMax=2,plotstyle=curve,linecolor=black!80,linewidth=1.25pt,linestyle=dashed,dash=2.0pt 1.0pt,xStart=0.0,xEnd=1.0,yStart=-0.04,yEnd=40]{\DataIIa}
\listplot[plotNoX=2,plotNo=2,plotNoMax=2,plotstyle=line,linecolor=red!80,linewidth=1.25pt,linestyle=dashed,dash=2.0pt 1.0pt,xStart=0.0,xEnd=1.0,yStart=-0.04,yEnd=40]{\DataIIb}
\listplot[plotNoX=2,plotNo=2,plotNoMax=2,plotstyle=line,linecolor=blue!80,linewidth=1.25pt,linestyle=dashed,dash=2.0pt 1.0pt,xStart=0.0,xEnd=1.0,yStart=-0.04,yEnd=40]{\DataIIc}
\listplot[plotNoX=2,plotNo=2,plotNoMax=2,plotstyle=line,linecolor=darkpastelgreen,linewidth=1.25pt,linestyle=dashed,dash=2.0pt 1.0pt,xStart=0.0,xEnd=1.0,yStart=-0.04,yEnd=0.06]{\DataIId}
%
%
\psset{xunit=10mm, yunit=10mm}
\rput[tl]{0.0}(0.10cm,4.80cm){
\psframe[linewidth=0.25pt,fillstyle=solid,fillcolor=white](0.0,0.0)(1.25,1.60)
\psline[linewidth=1.5pt,linestyle=solid,dash=2.0pt 1.0pt,linecolor=black!80]{-}(0.075,0.20)(0.40,0.20)
\rput[l](0.55,0.20){\footnotesize $\varepsilon_{\theta\theta}^{\mathrm{RA}}$}
\psline[linewidth=1.5pt,linestyle=solid,dash=0.5pt 0.75pt 2.0pt 0.75pt,linecolor=red!80]{-}(0.075,0.6)(0.40,0.6)
\rput[l](0.55,0.6){\footnotesize $\varepsilon_{\theta\theta}^{\mathrm{RC}}$}
\psline[linewidth=1.5pt,linestyle=solid,linecolor=blue!80]{-}(0.075,1.0)(0.40,1.0)
\rput[l](0.55,1.0){\footnotesize $\varepsilon_{\theta\theta}^{\mathrm{RP}}$}
\psline[linewidth=1.5pt,linestyle=solid,linecolor=darkpastelgreen]{-}(0.075,1.40)(0.40,1.40)
\rput[l](0.55,1.40){\footnotesize $\varepsilon_{rr}^{\mathrm{CM}}$}
}
}
}
}
%
%
\rput{0.0}(80.0mm,-3.0mm){
\rput{0.0}(-1.0mm,4.0mm){
\rput{0.0}(15mm,8.5mm){
%
\psset{xunit=13.0mm, yunit=13.0mm}
{\small
\psaxes[axesstyle=frame,linewidth=0.0pt,tickstyle=inner,ticksize=0 1.5pt,subticksize=0.75,xsubticks=10,ysubticks=10,xlabelsep=-3pt,ylabelsep=-3pt,
        xLabels={$$0.0$$,$$0.2$$,$$0.4$$,$$0.6$$,$$0.8$$,$$1.0$$},yLabels={$$-0.04$$,$$-0.02$$,$$0.0$$,$$0.02$$,$$0.04$$,$$0.06$$,$$0.08$$,$$1.00$$,$$1.20$$,}](5.0,5.0)
}
\psset{xunit=6.5mm, yunit=6.5mm}
\psaxes[linewidth=0.4pt,axesstyle=frame,tickstyle=inner,ticksize=0 1.5pt,subticksize=0.75,xsubticks=0,ysubticks=0,
        Dx=0.2,dx=2.0,Ox=0.0,Dy=0.02,dy=2.0,Oy=0.0,labels=none]{->}(0.0,0.0)(10.0,10.0)
%
\pstScalePoints(10.0,100.0){}{-0.04 sub}
%
\rput{0.0}(0.0,-8.0mm){
\rput{0.0}(5,0.0){\small $\xi_{k}$~$[\mathrm{-}]$}
}
\rput{0.0}(-11.0mm,0.0){
\rput{90.0}(0.0,5.0){\small $\varepsilon_{ij}^{(k)}$~$[\mathrm{-}]$}
}
\listplot[plotNoX=2,plotNo=2,plotNoMax=2,plotstyle=curve,linecolor=black!80,linewidth=1.25pt,linestyle=solid,xStart=0.0,xEnd=1.0,yStart=-0.04,yEnd=40]{\DataIe}
\listplot[plotNoX=2,plotNo=2,plotNoMax=2,plotstyle=line,linecolor=red!80,linewidth=1.25pt,linestyle=solid,xStart=0.0,xEnd=1.0,yStart=-0.04,yEnd=40]{\DataIf}
\listplot[plotNoX=2,plotNo=2,plotNoMax=2,plotstyle=line,linecolor=blue!80,linewidth=1.25pt,linestyle=solid,xStart=0.0,xEnd=1.0,yStart=-0.04,yEnd=40]{\DataIg}
\listplot[plotNoX=2,plotNo=2,plotNoMax=2,plotstyle=line,linecolor=darkpastelgreen,linewidth=1.25pt,linestyle=solid,xStart=0.0,xEnd=1.0,yStart=-0.04,yEnd=40]{\DataIh}
\listplot[plotNoX=2,plotNo=2,plotNoMax=2,plotstyle=curve,linecolor=black!80,linewidth=1.25pt,linestyle=dashed,dash=2.0pt 1.0pt,xStart=0.0,xEnd=1.0,yStart=-0.04,yEnd=40]{\DataIIe}
\listplot[plotNoX=2,plotNo=2,plotNoMax=2,plotstyle=line,linecolor=red!80,linewidth=1.25pt,linestyle=dashed,dash=2.0pt 1.0pt,xStart=0.0,xEnd=1.0,yStart=-0.04,yEnd=40]{\DataIIf}
\listplot[plotNoX=2,plotNo=2,plotNoMax=2,plotstyle=line,linecolor=blue!80,linewidth=1.25pt,linestyle=dashed,dash=2.0pt 1.0pt,xStart=0.0,xEnd=1.0,yStart=-0.04,yEnd=40]{\DataIIg}
\listplot[plotNoX=2,plotNo=2,plotNoMax=2,plotstyle=line,linecolor=darkpastelgreen,linewidth=1.25pt,linestyle=dashed,dash=2.0pt 1.0pt,xStart=0.0,xEnd=1.0,yStart=-0.04,yEnd=40]{\DataIIh}
%
%
\psset{xunit=10mm, yunit=10mm}
\rput[tl]{0.0}(0.10cm,4.80cm){
\psframe[linewidth=0.25pt,fillstyle=solid,fillcolor=white](0.0,0.0)(1.25,1.60)
\psline[linewidth=1.5pt,linestyle=solid,dash=2.0pt 1.0pt,linecolor=black!80]{-}(0.075,0.20)(0.40,0.20)
\rput[l](0.55,0.20){\footnotesize $\varepsilon_{\theta\theta}^{\mathrm{RA}}$}
\psline[linewidth=1.5pt,linestyle=solid,dash=0.5pt 0.75pt 2.0pt 0.75pt,linecolor=red!80]{-}(0.075,0.6)(0.40,0.6)
\rput[l](0.55,0.6){\footnotesize $\varepsilon_{\theta\theta}^{\mathrm{RC}}$}
\psline[linewidth=1.5pt,linestyle=solid,linecolor=blue!80]{-}(0.075,1.0)(0.40,1.0)
\rput[l](0.55,1.0){\footnotesize $\varepsilon_{\theta\theta}^{\mathrm{RP}}$}
\psline[linewidth=1.5pt,linestyle=solid,linecolor=darkpastelgreen]{-}(0.075,1.40)(0.40,1.40)
\rput[l](0.55,1.40){\footnotesize $\varepsilon_{rr}^{\mathrm{CM}}$}
}
}
}
}
%
\rput{0.0}(0.0cm,7.75cm){\small (\textbf{a})}
\rput{0.0}(8.30cm,7.75cm){\small (\textbf{b})}
%
\end{pspicture}
%
\end{center}
\caption{The distribution of the strain components in: \textbf{a} lateral and \textbf{b} medial menisci, and along the radial anterior, central, posterior and circumferential paths as defined in Fig.~\ref{fig:FEModel}(\textbf{c}). (where $k$ indicates the path with $k \equiv \mathrm{RA}$, $\mathrm{RC}$, $\mathrm{RP}$ and $\mathrm{CM}$, respectively). $\xi_{k}$ is the local coordinate normalised by the total length of the path that is measured from the white to the red for the radial paths and from the posterior to the anterior for the circumferential path. The solid lines represent the linear elastic model and the dashed line represent Ogden model. Ogden and linear elastic models show similar behaviour with differences in magnitudes. The maximum values of the strain components are in the same locations. Ogden model yields larger values of the strain components in the different paths.}
\label{fig:MenPlotI}
\end{figure}

It proves instructive to initially compare the choice of isotropic linear elastic model against the isotropic nonlinear hyperelastic models. We consider the circumferential direction and middle slice (CM) experimental data, see Fig.~\ref{fig:FitResI}. The uniaxial tensile test data are used to estimate the material parameters of the linear elastic and isotropic hyperelastic models in Sec.~\ref{subsubsec:IsoHyperElas}. Using a nonlinear least squares method, the Cauchy stress is fitted to the uniaxial tension data, see Appendix~\ref{sec:FitExpDa}. The fitting result shows that Ogden model with two parameters yields the best fit with a minimum root-mean-square error of $\approx 0.93$~$\mathrm{MPa}$. Hence, we choose Ogden model and compare it with the linear elastic model. The obtained parameter values are reported in Table~\ref{tab:FitParaI} and the fitting results are plotted in Fig.~\ref{fig:FitResI}.
\renewcommand{\arraystretch}{1.5}
\begin{table*} 
\caption{The fitting parameters of the linear elastic model and Ogden model ($N=1$).}
\label{tab:FitParaI}
\begin{center}
\begin{threeparttable}
\begin{tabular}{c c}
\hline
\multicolumn{2}{c}{(\emph{i}) Linear Elastic Model.} \\
\hline
$E$~[$\mathrm{MPa}$]        & $\nu$~[$-$] \\
\hline
$205.45$                    &  $0.49$   \\
\hline
\multicolumn{2}{c}{(\emph{ii}) Ogden Model.} \\
\hline
$\mu_{1}$~[$\mathrm{MPa}$]  & $\alpha_{1}$~[$-$] \\
\hline
$2.27$                      &  $25.23$   \\
\hline
\end{tabular}
\end{threeparttable}
\end{center}
\end{table*}
\renewcommand{\arraystretch}{1.0}
\\
To compare the two models, we consider the power walking activity case and take the average weight of an adult to be $\approx 63$~$\mathrm{kg}$ \cite{Walpole2012}. Using the information in Table~\ref{tab:LoadPara}, the average peak force becomes $\hat{F} \approx 1785.42$~$\mathrm{N}$. Quasi-static analysis is adopted such that the force is ramped up monotonically from $0$ to the maximum force $\hat{F}$. It should be noted that the knee force at rest is obtained from the initial loading step to be $F_{0} = 0.27~\mathrm{N}$.
\\
Firstly, we explore the stresses in the tibial cartilage. Figs.~\ref{fig:CPTibI}(\textbf{a}) and (\textbf{b}) illustrate the distributions of the contact pressure and $1^{\mathrm{st}}$ principal stress, respectively, at the maximum load. The contact pressure distribution implies that the femoral cartilage is in contact with both menisci ({i.e.}~in the anterior and posterior horns of the lateral meniscus and in the middle of medial meniscus) and with the tibial cartilage in the vicinity of the lateral meniscus. The distributions are similar for both models, with a higher value of the maximum contact pressure in the case of linear elastic model, {i.e.}~$p_{\mathrm{c}}^{\mathrm{max}}=37.7$ and $32.4$~$\mathrm{MPa}$ for the linear elastic and Ogden hyperelastic models, respectively. The contact area is larger in the case of Ogden model. Hence, the menisci are stiffer in the linear elastic case at lower stretches, {i.e.}~the instantaneous elastic modulus (the slope in Fig.~\ref{fig:FitResI}) is significantly larger than Ogden model, which yields harder contact; and therefore, higher values of the contact pressure and smaller contact area. On the other hand, in the case of Ogden model, the menisci deform more which results in a larger contact area and smaller contact pressure. The distribution of the $1^{\mathrm{st}}$ principal stress in Fig.~\ref{fig:CPTibI}(\textbf{b}) complies with the contact pressure such that it is compressive and takes its minimum value within the contact area. The $1^{\mathrm{st}}$ principal stress is tensile in the regions between contact areas which experience local stretching to balance the compression in the contact areas. The maximum value occurs in the region between contact areas of the femoral cartilage and lateral meniscus' anterior horn which is higher for the linear elastic model, {i.e.}~$\sigma_{1}^{\mathrm{max}}=1.43$ and $1.18$~$\mathrm{MPa}$ for the linear elastic and Ogden hyperelastic models, respectively. 
\\
Figs.~\ref{fig:MenResI}(\textbf{a}), (\textbf{b}) and (\textbf{c}) show the distributions of the radial, circumferential and axial strain components, $\varepsilon_{rr}$, $\varepsilon_{\theta\theta}$, and $\varepsilon_{zz}$, respectively, in the menisci's local coordinates. Results show that deformation is larger in the case of Ogden model. Further, the deformation is greater in the lateral meniscus (right) than the medial meniscus (left). The menisci deform mainly by the contact of the femoral cartilage, and therefore, the larger deformation of the lateral meniscus can be attributed to the larger contact area. The comparison between reference and current configurations of the menisci suggests that menisci move both radially and tangentially with a larger movement in the case of Ogden model as illustrated in Table~\ref{tab:MenMoveI}. The result show that the movements depend on the model and their magnitudes are larger in the case of Ogden model. The lateral meniscus movement is distributed over the tangential direction due to the freedom of movement, whereas the medial meniscus movement is localised ({i.e.} $-23.62^{\circ} \leq \theta \leq -63.08^{\circ}$) due to the constraint by its attachment to the tibial plateau.  
\renewcommand{\arraystretch}{1.5}
\begin{table}[!ht]
\caption{The minimum and maximum movement of the menisci at the maximum load for the linear elastic and hyperelastic Ogden models.}
\label{tab:MenMoveI}
\begin{center}
\begin{threeparttable}
\begin{tabular}{c c c c c}
\hline
\multicolumn{5}{c}{(\emph{i}) $\mathrm{Linear \ Elastic \ Model}$}  \\ 
\hline
$\mathrm{Meniscus}$ & \multicolumn{2}{c}{$u_{r}$~[$\mathrm{mm}$]} & \multicolumn{2}{c}{$u_{\theta}$~[$\mathrm{mm}$]} \\
\hline
$\mathrm{LM}$ & $-0.34$~($96.46^{\circ}$) & $0.53$~($32.84^{\circ}$) & $-0.32$~($69.33^{\circ}$) & $0.53$~($9.67^{\circ}$)  \\
$\mathrm{MM}$ & $-0.17$~($-1.42^{\circ}$) & $1.70$~($-28.16^{\circ}$) & $-0.43$~($2.86^{\circ}$) & $0.21$~($-55.08^{\circ}$)  \\
\hline
\multicolumn{5}{c}{(\emph{ii}) $\mathrm{Circumferential \ Ogden \ Model}$}  \\ 
\hline
$\mathrm{Meniscus}$ & \multicolumn{2}{c}{$u_{r}$~[$\mathrm{mm}$]} & \multicolumn{2}{c}{$u_{\theta}$~[$\mathrm{mm}$]} \\
\hline
$\mathrm{LM}$ & $-0.35$~($-11.62^{\circ}$) & $0.72$~($40.35^{\circ}$) & $-0.36$~($67.85^{\circ}$) & $0.72$~($7.61^{\circ}$)  \\
$\mathrm{MM}$ & $-0.17$~($0.0^{\circ}$) & $1.85$~($-28.72^{\circ}$) & $-0.45$~($-0.19^{\circ}$) & $0.21$~($-68.07^{\circ}$) \\
\hline
\end{tabular}
\begin{tablenotes}
    \item[$\dagger$] The circumferential position $\theta$ is provided in the parenthesis.
\end{tablenotes}
\end{threeparttable}
\end{center}
\end{table}
\renewcommand{\arraystretch}{1.0}
\\
To investigate the details of menisci deformation, we explore the strain component normal to the most common tearing paths. Hence, we study the hoop deformation, $\varepsilon_{\theta\theta}$, along the radial anterior, central and posterior paths; and the radial deformation, $\varepsilon_{rr}$, along the circumferential path, see Fig.~\ref{fig:FEModel}(\textbf{c}). Figs.~\ref{fig:MenPlotI}(\textbf{a}) and (\textbf{b}) illustrate the results in lateral and medial menisci, respectively. Along the central radial path, the hoop strain is maximum at the white zone (the avascular region which is more than $5~\mathrm{mm}$ from periphery) at $\xi_{\mathrm{RC}}=0.034$ in the lateral meniscus and $\xi_{\mathrm{RC}}=0.245$ in the medial meniscus; and decreases towards the red zone (the well-vascularised region which is within $3$-$5~\mathrm{mm}$ from periphery) at $\xi_{\mathrm{RC}}=1$ in both menisci. The hoop strain is minimum at the white region and increase towards the red region with a slight decrease at the end of the anterior and posterior radial paths. Results suggest that the hoop deformation is controlled by the radial movement of the menisci. Hence, in the lateral meniscus, the radial movement is larger and concentrated in the central part, which generates a tensile straining at the centre path and compressive or negligible straining in the anterior and posterior paths. In the medial meniscus, the radial movement is shifted towards the anterior due to the constraint which causes a tensile straining in the red zone of the anterior radial path. In the circumferential path, the radial strain is maximum in the lateral horns ($\xi_{\mathrm{CM}}=0$ and $1$) and in $\xi_{\mathrm{CM}}=0.44$ of the medial meniscus which can be explained by the location of the contact between the femoral cartilage and menisci as indicated in Figs.~\ref{fig:CPTibI} and~\ref{fig:MenResI}. It is worth noting that, in the medial meniscus' central part, radial strain changes sign from tensile to compressive and to tensile again due to the irregularity of the femoral cartilage. The maximum strains are larger in the medial meniscus due to the constraint. Ogden model predicts larger tensile and compressive strain than the linear elastic model. 
%
%
\subsection{Spatial dependency}
\label{subsec:SpatDepend}

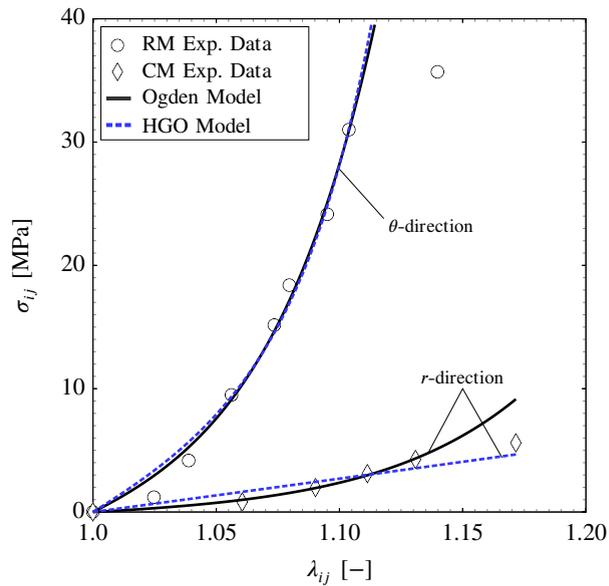
\begin{figure} 
\begin{center}
%
\readdata{\DataIa}{Stress_Stretch_Experimental_Data_Circumferential_Slice5.txt}
\readdata{\DataIb}{Stress_Stretch_Fitting_Results_Circumferential_Slice5_Isotropic_Model.txt}
\readdata{\DataIIa}{Stress_Stretch_Experimental_Data_Radial_Slice5.txt}
\readdata{\DataIIb}{Stress_Stretch_Fitting_Results_Radial_Slice5_Isotropic_Model.txt}
\readdata{\DataIII}{Stress_Stretch_Fitting_Results_Slice5_HGO_Anisotropic_Model.txt}
%
%
\begin{pspicture}[showgrid=false](0,0)(9.0,8.0)
%
\rput{0.0}(-1.0mm,4.0mm){
\rput{0.0}(15mm,8.5mm){
%
\psset{xunit=16.25mm, yunit=16.25mm}
{\small
\psaxes[axesstyle=frame,linewidth=0.0pt,tickstyle=inner,ticksize=0 1.5pt,subticksize=0.75,xsubticks=10,ysubticks=10,xlabelsep=-3pt,ylabelsep=-3pt,
        xLabels={$$1.0$$,$$1.05$$,$$1.10$$,$$1.15$$,$$1.20$$},yLabels={$$0$$,$$10$$,$$20$$,$$30$$,$$40$$,}](4.0,4.0)
}
\psset{xunit=6.5mm, yunit=6.5mm}
\psaxes[linewidth=0.4pt,axesstyle=frame,tickstyle=inner,ticksize=0 1.5pt,subticksize=0.75,xsubticks=0,ysubticks=0,
        Dx=0.05,dx=2.5,Ox=0.0,Dy=10.0,dy=2.5,Oy=0.0,labels=none]{->}(0.0,0.0)(10.0,10.0)
%
\pstScalePoints(50.0,0.25){1 sub}{}
%
\rput{0.0}(0.0,-8.0mm){
\rput{0.0}(5,0.0){\small $\lambda_{ij}$~$[\mathrm{-}]$}
}
\rput{0.0}(-9.0mm,0.0){
\rput{90.0}(0.0,5.0){\small $\sigma_{ij}$~$[\mathrm{MPa}]$}
}
%
\listplot[linecolor=black,fillcolor=white,plotstyle=dots,dotstyle=o,dotsize=5.0pt,linewidth=2.0pt]{\DataIa}
\listplot[linecolor=black,fillcolor=white,plotstyle=dots,dotstyle=diamond,dotsize=5.0pt,linewidth=2.0pt]{\DataIIa}
\listplot[plotNoX=1,plotNo=5,plotNoMax=5,plotstyle=line,linecolor=black!100,linewidth=1.0pt,linestyle=solid,yStart=0,yEnd=40]{\DataIb}
\listplot[plotNoX=1,plotNo=5,plotNoMax=5,plotstyle=line,linecolor=black!100,linewidth=1.0pt,linestyle=solid,yStart=0,yEnd=10]{\DataIIb}
\listplot[plotNoX=1,plotNo=1,plotNoMax=3,plotstyle=line,linecolor=blue!80,linewidth=1.0pt,linestyle=dashed,dash=2.0pt 1.0pt,yStart=0,yEnd=10]{\DataIII}
\listplot[plotNoX=3,plotNo=3,plotNoMax=3,plotstyle=line,linecolor=blue!80,linewidth=1.0pt,linestyle=dashed,dash=2.0pt 1.0pt,yStart=0,yEnd=40]{\DataIII}
%
\psline[linewidth=0.25pt,linestyle=solid,linecolor=black]{-}(5.0,6.95)(5.90,5.95)
\rput[l]{0.0}(6.0,5.8){\scriptsize $\theta$-$\mathrm{direction}$}
\psline[linewidth=0.25pt,linestyle=solid,linecolor=black]{-}(6.80,1.20)(7.50,2.5)(8.30,1.10)
\rput[b]{0.0}(7.50,2.6){\scriptsize $r$-$\mathrm{direction}$}
%
%
\psset{xunit=10mm, yunit=10mm}
\rput[tl]{0.0}(0.10cm,4.90cm){
%
%
\psframe[linewidth=0.25pt,fillstyle=solid,fillcolor=white](0.0,0.0)(2.75,1.5)
\psline[linewidth=1.5pt,linestyle=dashed,dash=2.0pt 1.0pt,linecolor=blue!80]{-}(0.075,0.20)(0.40,0.20)
\rput[l](0.55,0.20){\footnotesize $\mathrm{HGO} \ \mathrm{Model}$}
\psline[linewidth=1.5pt,linestyle=solid,dash=0.5pt 0.75pt 2.0pt 0.75pt,linecolor=black!80]{-}(0.075,0.55)(0.40,0.55)
\rput[l](0.55,0.55){\footnotesize $\mathrm{Ogden} \ \mathrm{Model}$}
\psdots[linecolor=black,fillcolor=white,dotstyle=diamond,dotscale=1.35](0.225,0.90)
\rput[l](0.55,0.90){\footnotesize $\mathrm{CM} \ \mathrm{Exp.} \ \mathrm{Data}$}
\psdots[linecolor=black,fillcolor=white,dotstyle=o,dotscale=1.35](0.225,1.25)
\rput[l](0.55,1.25){\footnotesize $\mathrm{RM} \ \mathrm{Exp.} \ \mathrm{Data}$}
}
}
}
%
\end{pspicture}
%
\end{center}
\caption{The comparison between uniaxial true stress-stretch data of RM and CM specimens (i.e.~ in the radial, $r$, and circumferential, $\theta$, directions, respectively) and Ogden and HGO material models. The fitting results imply that HGO model provides a good description of the material anisotropy assuming transverse isotropy whereas Ogden models yield good fit for the radial circumferential directions individually due to isotropy of the model.}
\label{fig:FitResII}
\end{figure}

\begin{figure}
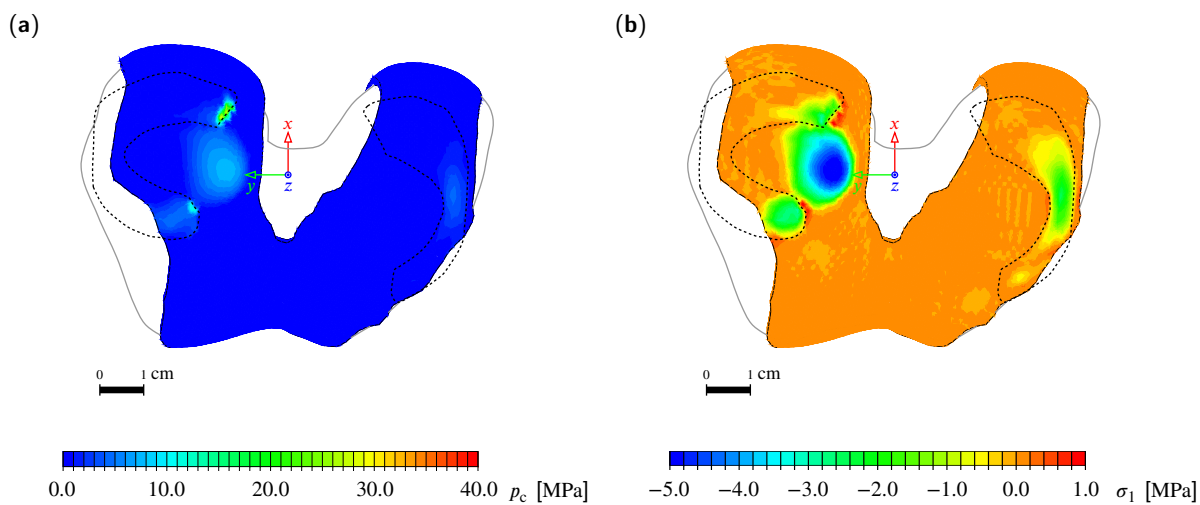
 
\begin{center}
\include{Tibial_Results_II}
\end{center}
\caption{The distributions are: \textbf{a} the contact pressure $p_{\mathrm{c}}$; and \textbf{b} the $1^{\mathrm{st}}$ principal stress $\sigma_{1}$, for the case of Holzapfel-Gasser-Ogden (HGO) model under quasi-static loading conditions. The dashed lines indicate the current locations of the lateral and medial menisci ({i.e.} left and right, respectively). HGO model yields similar contact pattern in comparison with Ogden and linear elastic models in Fig.~\ref{fig:CPTibI} such that the femoral cartilage is in contact with the anterior and posterior horns of both the lateral meniscus and the middle of medial meniscus. The maximum contact pressure is lower than Ogden model whereas the contact area is larger which implies that HGO model provides a softer response than isotropic circumferential Ogden model.}
\label{fig:CPTibII}
\end{figure}

\begin{figure}
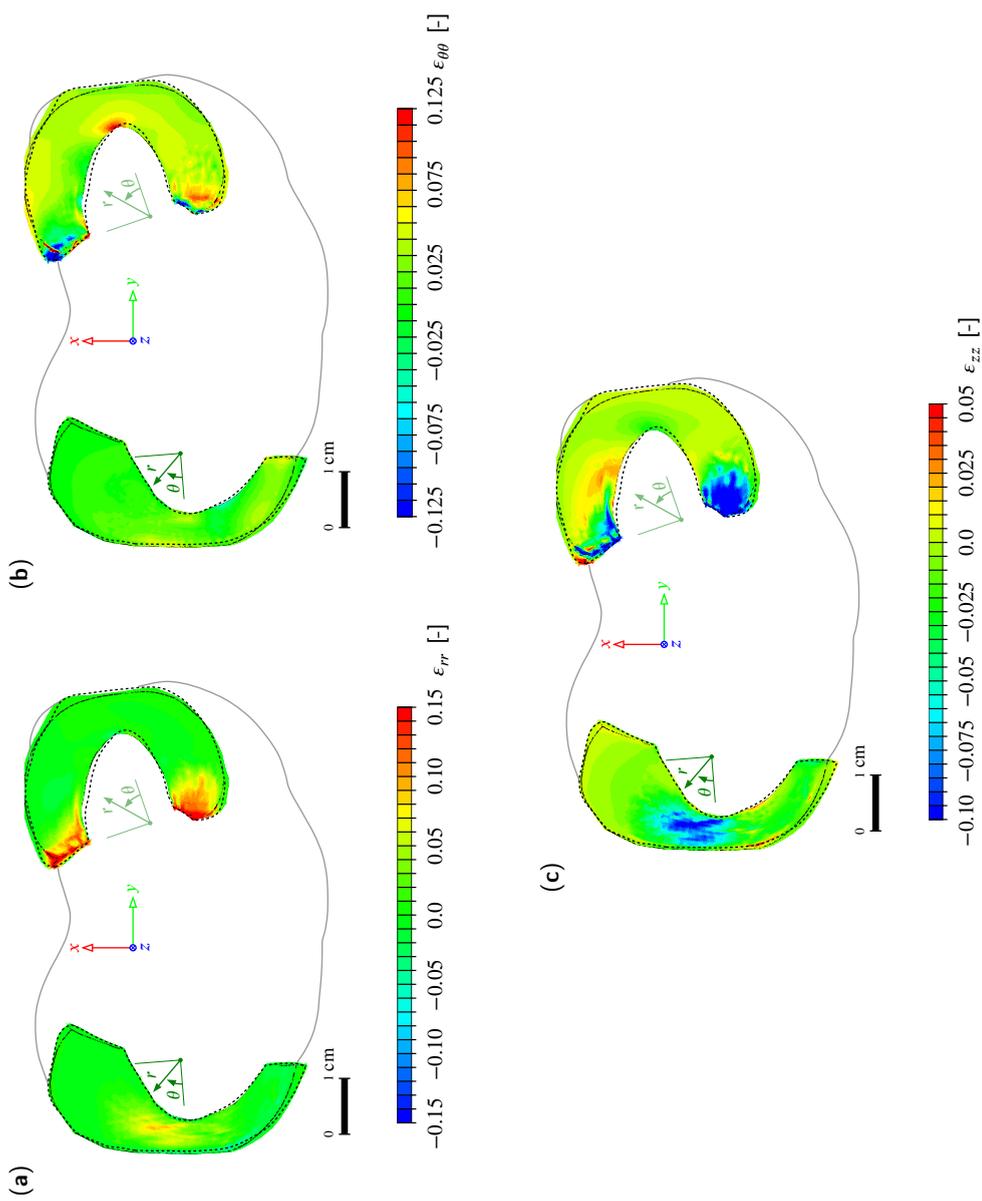
 
\begin{center}
\include{Menisci_Results_II}
\end{center}
\caption{The distribution of the strain in: \textbf{a} $r$-direction; \textbf{b} $\theta$-direction and \textbf{c} $z$-direction, for the case of Holzapfel-Gasser-Ogden (HGO) model under quasi-static loading conditions. The dashed lines indicate the reference positions of the medial and lateral menisci ({i.e.} left and right, respectively). Similar to isotropic Ogden and linear elastic models in Fig.~\ref{fig:MenResI}, the deformation is greater in the lateral meniscus (right) than the medial meniscus (left). HGO model yields larger deformation as well as menisci movement than Ogden model with movement that is more irregular in the lateral meniscus and more localised in the medial meniscus.}
\label{fig:MenResII}
\end{figure}

To study the spatial dependency, we compare the choice of isotropic vs anisotropic hyperelasticity. We consider isotropic Ogden and anisotropic HGO models in Sec.~\ref{subsec:HyperElas} and adopt the radial and circumferential directions of the middle slice experimental data in Sec.~\ref{subsec:MenExpda} ({i.e.} RM and CM, respectively, shown in Fig.~\ref{fig:FitResII}). Assuming \emph{isotropy}, two sets of Ogden model parameters are possible, that can be obtained by using the experimental data in radial or circumferential directions, separately. In the previous section, we obtained the set of parameters associated with the circumferential direction data (CM). Thus, the radial direction experimental data (RM) are fitted to isotropic Ogden model using the procedure in the previous section. Two parameters are found to be sufficient, reporting a root-mean-square error of $\approx 0.1$~$\mathrm{MPa}$. In the context of \emph{anisotropy}, menisci are modelled as transversely isotropic with circumferentially oriented and perfectly aligned collagen fibres, {i.e.} $\kappa=0$ in HGO model. Hence, the full set of experimental data is needed to obtain HGO model material parameters. Simultaneous fitting of Cauchy's stress is performed using a nonlinear least squares method, see Appendix~\ref{sec:FitExpDa}. HGO model is capable of fitting the data with a root-mean-square error of $\approx 0.93$~$\mathrm{MPa}$. The fitting results are plotted in Fig.~\ref{fig:FitResII} and the obtained parameters are reported in Table~\ref{tab:FitParaII}.
\renewcommand{\arraystretch}{1.5}
\begin{table}[!ht]
\caption{The fitting parameters of the Ogden model ($N=1$) and Holzapfel-Gasser-Ogden (HGO) model.}
\label{tab:FitParaII}
\begin{center}
\begin{threeparttable}
\begin{tabular}{c c c c}
\hline
\multicolumn{4}{c}{(\emph{i}) Ogden model ($N=1$) Model.} \\
\hline
$\mu_{1}$~[$\mathrm{MPa}$] & $\alpha_{1}$~[$-$] & {-} & {-} \\
\hline
$205.45$ & $0.49$ & {-} & {-} \\
\hline
\multicolumn{4}{c}{(\emph{ii}) Holzapfel-Gasser-Ogden (HGO) model.} \\
\hline
$\mu$~[$\mathrm{MPa}$] & $\kappa$~[$-$] & $\kappa_{1}$~[$\mathrm{MPa}$] & $\kappa_{2}$~[$-$]  \\
\hline
$4.50$  &  $0.0$   &  $24.49$  &  $6.21$   \\
\hline
\end{tabular}
\end{threeparttable}
\end{center}
\end{table}
\renewcommand{\arraystretch}{1.0}
\\
Figs.~\ref{fig:CPTibII}(\textbf{a}) and (\textbf{b}) show the distributions of the contact pressure and $1^{\mathrm{st}}$ principal stress, respectively, in the tibial cartilage at the maximum load of power walking case. Similar contact pattern is obtained such that the femoral cartilage is in contact with the anterior and posterior horns of both the lateral meniscus and the middle of medial meniscus. The maximum contact pressure is $p_{\mathrm{c}}^{\mathrm{max}}=31.20$~$\mathrm{MPa}$ which is lower than Ogden model in Sec.~\ref{subsec:LinVsNon}. The contact area is larger than Ogden model. Hence, HGO model provides a softer response than isotropic circumferential Ogden model which can be explained by the contribution of the radial direction to the total deformation, {i.e.}~the instantaneous elastic modulus is an order of magnitude higher in the circumferential direction ({e.g.} $E_{r}=49.5$~$\mathrm{MPa}$ and $E_{\theta}=525.6$~$\mathrm{MPa}$ at $\lambda_{r} = \lambda_{\theta} = \lambda_{\theta}=1.1$). Fig.~\ref{fig:CPTibII}(\textbf{b}) shows larger areas are under compressive $1^{\mathrm{st}}$ principal stress with a maximum tensile value of $\sigma_{1}^{\mathrm{max}}=1.03$~$\mathrm{MPa}$ that is lower than the case of circumferential Ogden model. 
\\
Similarly, the maximum value takes place in the region between the contact areas of the femoral cartilage with the tibial cartilage and lateral meniscus' anterior horn due to local stretching.  
\\
The deformation of the menisci is illustrated in Figs.~\ref{fig:MenResII}. Results show that the deformation is larger than Ogden model (see Fig.~\ref{fig:MenResI}) which can be attributed to the anisotropy. Similar to the isotropic case, the deformation is greater in the lateral meniscus. A comparison of the results in Tables~\ref{tab:MenMoveI} and~\ref{tab:MenMoveII} indicates that the menisci movement is larger with respect to Ogden model, more irregular in the lateral meniscus and more localised in the medial meniscus.
\renewcommand{\arraystretch}{1.5}
\begin{table}[!ht]
\caption{The minimum and maximum movement of the menisci at the maximum load for the Holzapfel-Gasser-Ogden (HGO) model.}
\label{tab:MenMoveII}
\begin{center}
\begin{threeparttable}
\begin{tabular}{c c c c c}
\hline
$\mathrm{Meniscus}$ & \multicolumn{2}{c}{$u_{r}$~[$\mathrm{mm}$]} & \multicolumn{2}{c}{$u_{\theta}$~[$\mathrm{mm}$]} \\
\hline
$\mathrm{LM}$ & $-0.57$~($-86.30^{\circ}$) & $1.18$~($-5.97^{\circ}$) & $-0.29$~($-86.30^{\circ}$) & $1.24$~($6.36^{\circ}$)  \\
$\mathrm{MM}$ & $-2.02$~($87.64^{\circ}$) & $2.05$~($-25.56^{\circ}$) & $-0.58$~($-3.38^{\circ}$) & $0.29$~($-55.15^{\circ}$)  \\
\hline
\end{tabular}
\begin{tablenotes}
    \item[$\dagger$] The circumferential position $\theta$ is provided in the parenthesis.
\end{tablenotes}
\end{threeparttable}
\end{center}
\end{table}
\renewcommand{\arraystretch}{1.0}
\\
Figs.~\ref{fig:MenPlotII}(\textbf{a}) and (\textbf{b}) show the different strain components along the radial and circumferential paths in the lateral and medial menisci, for isotropic Ogden models of the radial and tangential directions and for HGO model respectively. Similar behaviour is observed for radial and tangential isotropic Ogden models with small difference in magnitudes. HGO model shows slightly different distributions from Ogden models due to the anisotropy. The maximum values of the strain components are in the same locations and larger in the radial anterior and posterior paths and smaller in the radial central and circumferential paths in the case of Ogden isotropic models.
\begin{figure}[!ht]
\begin{center}
\include{Menisci_Plots_Results_II}
\end{center}
\caption{The distribution of the strain components in: \textbf{a} lateral and \textbf{b} medial menisci, and along the radial anterior, central, posterior and circumferential paths as defined in Fig.~\ref{fig:FEModel}(\textbf{c}). (where $k$ indicates the path with $k \equiv \mathrm{RA}$, $\mathrm{RC}$, $\mathrm{RP}$ and $\mathrm{CM}$, respectively). $\xi_{k}$ is the local coordinate normalised by the total length of the path that is measured from the white to the red for the radial paths and from the posterior to the anterior for the circumferential path. The solid, dashed and dash-dot lines represent the HGO, radial and circumferential Ogden models, respectively. HGO model shows slightly different distributions from Ogden models due to the anisotropy. The maximum values of the strain components are in the same locations and larger in the radial anterior and posterior paths and smaller in the radial central and circumferential paths in the case of Ogden isotropic models.}
\label{fig:MenPlotII}
\end{figure}
%
%
\subsection{Time-dependent behaviour}
\label{subsec:Timebehav}

\begin{figure}
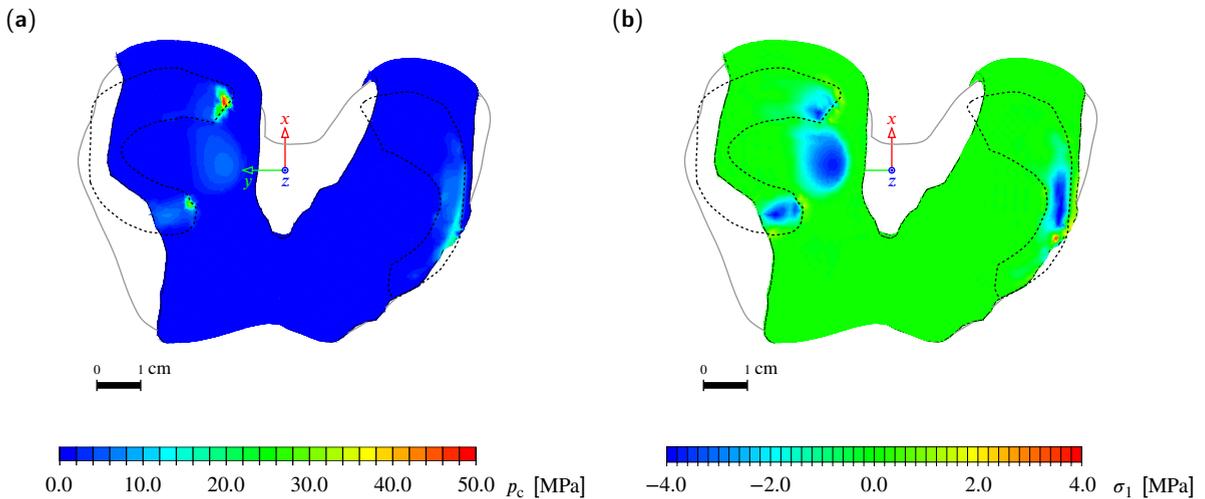
 
\begin{center}
\include{Tibial_Results_III}
\end{center}
\caption{The distributions of: \textbf{a} the contact pressure $p_{\mathrm{c}}$; and \textbf{b} the $1^{\mathrm{st}}$ principal stress $\sigma_{1}$, for the case of time-dependent Holzapfel-Gasser-Ogden (HGO) model under dynamic loading conditions. The dashed lines indicate the current locations of the lateral and medial menisci ({i.e.}~left and right, respectively). Time-dependent HGO model yields similar contact pattern with larger maximum pressure and contact area in comparison with the time-independent model in Fig.~\ref{fig:CPTibII}. Larger areas are under compressive $1^{\mathrm{st}}$ principal stress that is significantly larger than the case of the time-independent HGO model with the maximum value taking place in the region between the contact areas of the medial meniscus with the tibial cartilage close to the attachment due to local stretching.}
\label{fig:CPTibIII}
\end{figure}

\begin{figure}
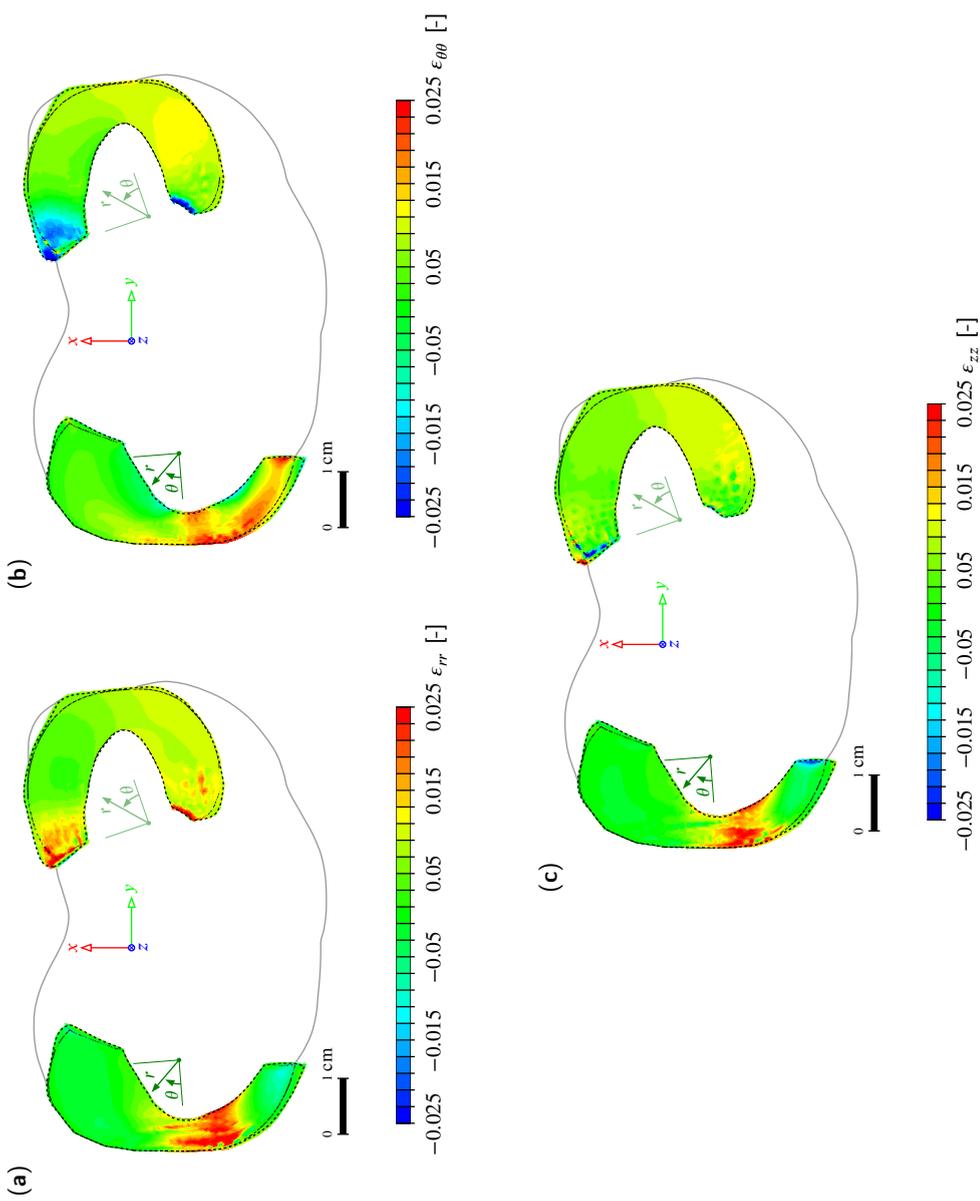
 
\begin{center}
\include{Menisci_Results_III}
\end{center}
\caption{The distribution of the strain in: \textbf{a} $r$-direction; \textbf{b} $\theta$-direction and \textbf{c} $z$-direction, for the case of time-dependent Holzapfel-Gasser-Ogden (HGO) model under dynamic loading conditions. The dashed lines indicate the reference positions of the medial and lateral menisci ({i.e.}~left and right, respectively). Results show that the deformation is significantly smaller than the time-independent models in Figs.~\ref{fig:MenResI} and ~\ref{fig:MenResII} due to the viscous contribution. The deformation is greater in the medial than lateral meniscus which is opposite to the case of the pure elastic models.}
\label{fig:MenResIII}
\end{figure}

\begin{figure} 
\begin{center}
%
%
\readdata{\DataXa}{HGO_Quasi_Static_Anterior_LM_Results.txt}
\readdata{\DataXb}{HGO_Quasi_Static_Central_LM_Results.txt}
\readdata{\DataXc}{HGO_Quasi_Static_Posterior_LM_Results.txt}
\readdata{\DataXd}{HGO_Quasi_Static_Circumferential_LM_Results.txt}
\readdata{\DataXe}{HGO_Quasi_Static_Anterior_MM_Results.txt}
\readdata{\DataXf}{HGO_Quasi_Static_Central_MM_Results.txt}
\readdata{\DataXg}{HGO_Quasi_Static_Posterior_MM_Results.txt}
\readdata{\DataXh}{HGO_Quasi_Static_Circumferential_MM_Results.txt}
\readdata{\DataXIIa}{HGO_P_Walking_Anterior_LM_Results.txt}
\readdata{\DataXIIb}{HGO_P_Walking_Central_LM_Results.txt}
\readdata{\DataXIIc}{HGO_P_Walking_Posterior_LM_Results.txt}
\readdata{\DataXIId}{HGO_P_Walking_Circumferential_LM_Results.txt}
\readdata{\DataXIIe}{HGO_P_Walking_Anterior_MM_Results.txt}
\readdata{\DataXIIf}{HGO_P_Walking_Central_MM_Results.txt}
\readdata{\DataXIIg}{HGO_P_Walking_Posterior_MM_Results.txt}
\readdata{\DataXIIh}{HGO_P_Walking_Circumferential_MM_Results.txt}
%
%
\begin{pspicture}[showgrid=false](0,0)(16.0,8.0)
%
\rput{0.0}(-3.0mm,-3.0mm){
\rput{0.0}(-1.0mm,4.0mm){
\rput{0.0}(15mm,8.5mm){
%
\psset{xunit=13.0mm, yunit=9.2857mm}
{\small
\psaxes[axesstyle=frame,linewidth=0.0pt,tickstyle=inner,ticksize=0 1.5pt,subticksize=0.75,xsubticks=10,ysubticks=10,xlabelsep=-3pt,ylabelsep=-3pt,
        xLabels={$$0.0$$,$$0.2$$,$$0.4$$,$$0.6$$,$$0.8$$,$$1.0$$},yLabels={$$-0.02$$,$$0.0$$,$$0.02$$,$$0.04$$,$$0.06$$,$$0.08$$,$$1.00$$,$$1.20$$,}](5.0,7.0)
}
\psset{xunit=6.5mm, yunit=6.5mm}
\psaxes[linewidth=0.4pt,axesstyle=frame,tickstyle=inner,ticksize=0 1.5pt,subticksize=0.75,xsubticks=0,ysubticks=0,
        Dx=0.2,dx=2.0,Ox=0.0,Dy=0.02,dy=1.42857,Oy=0.0,labels=none]{->}(0.0,0.0)(10.0,10.0)
%
\pstScalePoints(10.0,71.42857){}{-0.02 sub}
%
\rput{0.0}(0.0,-8.0mm){
\rput{0.0}(5,0.0){\small $\xi_{k}$~$[\mathrm{-}]$}
}
\rput{0.0}(-11.0mm,0.0){
\rput{90.0}(0.0,5.0){\small $\varepsilon_{ij}^{(k)}$~$[\mathrm{-}]$}
}
\listplot[plotNoX=2,plotNo=2,plotNoMax=2,plotstyle=line,linecolor=black!80,linewidth=1.25pt,linestyle=solid,xEnd=1.0,yStart=-0.04,yEnd=40]{\DataXa}
\listplot[plotNoX=2,plotNo=2,plotNoMax=2,plotstyle=line,linecolor=red!80,linewidth=1.25pt,linestyle=solid,xStart=0.0,xEnd=1.0,yStart=-0.04,yEnd=40]{\DataXb}
\listplot[plotNoX=2,plotNo=2,plotNoMax=2,plotstyle=line,linecolor=blue!80,linewidth=1.25pt,linestyle=solid,xStart=0.0,xEnd=1.0,yStart=-0.04,yEnd=40]{\DataXc}
\listplot[plotNoX=2,plotNo=2,plotNoMax=2,plotstyle=line,linecolor=darkpastelgreen,linewidth=1.25pt,linestyle=solid,xStart=0.0,xEnd=1.0,yStart=-0.04,yEnd=0.12]{\DataXd}
\listplot[plotNoX=2,plotNo=2,plotNoMax=2,plotstyle=line,linecolor=black!80,linewidth=1.25pt,linestyle=dashed,dash=2.0pt 1.0pt,xStart=0.0,xEnd=1.0,yStart=-0.04,yEnd=40]{\DataXIIa}
\listplot[plotNoX=2,plotNo=2,plotNoMax=2,plotstyle=line,linecolor=red!80,linewidth=1.25pt,linestyle=dashed,dash=2.0pt 1.0pt,xStart=0.0,xEnd=1.0,yStart=-0.04,yEnd=40]{\DataXIIb}
\listplot[plotNoX=2,plotNo=2,plotNoMax=2,plotstyle=line,linecolor=blue!80,linewidth=1.25pt,linestyle=dashed,dash=2.0pt 1.0pt,xStart=0.0,xEnd=1.0,yStart=-0.04,yEnd=40]{\DataXIIc}
\listplot[plotNoX=2,plotNo=2,plotNoMax=2,plotstyle=line,linecolor=darkpastelgreen,linewidth=1.25pt,linestyle=dashed,dash=2.0pt 1.0pt,xStart=0.0,xEnd=1.0,yStart=-0.04,yEnd=40]{\DataXIId}
%
\psset{xunit=10mm, yunit=10mm}
\rput[tl]{0.0}(0.10cm,4.80cm){
\psframe[linewidth=0.25pt,fillstyle=solid,fillcolor=white](0.0,0.0)(1.25,1.60)
\psline[linewidth=1.5pt,linestyle=solid,dash=2.0pt 1.0pt,linecolor=black!80]{-}(0.075,0.20)(0.40,0.20)
\rput[l](0.55,0.20){\footnotesize $\varepsilon_{\theta\theta}^{\mathrm{RA}}$}
\psline[linewidth=1.5pt,linestyle=solid,dash=0.5pt 0.75pt 2.0pt 0.75pt,linecolor=red!80]{-}(0.075,0.6)(0.40,0.6)
\rput[l](0.55,0.6){\footnotesize $\varepsilon_{\theta\theta}^{\mathrm{RC}}$}
\psline[linewidth=1.5pt,linestyle=solid,linecolor=blue!80]{-}(0.075,1.0)(0.40,1.0)
\rput[l](0.55,1.0){\footnotesize $\varepsilon_{\theta\theta}^{\mathrm{RP}}$}
\psline[linewidth=1.5pt,linestyle=solid,linecolor=darkpastelgreen]{-}(0.075,1.40)(0.40,1.40)
\rput[l](0.55,1.40){\footnotesize $\varepsilon_{rr}^{\mathrm{CM}}$}
}
}
}
}
%
%
\rput{0.0}(80.0mm,-3.0mm){
\rput{0.0}(-1.0mm,4.0mm){
\rput{0.0}(15mm,8.5mm){
%
\psset{xunit=13.0mm, yunit=9.2857mm}
{\small
\psaxes[axesstyle=frame,linewidth=0.0pt,tickstyle=inner,ticksize=0 1.5pt,subticksize=0.75,xsubticks=10,ysubticks=10,xlabelsep=-3pt,ylabelsep=-3pt,
        xLabels={$$0.0$$,$$0.2$$,$$0.4$$,$$0.6$$,$$0.8$$,$$1.0$$},yLabels={$$-0.04$$,$$-0.02$$,$$0.0$$,$$0.02$$,$$0.04$$,$$0.06$$,$$0.08$$,$$1.00$$,$$1.20$$,}](5.0,7.0)
}
\psset{xunit=6.5mm, yunit=6.5mm}
\psaxes[linewidth=0.4pt,axesstyle=frame,tickstyle=inner,ticksize=0 1.5pt,subticksize=0.75,xsubticks=0,ysubticks=0,        Dx=0.2,dx=2.0,Ox=0.0,Dy=0.02,dy=1.42857,Oy=0.0,labels=none]{->}(0.0,0.0)(10.0,10.0)
%
\pstScalePoints(10.0,71.42857){}{-0.04 sub}
%
\rput{0.0}(0.0,-8.0mm){
\rput{0.0}(5,0.0){\small $\xi_{k}$~$[\mathrm{-}]$}
}
\rput{0.0}(-11.0mm,0.0){
\rput{90.0}(0.0,5.0){\small $\varepsilon_{ij}^{(k)}$~$[\mathrm{-}]$}
}
\listplot[plotNoX=2,plotNo=2,plotNoMax=2,plotstyle=line,linecolor=black!80,linewidth=1.25pt,linestyle=solid,xStart=0.0,xEnd=1.0,yStart=-0.04,yEnd=40]{\DataXe}
\listplot[plotNoX=2,plotNo=2,plotNoMax=2,plotstyle=line,linecolor=red!80,linewidth=1.25pt,linestyle=solid,xStart=0.0,xEnd=1.0,yStart=-0.04,yEnd=40]{\DataXf}
\listplot[plotNoX=2,plotNo=2,plotNoMax=2,plotstyle=line,linecolor=blue!80,linewidth=1.25pt,linestyle=solid,xStart=0.0,xEnd=1.0,yStart=-0.04,yEnd=40]{\DataXg}
\listplot[plotNoX=2,plotNo=2,plotNoMax=2,plotstyle=line,linecolor=darkpastelgreen,linewidth=1.25pt,linestyle=solid,xStart=0.0,xEnd=1.0,yStart=-0.04,yEnd=40]{\DataXh}
\listplot[plotNoX=2,plotNo=2,plotNoMax=2,plotstyle=line,linecolor=black!80,linewidth=1.25pt,linestyle=dashed,dash=2.0pt 1.0pt,xStart=0.0,xEnd=1.0,yStart=-0.04,yEnd=40]{\DataXIIe}
\listplot[plotNoX=2,plotNo=2,plotNoMax=2,plotstyle=line,linecolor=red!80,linewidth=1.25pt,linestyle=dashed,dash=2.0pt 1.0pt,xStart=0.0,xEnd=1.0,yStart=-0.04,yEnd=40]{\DataXIIf}
\listplot[plotNoX=2,plotNo=2,plotNoMax=2,plotstyle=line,linecolor=blue!80,linewidth=1.25pt,linestyle=dashed,dash=2.0pt 1.0pt,xStart=0.0,xEnd=1.0,yStart=-0.04,yEnd=40]{\DataXIIg}
\listplot[plotNoX=2,plotNo=2,plotNoMax=2,plotstyle=line,linecolor=darkpastelgreen,linewidth=1.25pt,linestyle=dashed,dash=2.0pt 1.0pt,xStart=0.0,xEnd=1.0,yStart=-0.04,yEnd=40]{\DataXIIh}
%
%
\psset{xunit=10mm, yunit=10mm}
\rput[tl]{0.0}(0.10cm,4.80cm){
\psframe[linewidth=0.25pt,fillstyle=solid,fillcolor=white](0.0,0.0)(1.25,1.60)
\psline[linewidth=1.5pt,linestyle=solid,dash=2.0pt 1.0pt,linecolor=black!80]{-}(0.075,0.20)(0.40,0.20)
\rput[l](0.55,0.20){\footnotesize $\varepsilon_{\theta\theta}^{\mathrm{RA}}$}
\psline[linewidth=1.5pt,linestyle=solid,dash=0.5pt 0.75pt 2.0pt 0.75pt,linecolor=red!80]{-}(0.075,0.6)(0.40,0.6)
\rput[l](0.55,0.6){\footnotesize $\varepsilon_{\theta\theta}^{\mathrm{RC}}$}
\psline[linewidth=1.5pt,linestyle=solid,linecolor=blue!80]{-}(0.075,1.0)(0.40,1.0)
\rput[l](0.55,1.0){\footnotesize $\varepsilon_{\theta\theta}^{\mathrm{RP}}$}
\psline[linewidth=1.5pt,linestyle=solid,linecolor=darkpastelgreen]{-}(0.075,1.40)(0.40,1.40)
\rput[l](0.55,1.40){\footnotesize $\varepsilon_{rr}^{\mathrm{CM}}$}
}
}
}
}
%
\rput{0.0}(0.0cm,7.75cm){\small (\textbf{a})}
\rput{0.0}(8.30cm,7.75cm){\small (\textbf{b})}
%
\end{pspicture}
%
%
\end{center}
\caption{The distribution of the strain components in: \textbf{a} lateral and \textbf{b} medial menisci, and along the radial anterior, central, posterior and circumferential paths as defined in Fig.~\ref{fig:FEModel}(\textbf{c}). (where $k$ indicates the path with $k \equiv \mathrm{RA}$, $\mathrm{RC}$, $\mathrm{RP}$ and $\mathrm{CM}$, respectively). $\xi_{k}$ is the local coordinate normalised by the total length of the path that is measured from the white to the red for the radial paths and from the posterior to the anterior for the circumferential path. The solid and dashed lines represent the time-independent and dependent HGO models, respectively. In the lateral meniscus, the strain components are significantly lower and show less variation over the radial paths. Along the circumferential path, similar behaviour is observed. In the medial meniscus, the strain components vary slightly and differently. The maximum values of the strain components are in the same locations and difference in magnitudes can be attributed to the stiff behaviour of time-dependent HGO model with respect to the time-independent HGO model.}
\label{fig:MenPlotIII}
\end{figure}
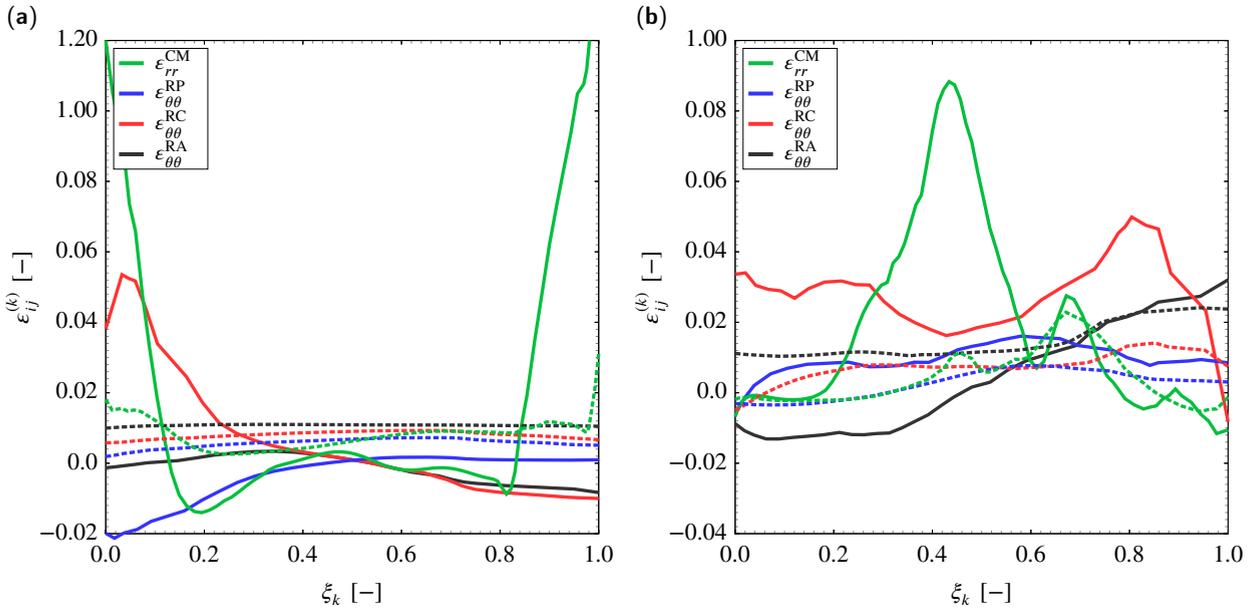

In this section, we explore the time-dependent behaviour of the menisci. We adopt the finite strain-linear viscoelaticity formulation in Sec~\ref{subsec:HyperElas}. The time-independent behaviour ({i.e.}~long-term response) is modelled using the anisotropic HGO model in Sec.~\ref{subsec:SpatDepend} and the parameters in Table~\ref{tab:FitParaII}. The linear viscoelastic model of generalised Maxwell type (Prony series) is taken to determine the time-dependent behaviour. To obtain the linear viscoelastic model parameters, we use the confined compression test data. Using a nonlinear least squares method, the generalised Kelvin model parameters are obtained by fitting the creep compliance to the experimental data, see Appendix~\ref{sec:ConfComp}. Consequently, the generalised Maxwell parameters are determined using the interconversion method in Appendix~\ref{sec:EstVisco}. It is worth mentioning that the retardation and relaxation times ($\eta_{i}$ and $\tau_{i}$, respectively) are taken to be equal. The fitting parameters of the generalised Kelvin and Maxwell models are reported in Table~\ref{tab:FitParaIII}.
\renewcommand{\arraystretch}{1.5}
\begin{table}[!ht]
\caption{The fitting parameters of the generalised Kelvin and Maxwell models.}
\label{tab:FitParaIII}
\begin{center}
\begin{threeparttable}
\begin{tabular}{c c c c}
\hline
$i$ & $\bar{j}_{i} = \bar{d}_{i}$~[$-$] & $\bar{k}_{i} = \bar{g}_{i}$~[$-$] & $\tau_{i} = \eta_{i}$~[$\mathrm{s}$]     \\
\hline
$1$   &  $4.41$   &  $0.777$  &  $1.12$     \\
$2$   &  $13.47$   &  $0.133$  &  $63.77$    \\
$3$   &  $18.27$    &  $0.052$  &  $837.18$   \\
$4$   &  $54.74$    &  $0.008$  &  $4899.07$  \\
\hline
\end{tabular}
\end{threeparttable}
\end{center}
\end{table}
\renewcommand{\arraystretch}{1.0}
\\
We consider the power walking activity case and the average loading parameters in Table~\ref{tab:LoadPara}, {i.e.}~$\hat{F} \approx 1782.42~\mathrm{N}$ and $\omega = 1.7~\mathrm{Hz}$. Dynamic analysis using implicit time integration is used wherein the sinusoidal force in Eq.~(\ref{eq:FII}) is applied. The finite strain-linear viscoelasticity model is then compared to the time-independent HGO model in Sec~\ref{subsec:SpatDepend}.
\\
Figs.~\ref{fig:CPTibIII}(\textbf{a}) and (\textbf{b}) show the distributions of the contact pressure and $1^{\mathrm{st}}$ principal stress, respectively, in the tibial cartilage at the maximum load. Similar contact pattern is obtained with larger maximum pressure of $p_{\mathrm{c}}^{\mathrm{max}}=51.23$~$\mathrm{MPa}$ and contact area with respect to the time-independent HGO model. Hence, the viscoelastic HGO model yields a stiffer response than the time-independent HGO model due to the viscous dissipation, {i.e.}~the characteristic time of the power walking is $\approx 0.59~\mathrm{s}$, which is comparable with the material relaxation time at ambient temperature ($\approx 1~\mathrm{s}$). Fig.~\ref{fig:CPTibIII}(\textbf{b}) shows that larger areas are under compressive $1^{\mathrm{st}}$ principal stress with a maximum tensile value of $\sigma_{1}^{\mathrm{max}}=4.0$~$\mathrm{MPa}$ that is considerably larger than the case of the time-independent HGO model. In contrary fashion, the maximum value takes place in the region between the contact areas of the medial meniscus with the tibial cartilage close to the attachment due to local stretching.  
\\
The deformation of the menisci is illustrated in Fig.~\ref{fig:MenResIII}. Results show that the deformation is significantly smaller than the time-independent  HGO model in Fig.~\ref{fig:MenResII} due to the viscous contribution. The deformation is greater in the medial than lateral meniscus which is opposite to the case of the time-independent models. The movement of the menisci is negligible, which can be explained by the difference in stiffness between menisci and tibial cartilage that limits the sliding. 
\\
Figs.~\ref{fig:MenPlotIII}(\textbf{a}) and (\textbf{b}) show the different strain components along the radial and circumferential paths in the lateral and medial menisci, respectively, for time-independent and time-dependent HGO models. In the lateral meniscus, the strain components are significantly lower and show less variation over the radial paths. Along the circumferential path, similar behaviour is observed. In the medial meniscus, the strain components vary slightly and differently. The maximum values of the strain components are in the same locations and difference in magnitudes can be attributed to the stiff behaviour of time-dependent HGO model with respect to the time-independent models. 
%
%
%
\section{Discussion on material models}
\label{Conc}

In the second part of this paper, we explored the impact of the choice of the material model for the menisci. In particular, we investigated the importance of linearity vs nonlinearity, isotropy vs anisotropy, as well as time-dependency vs independency. We considered the commonly used constitutive models in the literature, namely, isotropic linear elastic, isotropic and anisotropic hyperelastic, and anisotropic finite strain-linear viscoelatic models. The same sets of experimental data from the literature \citep{Proctor1989} were adopted to estimate the material parameters. A subject specific knee under quasi-static and dynamic loading conditions was analysed using the finite element method to compare the meniscus models. The knee model was evaluated in terms of the contact pressure and $1^{\mathrm{st}}$ principal stress of the tibial cartilage; the deformation of the menisci; and the maximum strain on menisci tearing paths ({i.e.}~the factors affect knee osteoarthritis (OA)). The main findings are summarised below:
\setlength\itemsep{0.5em}
\begin{enumerate}[leftmargin=*,align=left]
  \item The isotropic linear elastic and nonlinear hyperelastic models yield similar behaviour. The distributions of the contact pressure and $1^{\mathrm{st}}$ principal stress in the tibial cartilage as well as the strain of menisci are comparable.   
  \item The linear elastic model results in a stiffer response that the Ogden (non-linear elastic model). The contact pressure and $1^{\mathrm{st}}$ principal stress are larger ({i.e.}~the difference is $5.3~\mathrm{MPa}$ in the maximum contact pressure and $0.25~\mathrm{MPa}$ in the maximum $1^{\mathrm{st}}$ principal stress). Contact area is smaller in the tibial cartilage. Further, the menisci are less deformed.
  \item The anisotropic hyperelastic (HGO) model shows a softer behaviour than the isotropic Ogden models with similar distributions of the contact pressure and $1^{\mathrm{st}}$ principal stress in the tibial cartilage ({i.e.}~the difference is $1.2~\mathrm{MPa}$ in the maximum contact pressure and $0.15~\mathrm{MPa}$ in the maximum $1^{\mathrm{st}}$ principal stress). The deformation pattern of the menisci is different from the isotropic case due to anisotropy.
  \item The time-dependent anisotropic hyperelastic (HGO) model shows a remarkably stiffer response than the time-independent anisotropic hyperelastic (HGO). Time-dependent HGO yields larger and similar distributions of the contact pressure and $1^{\mathrm{st}}$ principal stress in the tibial cartilage ({i.e.}~the difference is $20.0~\mathrm{MPa}$ in the maximum contact pressure and $3.97~\mathrm{MPa}$ in the maximum $1^{\mathrm{st}}$ principal stress in comparison with the time-independent HGO) and lower deformation of the menisci.
  \item The measurements of menisci deformation imply that its magnitude depend on the location such that the circumferential is slightly larger than the radial strain ({i.e.}~$<3\%$) \citep{Freutel2014}. The time-dependent HGO model yields comparable values of the circumferential and radial strain components ({i.e.}~$<2.5\%$).
  \item The menisci move radially and circumferentially to accommodate the load ({i.e.}~stabilise the knee) due to the contact of the cartilage. This movement is reduced in case of a stiffer response. Moreover, it is considerably different when isotropic/anisotropic models are adopted. However, modelling the movement of the menisci under loading is highly influenced by the applied boundary conditions (i.e.~meniscal horns are fixed to the bones which is a simplification of what happens in reality). 
  \item The experimental studies suggest that the menisci movement is irregular and depends on location within the meniscus \citep{Kawahara2001,Patel2004,Shefelbine2006,Fowlie2011,Freutel2014}. The pars intermedia moves the most and the regions close to the horns move the least, which is strongly dependent on the loading level. Typical displacement is less than $3~\mathrm{mm}$, which is consistent with the time-dependent HGO model prediction. Nevertheless, in these experiments, a large flexion angle is employed ({i.e.}~$>30^{\circ}$) which is expected to cause larger movement in comparison with the adopted loading conditions ({i.e.}~zero flexion angle). Hence, the time-dependent HGO model seems to predict the movement as well as the irregularity.
  \item The strain at failure depends on location and direction within the meniscus \citep{Tissakht1995,Villegas2007,Peloquin2016}. The reported values are in the range $12$-$36\%$ for the circumferential direction and $17$-$44\%$ for the radial direction. The strain distribution along the tearing paths of time-independent models predict comparable values. The time-dependent HGO model yields significantly smaller values which is more realistic for a power walking activity. 
\end{enumerate}
These calculations provide an insight into effects of the choice of a material model for the menisci which determine how contact stresses develop within cartilage and how menisci deform under loading. The results show that the different models yield different contact stresses in the cartilages and deformation of the menisci. The time-dependent HGO model yields menisci deformation and movement comparable to the experimental values. Thus, this understanding can be used to motivate devising a physically based model which provide an accurate description of the time-independent, time-dependent and spatial variability of the meniscus tissue. 
%
%
\section{Identification of model parameters and model comparison based on Bayesian inference}
\label{sec:ParamIden}

It is evident that a central task within biomechanical/biomedical field is to build tools to identify and compare the models which describe the gathered data.
Often, two levels of inference are employed to model the data. At the first level, we assume that each proposed model is true, and then we identify the parameters of the individual model by fitting the model to the data. Fitting the model to the data includes inferring what values should the model parameters take. The results of this inference often are summarised by deterministic values (e.g.~the most probable parameter values) with some error bars on those parameters. This is repeated for each model. At the second level, the models are compared based on the data by assigning some preference or ranking to the models. For example, one can compare models based on the best fitting to the data (i.e.~minimum error) or based on the lowest number of parameters to be identified. Complex models, involving a large number of parameters, tend to have minimum errors but can lead to over-fitting issues. Therefore, it is important to have tools that can guide researchers in the choice of the models to adopt and in the rationale behind their comparison.

Here, we presented the knee meniscus as a case study. One can model the experimental data by any of the constitutive descriptions given in  Sec.~\ref{sec:ConstitDes}. At the first level of inference, we consider each constitutive model individually and we identify the relative parameters (e.g.~Young's modulus). At the second level of inference, the goal is to rank the models based on our particular set of data. The second level enables us to make statements such as:~``the neo-Hookean model describes our data the best'' or ``if the Ogden model fits our data the best, what order for the model should be used?''  

As previously reported, parameter identification frameworks (such as the ones presented in this work) are often based on the method of least squares, where the optimal parameter values are obtained by minimising the sum of squared residuals; an excellent review of the history of material parameter identification is provided in \cite{R_Mahnken_2014}.

Bayesian methods are able consistently and quantitatively to solve both these inference tasks. A framework based on Bayesian inference (BI) uses probability as its logic \cite{J_Beck_2010}. Bayesian frameworks enable the user to address uncertainties due to insufficient information alongside the identification of the mathematical model parameters. In a Bayesian setting, the user's uncertainty about the parameters is represented by probability distributions. The user's initial belief (about a parameter or model) is presented as probability distribution which later is updated with the information of the measurements according to Bayes’ rule \cite{A_Gelman_2003}. For example, the fact that Young's modulus is non-negative can be imposed by the user systematically by choosing a probability distribution which is only defined on $\mathbb{R\geq 0}$. As the user's initial belief gets updated by the measurements, this initial probability distribution will change.

\begin{figure}[h]
\begin{center}
\tikzset{block/.style= {draw, rectangle, align=center,minimum width=2cm,minimum height=1cm},}
\begin{tikzpicture}
\node [black,block] (a) {Bayes' rule\\$\pi(\textbf{p}|z)=\frac{\pi(\textbf{p})\pi(z|\textbf{p})}{\pi(z)}$};
\node [draw=none, fill=none,above=of a] (b) {};
\node [black,block, left=of b] (c) {Measurements};
\node [black,block, right=of b] (d) {Model};
\node [black,block, left=of a] (e) {Prior\\original belief\\($\pi(\textbf{p}$))};
\node [black,block, right=of a] (f) {Posterior\\($\pi(\textbf{p}|z$))};
\draw[black,thick,->] (d.south) -- (a);
\draw[black,thick,->] (c.south) -- (a);     
\draw[black,thick,->] (e.east)--(a.west);
\draw[black,thick,->] (a.east)--(f.west);
\draw[black,<-,bend right=45, thick] 
(e.south) to node[auto, swap] {Updating}(f.south); ;
\end{tikzpicture}
\end{center}
\caption{A schematic presentation of a Bayesian framework for parameter identification. Once the measurements are obtained and the model is selected, Bayes' rule is employed to update the user-selected prior. Note that $\pi(z|\textbf{p})$ is the likelihood function and describes the plausibility of obtaining the measurements $z$ given the parameter set $\textbf{p}$ \cite{H_Rappel_2018_3}.}
\label{Bayesian schematic}
\end{figure}
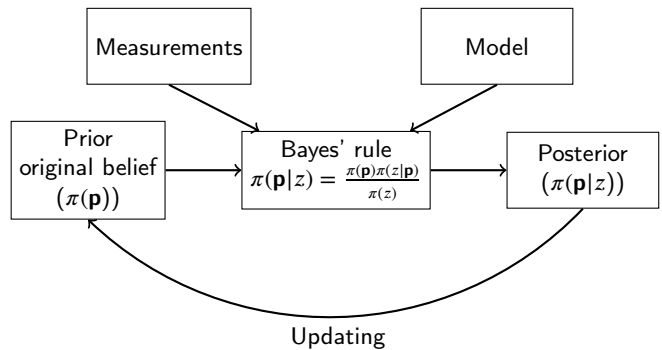

Fig.~\ref{Bayesian schematic} gives a schematic presentation of a Bayesian framework for parameter identification. First, the measurements and the model for which its parameters are to be identified are collected. Second, the likelihood function ($\pi(z|\textbf{p})$, describes the plausibility of a measurement set $z$, given the model parameters $\textbf{p}$ \cite{T_Catanach_2017}), is constructed by incorporating the measurements and model in the Bayes' rule block, and the prior ($\pi(\textbf{p})$, i.e.~the user's initial belief), is updated. Once the posterior ($\pi(\textbf{p}|z)$, i.e.~the result of the Bayes' rule block, is obtained for the first measurement, it can be used as the prior for the next measurement. 

As an example, consider the identification of Young's modulus. Let us assume that every measurement which is made (for a given strain as an input) is written as the linear elasticity law plus an error. This is a realisation from a normal distribution with zero mean and variance\footnote{Variance is a measure for the possible deviation of a random variable from its mean. Large variances indicate large possible differences and vice versa.} $s_{\omega}^{2}$ (i.e.~$N(0,s^{2}_{\omega})=\frac{1}{\sqrt{2\pi}s_{\omega}}\text{exp}\Big(-\frac{\omega^{2}}{2s^{2}_{\omega}}\Big)$), $z=E\epsilon+\omega$. Based on the definition of the likelihood function for a known strain ($\epsilon$), if Young's modulus is given, the only random variable on the right-hand side will be $\omega$. Consequently, the likelihood function, $\pi(z|E)$, which describes the plausibility of $z$ given Young's modulus, is similar to the distribution of the error. This means that one just needs to replace $\omega$ with its equivalent $z-E\epsilon$ (i.e.~$\pi(z|E)=\frac{1}{\sqrt{2\pi}s_{\omega}}\text{exp}\Big(-\frac{(z-E\epsilon)^{2}}{2s^{2}_{\omega}}\Big)$). Once the likelihood is constructed, it can be combined with the prior for Young's modulus (e.g.~truncated normal distribution to impose the fact that Young's modulus can not be negative).  
It is important to note that:
\renewcommand{\labelenumi}{\Roman{enumi}.}
\setlength\itemsep{0.5em}
\begin{enumerate}[leftmargin=*,align=left]
  \item Bayesian analysis directly addresses uncertainty by probability.  
  \item BI makes systematic incorporation of the user's/expert's knowledge in an identification process possible through combining prior information with the measurements.
  \item BI leads to the posterior distribution on which all the analyses are done. There is no need for separate estimation theories, testing and multiple comparisons \cite{P_B_Carlin_2000}. For example, for the unknown parameter $p$, one not only can estimate it, but with little additional effort can obtain an $n\%$ credible region \cite{W_Edwards_1963} in which it is believed the true parameter values are present with $n\%$ certainty \cite{J_Berger_1985}.
\end{enumerate}
Moreover, inverse problems based on least squares methods must be artificially regularised so that a unique solution is found. However, the Bayesian paradigm naturally leads to a regularisation \cite{G_Celeux_2012}, where the solution is made unique based on our a-priori knowledge of the solution, embedded within a \emph{prior}. This is critical when information is scarce, as approaches based on the method of least squares (or generally frequentist approaches) require a \emph{large} amount of data, which is usually not available in biomechanics. 

Furthermore, model comparison is usually a difficult task. Choosing a model that best describes the data is not a straightforward task. More complex models - involving a large number of parameters- always tend to fit the data better. As a result, the model choice based on the fit to data alone (e.g.~maximum likelihood model choice) leads us to select complex implausible models. Bayesian methodology, on the other hand, naturally and quantitatively embodies Occam's razor (i.e.~the principle that states the simplest model/explanation that fits the data should be preferred) \cite{D_MacKay_1992}. Moreover, Bayesian methodology penalises complex models automatically with no need of introducing artificial penalising terms into our model comparison criterion \cite{D_MacKay_1992}. A more detailed discussion about model selection using BI is presented in Sec.~\ref{subsection: model selection}. 

Analysing the posterior (e.g.~calculating mean value, covariance matrix\footnote{A square symmetric matrix in which the off-diagonal components are the covariances of pairs of random variables and diagonal components are the variances of the random variables. The covariance of two random variables $X$ and $Y$ is a measure of their joint variability \cite{J_A_Rice_1995} and is defined as the expected value (the mean) of the product of the deviations of two random variables $X$ and $Y$ from their means $m_{X}$ and $m_{Y}$, i.e.~$\mathrm{cov}(X,Y)=\big<(X-m_{X})(Y-m_{Y})\big>$. The variance of the random variable $X$ is the covariance of the random variable with itself. Readers who are interested in more definitions of statistical concepts are referred to Ref.~\cite{B_Everitt_2010}.}, posterior predictions i.e.~calculating the new measurements based on the current measurements) is not often possible without introducing (numerical) approximations. A relatively easy to follow chapter on Bayesian computation is provided in Ref.~\cite{P_B_Carlin_2000}. The statistical summaries and analyses of the posterior distribution can be approximated by drawing random samples of the unknown parameter from the posterior distribution \cite{D_Calvetti_2007}. Various sampling approaches are given in \cite{P_B_Carlin_2000} and \cite{M_C_Bishop_2006}. Often Markov chain Monte Carlo algorithms (MCMC) are employed to draw samples from the posterior distribution and approximate posterior quantities of interest \cite{W_R_Gilks_1995, S_Brooks_2011}. An important advantage of MCMC techniques is that they are relatively simple to implement, even if the posterior distribution is complex \cite{W_A_Link_2012}.

\paragraph{(Elasticity)} An early study in the field of parameter identification for mechanical models was done by Isenberg \cite{J_Isenberg_1979}, in which a Bayesian framework is employed for the identification of elastic parameters. Subsequently, other studies used BI for the identification of elastic parameters based on dynamic responses \cite{K_Alvin_1997,J_Beck_1998,T_Marwala_2005}. Gogu et al.~\cite{C_Gogu_2010} furthermore provided an introduction to the identification of Young's moduli using BI. Lai and Ip \cite{T_Lai_1996}, Daghia et al.~\cite{F_Daghia_2007}, Nichols et al.~\cite{J_Nichols_2010} and Gogu et al.~\cite{C_Gogu_2013} employed BI to identify the elastic constants of composite and laminate plates. Mohamedou et al.~\cite{M_Mohamedou_2019} used BI for the identification of the resin's Young's modulus in non-aligned short fibre composites.

\paragraph{(Elastoplasticity and hysteretic models)} In elastoplasticity and plasticity BI is used by Most \cite{T_most_2010}, Rappel et al.~\cite{H_Rappel_2018}, Zhang et al.~\cite{Y_Zhang_2019} and Zhang and Needleman \cite{Y_Zhang_2021}. A tutorial on the identification of material parameters in solid mechanics with a focus on elastic and elastoplastic material models is presented in \cite{H_Rappel_2018_2}. Furthermore, Muto and Beck \cite{M_Muto_2008}, Liu and Au \cite{P_Liu_2013} applied the Bayesian approach to hysteretic models.

\paragraph{(Creep, viscoelasticity and viscoplasticity)} Fitzenz et al.~\cite{D_Fitzenz_2007} used a Bayesian framework to identify parameters of a creep model of quartz. Examples of using BI for parameter identification in viscoelasticity are Zhang et al.~\cite{E_Zhang_2013} to identify Young's modulus of a viscoelastic polymer layer in a laminated structure, Mehrez et al.~\cite{L_Mehrez_2015} to identify the viscoelastic properties of aged and unaged asphalt, Miles et al.~\cite{P_Miles_2015} to identify the viscoelastic parameters of a dielectric elastomer, Hernandez et al.~\cite{W_Hernandez_2015} for identification of five viscoelastic parameters and Rappel et al.~\cite{H_Rappel_2017} to identify parameters of the standard linear solid (SLS) model. Moreover, Bayesian updating schemes are employed in some studies to identify material parameters in viscoplasticity \cite{J_Gang_2010,E_Janouchova_2018,E_Adeli_2020,A_Chakraborty_2021}.   

\paragraph{(Biomechanics and biomedical applications)} In biomechanics, BI (probability logic) is employed to identify parameters of constitutive models. For example, Madireddy et al.~\cite{S_Madireddy_2016} states that for constitutive models of soft tissues, due to high variability of experimental data, the parameters of these models have large uncertainties (e.g.~$50\%$ of mean values are reported as standard deviations of soft tissue material parameters in \cite{G_N_Maksym_1997}). This large variability in experimental data is caused by:

\renewcommand{\labelenumi}{\Roman{enumi}.}
\setlength\itemsep{0.5em}
\begin{enumerate}[leftmargin=*,align=left]
\item the complexity of soft tissues and their heterogeneous nature;
\item large variation of material response from patient to patient;
\item correct sample preparation; and
\item wide variation of experimental protocols between different laboratories due to lack of standards.      
\end{enumerate} 

Moreover, in most studies for parameter identification with biomedical/biomechanical applications, a small set of experimental data is available. This entails the use of a rigorous probabilistic framework for the identification of parameters in biomedical and biomechanical applications. In addition to addressing uncertainties, a Bayesian paradigm naturally regularises an inverse problem using the prior. In an inverse problem with ill-posed\footnote{An ill-posed problem does not meet the following conditions:~(\emph{i}) has a solution; 
(\emph{ii})  the solution is unique; and
(\emph{iii})  the unique solution's behaviour changes continuously with changes of the conditions.
} behaviour a form of regularisation is required to improve this behaviour \cite{A_M_Stuart_2010}. In other words, if a limited number of measurements is available, the prior in Bayesian inference improves the identifiability of the unknown parameters.

Oden and his collaborators have employed BI for identification and model assessment (see Sec.~\ref{subsection: model selection}) of diffuse-interface models for tumor growth \cite{T_J_Oden_2010,A_Hawkins-Daarud_2013, T_J_Oden_2013}. Madireddy et al.~\cite{S_Madireddy_2015,S_Madireddy_2016} employed BI to assess and rank different models that describe the hyperelastic response of soft tissues. The three models that are used in these studies are:~Ogden, Mooney–Rivlin and exponential models. Doraiswamy et al.~\cite{S_Doraiswamy_2016} presented a Bayesian framework to classify tissues (e.g.~healthy and unhealthy tissues). Some other parameter/model identification studies with biomedical/biomechanical applications in which BI is employed are:~Zhao and Pelegri \cite{X_Zhao_2016} for identification of the time constant of a Voigt-based tissue model, Brewick and Teferra \cite{P_T_Brewick_2018} to identify the parameters of one and two-term Ogden models based on testing data from porcine brain tissue, Teferra and Brewick \cite{K_Teferra_2019} to identify the parameters of hyper-viscoelastic model (one-term Ogden model coupled with Maxwell elements) describing the mechanical response of brain tissue in a finite deformation setting, and Zeraatpisheh et al.~\cite{M_Zeraatpisheh_2021} to identify the of neo-Hookean incompressible hyperelastic model based on tensile/compression test data. Zeraatpisheh et al.~\cite{M_Zeraatpisheh_2021} studied the effect of model uncertainty (i.e.~uncertainty due to simplification and idealisations made in mathematical models for physical systems see Ref.~\cite{M_Kennedy_2001}). 

Choosing an appropriate prior can be a critical part of a Bayesian identification framework. The effect of the prior is more significant when a small number of measurements is available (see Ref.~\cite{H_Rappel_2018_2} and \cite{H_Rappel_2017}), which is often the case in biomedical/biomechanical applications. A systematic approach to construct the prior distributions of material parameters for a computational model of the abdominal aortic wall is presented by \citet{S_Seyedsalehi_2015}.      

\paragraph{(Parameter fields)} BI and probability logic have often been employed to model and characterise parameter fields in applications dealing with spatially varying parameters. A common approach utilised in the literature is to model the parameter fields as random fields\footnote{A series of random variables, $\{X_{t}\}$, with $t$ having values in a certain range $T$. If the index set $T$ is continuous, we have a continuous random process \cite{B_Everitt_2010}.} \footnote{A random field is a random (stochastic) process in (Euclidean) space and, using the term ``field'' or``process'' will not change our definitions.} and identify the parameters that characterise these random fields. For example, Koutsourelakis used this approach to identify spatially varying parameters for perfect plasticity model \cite{P_Koutsourelakis_2009}, Rappel et al.~\cite{H_Rappel_2019_2} employed a random field (copula Gaussian process) to model the spatially varying homogenised Young's modulus field of a columnar polycrystalline material, and Savvas et al.~\cite{D_Savvas_2020} have proposed a Bayesian framework to identify the parameters fields of apparent material properties of two-phase composites.

Alternatively, \citet{P_Koutsourelakis_2012} applies BI directly on the finite element (FE) discretisation in which the stiffness tensor (i.e~the spatially varying properties) is constant within each finite element. Hence, the components of the element-wise tensors are the random variables of the posterior (i.e.~the number of dimensions of the posterior scales with the number of finite elements). Vigliotti et al.~\cite{A_Vigliotti_2018}, furthermore, have modelled spatially fluctuating fields of Young's moduli and Poisson's ratios by B-splines. In this work, a Bayesian framework is used to select the order of the B-splines. In the following section we provide a brief description of Bayesian inference.

\subsection{Bayes' theorem}    
\label{subsection:BI}
Let $\textbf{z}$ be the set of $n_{\mathrm{m}}$ measurements and $\textbf{p}$ be the set of  $n_{\mathrm{p}}$ parameters to be identified. According to Bayes' formula, we have:

\begin{equation}
\label{eq:BI}
\pi(\textbf{p}|\textbf{z})=\frac{\pi(\textbf{p})\pi(\textbf{z}|\textbf{p})}{\pi(\textbf{z})}=\frac{1}{\zeta}\pi(\textbf{p})\pi(\textbf{z}|\textbf{p}),
\end{equation}

\noindent where $\pi(\textbf{p})$ denotes the prior probability density function (PDF, prior is a PDF that describes the users' initial knowledge about the parameters, e.g.~Young's modulus cannot be negative), $\pi(\textbf{z}|\textbf{p})$ denotes the likelihood function (i.e.~the PDF that describes the plausibility of the measurement set $\textbf{z}$, for the given set of parameters $\textbf{p}$), $\pi(\textbf{p}|\textbf{z})$ denotes the posterior PDF (i.e.~the PDF that describes the plausibility of parameter set $\textbf{p}$, for the given set of measurements, $\textbf{z}$) and $\pi(\textbf{z})$ is called evidence. After the measurements are made, the value of evidence is considered to be known, and for this reason it equals a constant number ($\pi(\textbf{z})= \zeta \in \mathbb{R}^{+}$). The constant $\zeta$ in Eq.~(\ref{eq:BI}) is calculated by the following integral:

\begin{equation}
\label{eq:zeta}
\zeta=\int\pi(\textbf{p})\pi(\textbf{z}|\textbf{p})d\textbf{p},
\end{equation} 

\noindent with $d\textbf{p}=dp_{1}\cdots dp_{n_{\mathrm{p}}}$. 

Evaluating Eq.~(\ref{eq:zeta}) either analytically or numerically is difficult. This is particularly the case when we are dealing with high dimensional parameter spaces. However, often the unscaled form of Bayes' formula (Eq.~(\ref{eq:unscaled BI})) which gives the shape of the posterior but does not give the exact posterior is considered:

\begin{equation}
\label{eq:unscaled BI}
\pi(\textbf{p}|\textbf{z})\propto\pi(\textbf{p})\pi(\textbf{z}|\textbf{p}).
\end{equation}

This is because often the point in which the posterior is maximum (maximum a posteriori probability, i.e.~MAP estimate) is used as an estimator\footnote{A statistic whose value is used to estimate a parameter based on observed data. For example, the sample mean is an estimator of the population mean.}. MAP estimate defines the most probable set of parameters and is not dependent on $\zeta$. In this case, the posterior is approximated by a Gaussian distribution centred at the MAP (i.e.~Laplace approximation \cite{D_MacKay_2003}). Note that using Eq.~(\ref{eq:unscaled BI}), one can find where the modes (i.e.~any point where $\pi(\textbf{p}|\textbf{z})$ is maximal) are. However, since Eq.~(\ref{eq:unscaled BI}) is not a density, it will not give moments (e.g.~mean and variance) or probabilities.

Computational Bayesian statistics, on the other hand, is based on developing algorithms that simulate samples from the posterior when only the unscaled form of the posterior (i.e.~Eq.~(\ref{eq:unscaled BI})) is known \cite{W_M_Bolstad_2009}. Markov chain Monte Carlo algorithms \cite{W_R_Gilks_1995, S_Brooks_2011} are examples of these approaches that are frequently employed in studies related to Bayesian inference. Once the samples are drawn from the posterior, one can approximate statistical summaries such as mean and variance (or covariance matrix) see Chapter five in Ref.~\cite{D_Calvetti_2007} and Ref.~\cite{H_Rappel_2018_2}.

As an example, once again, consider the knee meniscus problem. For simplicity, we assume that the meniscus is modelled by linear elasticity law. The only parameter which we need to identify here is Young's modulus (assuming nearly incompressibility). Once the unscaled form of the posterior is obtained, as given in Eq.~(\ref{eq:unscaled BI}), we can find the value in which Eq.~(\ref{eq:unscaled BI}) is maximum  (MAP estimate). This parameter value denotes the most plausible value of Young's modulus in light of the data. Another option is to use sampling approaches to approximate the posterior and its statistical summaries.

\subsection{Posterior predictive distribution}
\label{subsection:posterior predictions}
An important application of a Bayesian updating framework is to make predictions about new measurements based on the current measurements. In a Bayesian framework, the information about the unknown parameter set $\textbf{p}$ is contained in the posterior density ($\pi(\textbf{p}|\textbf{z})$), and consequently, predictions about the new measurement $z^{\text{new}}$ can be made by averaging $\pi(z^{\text{new}}|\textbf{p},\textbf{z})$ over the posterior distribution:

\begin{equation}
\label{eq:ppd}
\pi(z^{\text{new}}|\textbf{z})=\int\pi(z^{\text{new}}|\textbf{p},\textbf{z})\pi(\textbf{p}|\textbf{z})d\textbf{p},
\end{equation}

\noindent where $\textbf{z}$ denotes the set of $n_{\mathrm{m}}$ already measured data. If the measurements are independent Eq.~(\ref{eq:ppd}) can be written as follows:

\begin{equation}
\label{eq:ppd2}
\pi(z^{\text{new}}|\textbf{z})=\int\pi(z^{\text{new}}|\textbf{p})\pi(\textbf{p}|\textbf{z})d\textbf{p}.
\end{equation} 

The resulting PDFs in Eqs.~(\ref{eq:ppd}) and~(\ref{eq:ppd2}) are known as the posterior predictive distribution (PPD). A key application of the PPD is model checking (see Chapter six in Ref.~\cite{A_Gelman_2003}). If a model fits properly, the replicated measurements generated using the model and the measured data should look similar. In other words, the measured data should be plausible under the posterior predictive distribution.

Evaluating the above integral is usually difficult especially for high dimensional problems. However, in practice, one can draw simulated values from the PPD as replicated measurements and compare them with the already measured data. If the replicated measurements and the measured data differ substantially, it indicates potential failings of the model. This can be achieved by employing a sampling procedure twice. First, samples are drawn from the posterior distribution for the parameters, given the measurements ($\pi(\textbf{p}|\textbf{z})$). Note that this is already done during the (numerical) analysis of the posterior, and hence, this step does not have to be applied again. Second, the $i~{\text{th}}$ sample drawn in the first step is replaced in $\pi(z^{\text{new}}|\textbf{p}_{i})$ and subsequently is used to generate a sample for new measurement.

\subsection{Model comparison using Bayes factors}
\label{subsection: model selection}
Using Bayes' theorem one can assess the plausibility of several models which are compatible with measurements. Let $\mathcal{M}_{1}$ and $\mathcal{M}_{2}$ be two models that we wish to compare given the measured data set $\textbf{z}$. We would like to compare the plausibility of these two models given a set of measured data (i.e.~$\pi(\mathcal{M}_{1}|\textbf{z})$ and $\pi(\mathcal{M}_{2}|\textbf{z})$). According to Bayes' formula we have:

\begin{equation}
\label{eq:model selection}
\pi(\mathcal{M}_{i}|\textbf{z})=\frac{\pi(\mathcal{M}_{i})\pi(\textbf{z}|\mathcal{M}_{i})}{\pi(\textbf{z})},
\end{equation} 

\noindent where $\pi(\mathcal{M}_{i})$ denotes the user's initial knowledge about model $\mathcal{M}_{i}$ and $\pi(\textbf{z})=\pi(\mathcal{M}_{1})\pi(\textbf{z}|\mathcal{M}_{1})+\pi(\mathcal{M}_{2})\pi(\textbf{z}|\mathcal{M}_{2})$. Furthermore, for an individual model the model likelihood (i.e.~$\pi(\mathcal{M}_{i}|\textbf{z})$) is obtained by taking the expectation of the \emph{data} likelihood with respect to the \emph{parameter} prior:

\begin{equation}
\label{eq:model likelihood}
\pi(\textbf{z}|\mathcal{M}_{i})=\int\pi(\textbf{z}|\textbf{p}_{i},\mathcal{M}_{i})\pi(\textbf{p}_{i}|\mathcal{M}_{i})d\textbf{p}_{i}.
\end{equation}

\noindent Note that $\textbf{p}_{i}$ is the parameter set for the individual model $\mathcal{M}_{i}$. Pairwise comparison of the models is possible by dividing the posteriors in Eq.~(\ref{eq:model likelihood}) for each individual model:

\begin{equation}
\label{eq:ratio}
\frac{\pi(\mathcal{M}_{1}|\textbf{z})}{\pi(\mathcal{M}_{2}|\textbf{z})}=\frac{\pi(\mathcal{M}_{1})}{\pi(\mathcal{M}_{2})} \times \frac{\pi(\textbf{z}|\mathcal{M}_{1})}{\pi(\textbf{z}|\mathcal{M}_{2})}.
\end{equation}

\noindent The second ratio on the right-hand side is known as the Bayes factor. Eq.~(\ref{eq:ratio}) shows how the prior odds are updated through the measurements and the Bayes factor yielding the posterior odds. Eq.~(\ref{eq:ratio}), furthermore, embodies Occam's razor (i.e.~the simplest model/explanation that fits the measurements should be preferred) \cite{D_MacKay_2003}. 

Let, without loss of generality, $\mathcal{M}_{1}$ be a simpler model that explains the data compared to $\mathcal{M}_{2}$. The prior odds $\frac{\pi(\mathcal{M}_{1})}{\pi(\mathcal{M}_{2})}$, make it possible for the user to insert a bias in favour of $\mathcal{M}_{1}$ based on his/her knowledge. The prior bias corresponds to the aesthetic (``A theory with mathematical beauty is
more likely to be correct than an ugly one that fits some experimental data'' (Paul A.M. Dirac)) or empirical motivation for Occam's razor (i.e.~empirically Occam's razor has been successful). However, using a prior bias is not necessary as the Bayes factor automatically embodies Occam's razor. This is because simple models tend to make precise predictions, whereas complex models due to their natural flexibility are capable of making a greater variety of predictions. In other words, if $\mathcal{M}_{2}$ is more complex than $\mathcal{M}_{1}$, its predictive probability $\pi(\textbf{z}|\mathcal{M}_{2})$ spreads substantially wider over the data space than $\pi(\textbf{z}|\mathcal{M}_{1}$). Therefore, if the data are compatible with both models, the simpler model $\mathcal{M}_{1}$ will have a higher likelihood ($\pi(\textbf{z}|\mathcal{M}_{1})$) compared to $\mathcal{M}_{2}$. This means in this case without the need for the user to express any subjective favour for the simpler model, $\mathcal{M}_{1}$ will be more probable than $\mathcal{M}_{2}$.

It is clear from the above discussion that if the user assigns equal priors to all the alternative models which are compatible with the measurements, model assessment can be done only by evaluating Bayes factors. A common interpretation of Bayes factors is provided by Jeffreys \cite{H_Jeffreys_1983}. For further reading on model comparison using BI, readers are referred to Refs.~\cite{D_MacKay_2003} and~\cite{A_Gelman_2003}. Furthermore, a practical guide to Bayesian model selection is provided in Ref.~\cite{H_Chipman_2001}.  

\section{Conclusion}
\label{Discussion}
At present, computational biomechanics represents a rather old field of research; in fact, users and researchers have accumulated a wealth of knowledge and understanding on the behaviour of soft tissues and how to model it.
In this work, we provided a review of the state of the art of the meniscal tissue, highlighting that this tissue is functionally graded and its mechanical and poromechanical properties are region-dependent and anisotropic. We can conclude that a full regional/directional investigation of the properties of this tissue is still lacking and a high variation of values of tissue parameters is found in the literature. This is mainly due to the fact that there are no experimental standards for sample preparation and testing of soft tissue. Furthermore, there is not a consensus on the choice of material models to adopt in order to recover the associated parameters. A number of approaches are put forward in the literature, which we have implemented and compared within a subject-specific finite element model of the knee. We conclude that the choice of the material model has a considerable impact on the results (i.e.~contact pressure in the cartilages and strain patterns). 
Within this work, we discussed how such knowledge and understanding can be used in order to rationally and systematically improve our modelling, simulation and predictive abilities of the behaviour of soft tissue through Bayesian model selection. We further analysed the difficulties associated with uncertainty quantification - both forward and inverse - and chose the knee meniscus as a paradigmatic example to show how the micro/meso-structure of tissue can be used to predict macroscopic behaviour of organs. 

One of the key discussion points we underline is the scarcity of data and inter/intra-subject variability, which calls for the use of Bayesian (as opposed to frequentist) approaches. We believe that rich and fertile grounds lie at the interface between equation-based modelling and data-driven simulations. Biomechanics data are inherently subject-specific, in vitro models are unable to reproduce in vivo conditions, because of (\emph{i}) the impossibility to measure boundary and initial conditions and (\emph{ii}) the state of a given patient and its material properties, which are seen for the first time when surgery is actually performed. 

Two complementary options seem to emerge. The first is to enable multi-scale models of organs through acceleration methods, e.g.~algebraic model order reduction \cite{P_Kerfriden_2010,P_Kerfriden_2011,P_Kerfriden_2012,P_Kerfriden_2013,P_Kerfriden_2013_2,P_Kerfriden_2014,goury2016automatised,K_C_Hoang_2015,K_C_Hoang_2016} or machine learning approaches \cite{S_Vijayaraghavan_2021}, which could enable building up macroscopic complexity from the bottom up.

The second lies in using data ``on the fly'', as it is being acquired (e.g.~through interventional medical imaging) to identify the best model for the patient and the associated parameters, or to learn the behaviour during the intervention itself. Such hybrid data/model approaches appear to be the most viable approach to lay the foundations for the patient-specific optimisation of medical interventions \cite{H_Talbot_2015,I_Peterlik_2017,S_H_Kong_2017,H_P_Bui_2018,C_J_Paulus_2019,A_Mendizabal_2019,S_Nikolaev_2020,A_Mendizabal_2020,N_Haouchine_2020,A_Odot_2021,N_Golse_2021}. 


%
%
\section*{Acknowledgements}
O.B would like to acknowledge the European Union’s Horizon 2020 - EU.1.3.2. - Nurturing excellence by means of cross-border and cross-sector mobility under the Marie Sklodowska-Curie individual fellowship {MSCA-IF-2017}, MetaBioMec, Grant agreement {ID:796405}. The authors would like to thank Mrs. Silvia Bassini for iconographic material. Hussein Rappel was supported by Wave 1 of The UKRI Strategic Priorities Fund under the EPSRC Grant EP/T001569/1 and EPSRC Grant EP/W006022/1, particularly the ``Digital twins for complex systems engineering'' theme within those grants and The Alan Turing Institute. Mark Girolami acknowledges support from the UK Engineering and Physical Sciences Research Council (grant nos.~EP/T000414/1, EP/R018413/2, EP/P020720/2, EP/R034710/1, EP/R004889/1), as well as a Research Chair supported by the Royal Academy of Engineering and Lloyds Register Foundation.
%
%
\clearpage 
\numberwithin{equation}{section}
\numberwithin{figure}{section}
\numberwithin{table}{section}
\appendix
%
%
\section{Fitting of hyperelastic models.}
\label{sec:FitExpDa}
The fitting problem of the hyperelastic models in section~\ref{sec:AnalyRes} is provided in this Appendix. 
We consider the stretch and invariant-based incompressible strain energy functions $\bar{\Psi}$ in section~\ref{subsec:HyperElas}. 
The experimental data are limited to the uniaxial case in which a specimen is taken to align with the circumferential or radial directions of the menisci (see section~\ref{subsec:MenExpda}). 
\\
Consider a pure homogeneous deformation that is assumed to take the following form
\begin{equation}
\Fbf =  {\lambda}_{r} \, {\ebf}_{r}\otimes{\ebf}_{r} + {\lambda}_{\theta} \, {\ebf}_{\theta}\otimes{\ebf}_{\theta} + {\lambda}_{z} \, {\ebf}_{z}\otimes{\ebf}_{z},
\label{eq:DefGra}
\end{equation}
where ${\ebf}_{i}$, $i=r,\theta,z$, denote the principal directions of the test specimen which are associated with the principal stretches ${\lambda}_i$. 
The kinematic of uniaxial deformation in $r$-direction is defined by ${\lambda}_{\theta} = {\lambda}_{z}$ where the stress state for is given by $\sigma_{r} \neq 0$, $\sigma_{\theta} = \sigma_{z} = 0$ and $\sigma_i$ are principal Cauchy stresses. 
Similarly, for uniaxial loading in $\theta$-direction, ${\lambda}_{r} = {\lambda}_{z}$, $\sigma_{\theta} \neq 0$ and $\sigma_{r} = \sigma_{z} = 0$.
\\
The principal Cauchy stress $\sigma_i$ can be determined, from Eq.~(\ref{eq:S0I}) using a standard push-forward operation, in terms of the principal stretches $\lambda_i$ as
\begin{equation}
\sigma_i =  {\lambda}_{i} \, \dfrac{\partial \bar{\Psi}}{\partial \lambda_{i}}+p,
\label{eq:CauchyI}
\end{equation}
where no summation is taken over $i$. For invariant-based strain energy functions, Cauchy stress is given by
\begin{equation}
\sigma_i =  {\lambda}_{i} \, \left[ \dfrac{\partial \bar{\Psi}}{\partial I_{1}} \, \dfrac{\partial I_{1}}{\partial \lambda_{i}} + \dfrac{\partial \bar{\Psi}}{\partial I_{2}} \, \dfrac{\partial I_{2}}{\partial \lambda_{i}}  \right]+p,
\label{eq:CauchyII}
\end{equation}
where $I_{1} = \lambda_{r}^{2}+\lambda_{\theta}^{2}+\lambda_{z}^{2}$ and $I_{2} = \lambda_{r}^{2} \, \lambda_{\theta}^{2} +\lambda_{\theta}^{2} \, \lambda_{z}^{2}+\lambda_{r}^{2} \, \lambda_{z}^{2}$. It should be mentioned that the unknown stretches and hydrostatic pressure can be determined using the boundary conditions.
\\
The isotropic models are fitted to individual experimental data using a nonlinear least-square method of error function:
\begin{equation}
E^2  =  \frac{1}{N_{\mathrm{exp}}} \, \sum\limits^{N_{\mathrm{exp}}}_{j = 1} \left({\sigma}_{i}({\lambda}^{(j)}_{i},\qbf)- \hat{\sigma}^{(j)}_{i}\right)^2
\label{eq:ErrI}
\end{equation}
where $\qbf$ are the unknowns, $\hat{\sigma}^{(j)}_{i}$ are the experimental observations and $N_{\mathrm{exp}}$ is the number of observations. Simultaneous fitting procedure, using a weighted nonlinear least-square method (due to unequal error values of the different data), is adopted in the case anisotropic model (HGO) model. 
The error function is then defined according to,
\begin{equation}
E^2  = \frac{1}{N_{\mathrm{exp}}} \, \Bigg[ w_{r} \, \sum\limits^{N_{\mathrm{exp}}, r}_{i = 1} \left({\sigma}_{r}({\lambda}^{(i)}_{r},\qbf)- \hat{\sigma}^{(i)}_{r}\right)^2 + w_{\theta} \, \sum\limits^{N_{\mathrm{exp}}, \theta}_{i = 1} \left({\sigma}_{\theta}({\lambda}^{(j)}_{\theta},\qbf)- \hat{\sigma}^{(j)}_{\theta}\right)^2 \Bigg]
\label{eq:ErrII}
\end{equation}
where $N_{\mathrm{exp},r}$ and $N_{\mathrm{exp},\theta}$ are the experimental observations for the Cauchy stress components in $r$ and $\theta$-directions, respectively, with $N_{\mathrm{exp}} =N_{\mathrm{exp},r}+N_{\mathrm{exp},\theta}$. 
The fitting weights are $w_{r}=1/\mathrm{max}( \hat{\sigma}^{(i)}_{r})$ and $w_{\theta}=1/\mathrm{max}(\hat{\sigma}^{(j)}_{\theta})$.
%
%
\section{Fitting of uniaxial confined compression experiment.}
\label{sec:ConfComp}

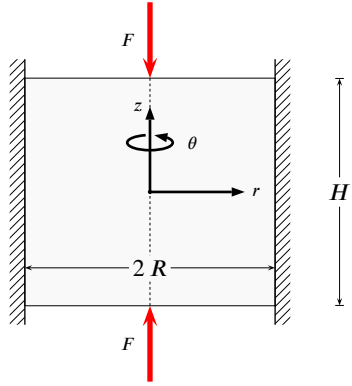
\begin{figure} 
\begin{center}
\begin{pspicture}[showgrid=false](0.0,0.0)(6.0,5.0)
\psset{unit=1.0cm}
\rput[bl]{0.0}(3.0,2.5){
%
%
%
%
\psframe[linewidth=0.25pt,linestyle=solid,linecolor=black,fillstyle=solid,fillcolor=lightlightgray,dimen=middle,framearc=0.0](-1.65,-1.50)(1.65,1.50)
%
\psline[linewidth=0.5pt,linestyle=solid,linecolor=black,dimen=middle]{-}(-1.65,-1.75)(-1.65,1.75)
\psframe[linestyle=none,fillstyle=hlines,hatchsep=2pt,hatchwidth=0.25pt,dimen=middle](-1.85,-1.75)(-1.65,1.75)
\psline[linewidth=0.5pt,linestyle=solid,linecolor=black,dimen=middle]{-}(1.65,-1.75)(1.65,1.75)
\psframe[linestyle=none,fillstyle=hlines,hatchsep=2pt,hatchwidth=0.25pt,dimen=middle](1.65,-1.75)(1.85,1.75)
%
%
\psline[linewidth=2.0pt,linecolor=red,dimen=middle]{->}(0.00,2.5)(0.00,1.5)
\rput[r](-0.2,2.0){\scriptsize $F$}
\psline[linewidth=2.0pt,linecolor=red,dimen=middle]{->}(0.00,-2.5)(0.00,-1.5)
\rput[r](-0.2,-2.0){\scriptsize $F$}
%
\psline[linewidth=0.125pt,linestyle=dashed,dash=1pt 1pt,linecolor=black,dimen=middle]{-}(0.0,-1.5)(0.0,1.5)
%
\psellipticarc[linewidth=1.0pt,linestyle=solid,linecolor=black,dimen=middle,ArrowFill=true,arrowinset=0.0]{->}(0.0,0.650)(0.30,0.1){120}{60}
\rput[l]{0.0}(0.50,0.650){\scriptsize $\theta$}
\psline[linewidth=1.0pt,linestyle=solid,linecolor=black,dimen=middle,ArrowFill=true,arrowinset=0.0]{->}(0.0,0.0)(1.25,0.0)
\rput[l]{0.0}(1.35,0.0){\scriptsize $r$}
\psline[linewidth=1.0pt,linestyle=solid,linecolor=black,dimen=middle,ArrowFill=true,arrowinset=0.0]{->}(0.0,0.0)(0.0,1.125)
\rput[r]{0.0}(-0.10,1.125){\scriptsize $z$}
\qdisk(0.0,0.0){0.80pt} 
%
\psline[linewidth=0.25pt,linestyle=solid]{<->}(-1.65,-1.0)(1.65,-1.0)
\rput(0.0,-1.0){\psframebox[linewidth=0,linestyle=none,framesep=1.5pt,fillstyle=solid,fillcolor=lightlightgray,dimen=middle]{\small $2 \, R$}}
\psline[linewidth=0.25pt,linestyle=solid]{|<->|}(2.5,1.5)(2.5,-1.5)
\rput(2.5,0.0){\psframebox[linewidth=0,linestyle=none,framesep=2.5pt,fillstyle=solid,fillcolor=white,dimen=middle]{\small $H$}}
%
%
}
%
%
\end{pspicture}
\end{center}
\caption{The schematic of a cylindrical specimen subjected to confined compression of a constant axial force $F$ in $z$-direction.}
\label{fig:Cylin}
\end{figure}

Consider a uniaxial confined compression experiment using the cylindrical specimen in Fig.~\ref{fig:Cylin}. The radius and height of the specimen are denoted by $R$ and $H$, respectively, and the cylindrical coordinates $(r,\theta,z)$ are used to describe the reference and deformed configurations. The specimen is subjected to constant axial force $F$ in $z$-direction. It should be mentioned that small deformation is assumed
The kinematics of deformation is defined by the strain components ${\varepsilon}_{z} \neq 0$ and ${\varepsilon}_{r} = {\lambda}_{\theta} = 0$. 
The stress state is given by $\sigma_{z} = F/(\pi \, R^{2})$, $\sigma_{r} = \sigma_{\theta} neq 0$ and other $\sigma_{ij} = 0$. 
The elastic solution yields the stress in $z$-direction as
\begin{equation}
\sigma_{z}  =  \left[ \dfrac{E}{\left( 1-\nu^{2}\right)\left( 1 - 2 \, \nu\right)} \right] \, \varepsilon_{z} =  \dfrac{3 \, K}{\left( 1 - \nu^{2} \right)}  \, \varepsilon_{z},
\label{eq:SigzI}
\end{equation}
where $E$ is Young's modulus, $K$ is the bulk modulus and $\nu$ is Poisson's ratio. 
\\
The linear viscoelastic solution can be determined using the viscoelastic correspondence principle. For simplicity, we assume that the material is incompressible and the change in volume is due to the bulk and shear behaviour ({i.e.} $\nu$ remains elastic whereas $E$ is a function of time). Hence, the viscoelastic solution becomes
\begin{equation}
\varepsilon_{z} \left( t \right) =  J \left( t \right) \, \sigma_{z},
\label{eq:EpszI}
\end{equation}
where $J \left( t \right)$ is the creep compliance that is defined as
\begin{equation}
\dfrac{3 \, \tilde{J} \left( s \right) \, \tilde{K} \left( s \right)}{\left( 1 - \nu^{2} \right)} = \, \tilde{J} \left( s \right) \, \tilde{K}' \left( s \right) = 1,
\label{eq:CreepCompI}
\end{equation}
where $\tilde{f}\left( s \right)$ is $s$-Laplace transform ({i.e.} Carson transform) of a function ${f}\left( t \right)$ and $s$ is Laplace transformation variable. 
Hence, Eq.~(\ref{eq:CreepCompI}) implies that the relaxation modulus $ K' \left( t \right)$ (scaled $K \left( t \right)$) can be determined from inversion of $J \left( t \right)$. 
More specifically, assuming Kelvin model, the compliance function in Eq.~(\ref{eq:DRI}) can be obtained from fitting the creep experimental data, and the relaxation function in Eq.~(\ref{eq:kRI}) can then be determined using the interconversion method in Appendix~\ref{sec:EstVisco}. 
Additionally, the shear relaxation and compliance functions $g_{\mathrm{R}}\left( t \right)$ and $d_{\mathrm{C}}\left( t \right)$ are identical to the bulk functions $k_{\mathrm{R}}\left( t \right)$ and $j_{\mathrm{C}}\left( t \right)$, respectively.
The creep compliance is fitted to creep experimental data using a nonlinear least-square method of error function:
\begin{equation}
E^2  =  \frac{1}{N_{\mathrm{exp}}} \, \sum\limits^{N_{\mathrm{exp}}}_{i = 1} \left(J(t_{i},\pbf)- \hat{J}_{i}\right)^2
\label{eq:ErrIII}
\end{equation}
where $\pbf$ are generalised Voigt (Kelvin) parameters, $\hat{J}_{i}$ are the experimental observations at time $t_{i}$ and $N_{\mathrm{exp}}$ is the number of observations.
%
%
\section{Estimation of linear viscoelastic parameters using interconversion between viscoelastic material functions.}
\label{sec:EstVisco}
Consider a generalised Maxwell model that consists of $M$ Maxwell elements such that each element is a combination of spring and dashpot connected in series. 
The relaxation modulus is
\begin{equation}
E \left( t \right) = E_{\infty}+ \sum\limits_{i = 1}^{M} E_{i} \, e^{-\frac{t}{\tau_{i}}},
\label{eq:ERI}
\end{equation}
where $E_{\infty}$ is the long-term modulus, $E_{i}$ are relaxation strengths and $\tau_{i}$ are the characteristic relaxation times. 
The creep compliance can be written in the form of a generalised Voigt (Kelvin) model of $Q$ Voigt elements ({i.e.} a combination of spring and dashpot connected in parallel) that are connected in series as: 
\begin{equation}
D \left( t \right) = D_{0}+ \sum\limits_{i = 1}^{Q} D_{i} \, \left( 1- e^{-\frac{t}{\eta_{i}}} \right),
\label{eq:DCI}
\end{equation}
where $D_{0}$ is the glassy compliance, $D_{i}$ are retardation strengths and $\eta_{i}$ are the characteristic retardation times. 
It should be noted that the Eqs.~(\ref{eq:ERI}) and~(\ref{eq:DCI}) can be rewritten in the Prony series forms in Sec.~\ref{subsubsec:ViscoElas}.
The relationship between the relaxation and creep modulli can be established as
\begin{equation}
\int\limits_{0}^{t} E \left( t - \tau \right) \, \dfrac{\partial D \left( \tau \right)}{\partial \tau}  \, \mathrm{d} \tau = 1.
\label{eq:ERDCI}
\end{equation}
\\
Following \citet{Park1999}, using Eqs.~(\ref{eq:ERI}),~(\ref{eq:DCI}) and~(\ref{eq:ERDCI}), we can obtain a linear system of algebraic equations that can be solved for $E\left(t \right)$ or $D\left(t \right)$. 
If $D\left(t \right)$ is to be determined from $E\left(t \right)$, the system of equations become:
\begin{equation}
A_{kj} \, D_{j} = B_{k},
\label{eq:SySEqI}
\end{equation}
where
\begin{align}
A_{kj} & = \begin{dcases}
            E_{\infty} \left( 1 - e^{-\frac{t_{k}}{\eta_{j}}} \right) +  \sum\limits_{i = 1}^{M} \dfrac{\tau_{i} \, E_{i}}{\tau_{i}-\eta_{j}} \, \left( e^{-\frac{t_{k}}{\tau_{i}}} - e^{-\frac{t_{k}}{\eta_{j}}} \right), & \text{if} \quad \tau_{i} \neq \eta_{j},  \nonumber \\ 
            E_{\infty} \left( 1 - e^{-\frac{t_{k}}{\eta_{j}}} \right) +  \sum\limits_{i = 1}^{M} \dfrac{t_{k} \, E_{i}}{\tau_{j}} \, e^{-\frac{t_{k}}{\eta_{i}}},      & \text{if} \quad \tau_{i} = \eta_{j},  \nonumber \\ 
           \end{dcases} \\
B_{k} & =  1-D_{0} \left( E_{\infty}+ \sum\limits_{i = 1}^{M} E_{i} \, e^{-\frac{t_{k}}{\tau_{i}}} \right),  \nonumber
\end{align} 
and the glassy compliance is given by
\begin{equation}
D_{0} = \dfrac{1}{E_{\infty}+\sum\limits_{i = 1}^{M} E_{i}}.
\label{eq:D0I}
\end{equation}
If $E\left(t \right)$ is to be determined from $D\left(t \right)$, the system of equations become:
\begin{equation}
A_{ki} \, E_{i} = B_{k},
\label{eq:SySEqII}
\end{equation}
where
\begin{align}
A_{ki} & = \begin{dcases}
            D_{0} \left( e^{-\frac{t_{k}}{\tau_{i}}} -1 \right) +  \sum\limits_{j = 1}^{Q} \dfrac{\eta_{i} \, D_{i}}{\tau_{i}-\eta_{j}} \, \left( e^{-\frac{t_{k}}{\tau_{i}}} - e^{-\frac{t_{k}}{\eta_{j}}} \right), & \text{if} \quad \tau_{i} \neq \eta_{j},  \nonumber \\ 
            D_{0} \left( e^{-\frac{t_{k}}{\tau_{i}}} -1 \right) +  \sum\limits_{j = 1}^{Q} \dfrac{t_{k} \, D_{j}}{\tau_{i}} \, e^{-\frac{t_{k}}{\tau_{j}}},      & \text{if} \quad \tau_{i} = \eta_{j},  \nonumber \\ 
           \end{dcases} \\
B_{k} & =  1-E_{\infty} \left[ D_{0}+ \sum\limits_{j = 1}^{Q} D_{j} \, \left( 1-e^{-\frac{t_{k}}{\eta_{j}}}\right) \right],  \nonumber
\end{align} 
and the equilibrium modulus is given by
\begin{equation}
E_{\infty} = \dfrac{1}{D_{0}+\sum\limits_{i = 1}^{Q} D_{i}}.
\label{eq:EinftyI}
\end{equation}
%
%
%
%
\nocite{*}
%

%
%
\end{document}